\documentclass{article}
\usepackage{arxiv}

\usepackage{amssymb,amsmath,amsbsy,bm}
\usepackage[T1]{fontenc}
\usepackage[utf8]{inputenc}
\usepackage{microtype}
\usepackage[numbers]{natbib}
\usepackage{algorithm}
\usepackage{algpseudocode}
\usepackage{graphicx}
\usepackage{caption}
\usepackage{subfig}
\usepackage[unicode=true]{hyperref}
\hypersetup{breaklinks=true,
            pdfauthor={Ali Siahkoohi, Kamal Aghazade, Ali Gholami},
            pdftitle={Dual-space posterior sampling for Bayesian inference in constrained inverse problems},
            colorlinks=true,
            citecolor=black,
            urlcolor=black,
            linkcolor=black,
            pdfborder={0 0 0}}
\DeclareMathOperator*{\argmin}{arg\,min}
\DeclareMathOperator*{\minimize}{minimize}
\DeclareMathOperator*{\maximize}{maximize}

\title{Dual-space posterior sampling for Bayesian inference in constrained inverse problems}

\author{
  Ali Siahkoohi \\
  Department of Computer Science\\
  University of Central Florida\\
  Orlando, FL, USA \\
  \texttt{alisk@ucf.edu} \\
  \And
  Kamal Aghazade \\
  Institute of Geophysics\\
  Polish Academy of Sciences\\
  Warsaw, Poland \\
  \texttt{aghazade.kamal@igf.edu.pl} \\
  \And
  Ali Gholami \\
  Institute of Geophysics\\
  Polish Academy of Sciences\\
  Warsaw, Poland \\
  \texttt{agholami@igf.edu.pl} \\
}

\date{}

\usepackage{booktabs}
\usepackage{multirow}
\usepackage{makecell}
\usepackage{float}
\usepackage{listings}
\usepackage{upquote}

\begin{document}
\maketitle

\begin{abstract}
Inverse problems constrained by partial differential equations are often
ill-conditioned due to noisy and incomplete measured data or inherent
non-uniqueness of the problem. A prominent example is full waveform
inversion (FWI), which estimates Earth's subsurface properties by fitting
seismic measurements subject to the wave equation, where the
ill-conditioning is inherent to noisy, band-limited, finite-aperture
measured data and the presence of complex geological structures. Casting the inverse
problem into a Bayesian framework allows for a more comprehensive
description of its solution, where instead of a single estimate, a
distribution of plausible solutions---the posterior
distribution---characterizes the non-uniqueness and can be sampled to
quantify uncertainty. However, there is no clear procedure for
translating hard physical constraints, such as the wave equation, into
prior distributions that are amenable to existing sampling techniques.
To address this challenge, we perform posterior sampling in the dual
space using an augmented Lagrangian formulation, which translates hard
constraints into penalties amenable to sampling algorithms, while
enforcing them progressively through the multiplier updates so that they
are satisfied in the limit. We achieve this by seamlessly
integrating the alternating direction method of multipliers (ADMM) with
Stein variational gradient descent (SVGD)---a particle-based sampling
method---where the constraint is relaxed at each iteration and the
multiplier updates progressively enforce constraint satisfaction.
This enables posterior sampling in the presence of hard constraints
while inheriting the favorable conditioning properties of dual-space
solvers, where partial constraint relaxation allows productive updates
even when the current model is far from the true solution. We validate
the method on a stylized Rosenbrock conditional inference problem and on
frequency-domain FWI for a Gaussian anomaly model
and the Marmousi~II benchmark, demonstrating physically consistent
uncertainty estimates and posterior contraction with increasing data coverage.
\end{abstract}

\section{Introduction}\label{sec:introduction}

Quantifying uncertainty in the solution of inverse problems is of fundamental importance in many areas of science and engineering. In geophysical applications such as seismic imaging \citep{VirieuxOperto2009}, the goal is to determine the subsurface structure of the Earth from surface measurements of seismic waves. Due to the inherent ill-posedness of the problem---arising from noisy, band-limited, finite-aperture measured data and the presence of complex geological structures---different models of the subsurface may explain the observed data equally well. Relying on a single estimate of the subsurface properties ignores this non-uniqueness and may lead to overconfident interpretations that fail to capture the true range of plausible solutions.

Bayesian inference provides a principled framework for characterizing this non-uniqueness by representing the solution as a probability distribution---the posterior---conditioned on the observed data \citep{Tarantola2005, Stuart2010}. Sampling from the posterior yields an ensemble of models that collectively describe the range of plausible solutions and enable quantification of uncertainty in derived quantities such as credible intervals and event probabilities. However, generating samples from the posterior in high-dimensional geophysical inverse problems remains a computational challenge \citep{CurtisLomax2001}, as each sample requires evaluating the forward model, which for wave-based problems involves solving a partial differential equation (PDE) \citep{MartinEtAl2012, FichtnerSimute2018, RayEtAl2017, ElyMalcolmPoliannikov2018, ZhaoSen2021}.

A further challenge arises when the inverse problem is naturally posed as a constrained optimization problem---where the PDE relating model parameters to data must be satisfied as a hard constraint. Full waveform inversion (FWI) is a paradigmatic example: the seismic wavefields are governed by the wave equation, and any model estimate must be consistent with this physics.
Wavefield reconstruction methods \citep{vanLeeuwenHerrmann2013, vanLeeuwenHerrmann2016, RizzutiEtAl2021} exploit this structure by reformulating the constrained problem using a penalty approach, in which both the model parameters and the wavefields are treated as independent variables. These methods often yield better-conditioned subproblems and improved convergence behavior compared to conventional reduced-space formulations that enforce the PDE exactly at every iteration. However, achieving accurate satisfaction of the wave equation at convergence remains challenging in purely penalty-based approaches, as large penalty parameters are typically required, which can introduce severe ill-conditioning. In contrast, augmented Lagrangian (AL) formulations \citep{AghamiryGholamiOperto2019, GholamiAghamiryOperto2022,Operto_2023_FWI} alleviate this difficulty by combining penalty terms with Lagrange multipliers, enabling progressive and precise enforcement of the PDE constraint while maintaining improved numerical stability.
These dual approaches yield better conditioned subproblems and faster convergence compared to conventional reduced-space methods that enforce the PDE exactly at each iteration. However, extending these favorable properties from deterministic inversion to Bayesian inference requires a systematic procedure for translating hard PDE constraints into a sampling framework---a procedure that has remained largely unexplored.

The idea of relaxing PDE constraints has proven fruitful in other fields as well---in weather forecasting, weak-constraint four-dimensional variational data assimilation \citep{FisherEtAl2005} similarly relaxes the dynamical model constraint to improve conditioning and enable uncertainty estimation. In the context of seismic inverse problems, \citet{FangEtAl2018} relaxed the PDE constraint by introducing the wavefields as auxiliary variables and showed that the resulting extended posterior can be well-approximated by a Gaussian distribution centered at the maximum a posteriori (MAP) estimate. While this Gaussian approximation enables efficient sampling via the randomize-then-optimize method \citep{BardsleyEtAl2014} and avoids the cost of Markov chain Monte Carlo (MCMC) \citep{Hastings1970}, it has two fundamental limitations: (i) it relies on the posterior being approximately Gaussian, which precludes the characterization of multimodal or strongly non-Gaussian uncertainty that is common in complex geological settings; and (ii) it requires computing the MAP estimate and implicitly constructing the posterior covariance, which couples the sampling quality to the accuracy of a local quadratic approximation around a single mode. Recent variational approaches based on normalizing flows \citep{SiahkoohiEtAl2023, YinOrozcoHerrmann2025} or structured variational families \citep{ZhaoCurtis2024a, ZhaoCurtis2025} address some of these limitations but require either training neural networks or restricting the variational family to a parametric form.

In this work, we address these limitations by proposing ADMM-SVGD, a method that combines the ADMM with Stein variational gradient descent (SVGD) to perform posterior sampling in the dual space. Like \citet{FangEtAl2018}, we relax the PDE constraint and sample in the resulting extended formulation; in contrast to their Gaussian approximation of the posterior, however, we represent it nonparametrically through an ensemble of interacting particles evolved via SVGD, which makes no distributional assumptions and can capture multimodal posteriors, non-Gaussian tails, and complex correlation structures that a Gaussian approximation would miss. The AL formulation admits a natural probabilistic interpretation: for fixed Lagrange multipliers, it defines an unnormalized distribution over the model that evolves as the multipliers are updated, converging to the true posterior as the constraint is progressively enforced.

Standard FWI is typically formulated as a reduced-space problem, in which the state variables---the wavefields---are eliminated by strictly enforcing the wave-equation constraint at every iteration, leaving a search space over the model parameters alone on which the posterior is defined. While this reduction limits the number of optimization variables, the resulting objective function inherits the full nonlinearity of the underlying PDE solve, and the tight coupling between the model and the wavefields leads to the pathological multimodality and severe non-convexity known as the cycle-skipping pathology \citep{VirieuxOperto2009}. In contrast, the AL framework on which ADMM-SVGD is built operates in an extended space---or full space---treating the model parameters, the wavefields, and the Lagrange multipliers as independent optimization variables, so that the wavefield and the model are decoupled and the wavefield may prioritize fitting the observed data in the early stages of the algorithm, even where it remains physically inconsistent with the current model.

This brings about two main advantages for FWI. First, by relaxing the wave equation into a soft constraint, this approach decouples the tight model--wavefield coupling \citep{vanLeeuwenHerrmann2013, FangEtAl2018}; this decoupling is central to the efficacy of the ADMM-SVGD algorithm, as it provides a better-conditioned landscape for posterior sampling by allowing the particle ensemble to explore paths along which the physics is only approximately satisfied in early stages, thereby circumventing the high-energy barriers and spurious local modes of the reduced space. Second, the Lagrange multipliers act as a persistent error-correcting mechanism, improve the conditioning, and decrease the sensitivity to the penalty parameter choice: although the extended space initially facilitates exploration through constraint relaxation, the iterative updates of the dual variables progressively enforce the PDE constraint \citep{AghamiryGholamiOperto2019, GholamiAghamiryOperto2022}, so that the evolving ensemble of target distributions approaches the true posterior and the final samples are both statistically diverse and physically consistent. Because the per-particle and per-source wavefield solves are mutually independent, ADMM-SVGD realizes these advantages within an embarrassingly parallel structure.

To validate the proposed framework, we consider three problems of increasing complexity. First, we consider conditional inference on the Rosenbrock distribution---a stylized nonlinear constrained problem where we compare ADMM-SVGD against reduced-space SVGD (without the ADMM decomposition) to isolate the effect of the constraint splitting. Second, we apply the method to frequency-domain FWI for a Gaussian anomaly model, where ADMM-SVGD reaches a lower model error and data misfit than reduced-space SVGD, and the posterior uncertainty concentrates where the model is hardest to recover. Third, we present results on the Marmousi~II benchmark, showing posterior contraction with increasing data coverage and the emergence of multimodal uncertainty at geologically complex locations.

The remainder of the paper is organized as follows. We begin by formulating the inverse problem and its Bayesian interpretation, then review SVGD as a particle-based sampling method. We next introduce the constrained formulation and its dual-space extension via the AL, derive the ADMM-SVGD algorithm, and discuss convergence considerations. Finally, we present numerical results and conclude with a discussion of the method's strengths, limitations, and directions for future work.

\section{Inverse problems}\label{sec:inverse-problems}

We are concerned with estimating an unknown model $\bm{m}^* \in \mathcal{M}$---often referred to as the unknown---from $N_s$ noisy and indirect observed data $\bm{d} = \{\bm{d}_i\}_{i=1}^{N_s}$ with $\bm{d}_i \in \mathcal{D}$. Here, $\mathcal{M}$ and $\mathcal{D}$ denote the space of unknown models and data, respectively. The physical underlying data generation process is assumed to be encoded in forward modeling operators, $\mathcal{F}_i : \mathcal{M} \to \mathcal{D}$, which relate the unknown model to the observed data via the forward model
\begin{equation}
\bm{d}_i = \mathcal{F}_i(\bm{m}^*) + \bm{n}_i, \quad i = 1, \ldots, N_s.
\label{eq:forward}
\end{equation}
In the above expression, $\bm{n}_i$ is a vector of measurement noise, which might also include errors in the forward modeling operator. Solving ill-posed inverse problems is challenged by noise in the observed data, potential errors in the forward modeling operator, and the intrinsic nontrivial nullspace of the forward operator \citep{Tarantola2005}. These challenges can lead to non-unique solutions where different estimates of the unknown model may fit the observed data equally well.

A common approach to solving inverse problems is to formulate an optimization problem that minimizes the data misfit with an added regularization term:
\begin{equation}
\hat{\bm{m}} = \argmin_{\bm{m}} \frac{1}{2} \sum_{i=1}^{N_s} \|\bm{d}_i - \mathcal{F}_i(\bm{m})\|_2^2 + \mathcal{R}(\bm{m}),
\label{eq:deterministic-inverse}
\end{equation}
where $\mathcal{R}(\bm{m})$ is a regularization term that incorporates prior knowledge about the model, e.g., smoothness or sparsity. While this deterministic approach yields a single point estimate $\hat{\bm{m}}$, it ignores the intrinsic variability within inverse problem solutions. Under such conditions, the use of a single model estimate increases the risk of overfitting the data. Therefore, not only does solving inverse problems require regularization, but it also calls for a statistical inference framework that allows us to characterize the variability among the solutions by quantifying the solution uncertainty.

\section{Bayesian inference for inverse problems}\label{sec:bayesian-inverse-problems}

To systematically quantify the uncertainty, we cast the inverse problem into a Bayesian framework \citep{Tarantola2005, KaipioSomersalo2006, Stuart2010}. In this framework, instead of having a single estimate of the unknown, the solution is characterized by a probability distribution over the solution space $\mathcal{M}$ that is conditioned on data---namely the posterior distribution \citep{MalinvernoBriggs2004}. This conditional distribution, denoted by $p_\text{post}(\bm{m} | \bm{d})$, can according to Bayes' rule be written as follows:
\begin{equation}
p_\text{post}(\bm{m} | \bm{d}) = \frac{p_\text{like}(\bm{d} | \bm{m}) \, p_\text{prior}(\bm{m})}{p_\text{data}(\bm{d})},
\label{eq:bayes}
\end{equation}
where $p_\text{like}(\bm{d} | \bm{m})$ is the likelihood function, which quantifies how well the predicted data fits the observed data given the PDF of the noise distribution; $p_\text{prior}(\bm{m})$ is the prior distribution, which encodes prior beliefs on the unknown and can also be interpreted as a regularizer for the inverse problem; and $p_\text{data}(\bm{d})$ denotes the data PDF, which is a normalization constant that is independent of $\bm{m}$.

Under the assumption that the observed data $\bm{d} = \{\bm{d}_i\}_{i=1}^{N_s}$ are independent conditioned on the unknown model $\bm{m}$, and assuming Gaussian noise $\bm{n}_i \sim \mathcal{N}(\bm{0}, \sigma^2 \bm{I})$, the negative log-posterior can be expressed as
\begin{align}
-\log p_\text{post}(\bm{m} | \bm{d}) &= -\sum_{i=1}^{N_s} \log p_\text{like}(\bm{d}_i | \bm{m}) - \log p_\text{prior}(\bm{m}) + \log p_\text{data}(\bm{d}) \nonumber \\
&= \frac{1}{2\sigma^2} \sum_{i=1}^{N_s} \|\bm{d}_i - \mathcal{F}_i(\bm{m})\|_2^2 - \log p_\text{prior}(\bm{m}) + \text{const}.
\label{eq:neg-log-posterior}
\end{align}

In equations~\ref{eq:bayes} and~\ref{eq:neg-log-posterior}, the likelihood function $p_\text{like}(\bm{d}_i | \bm{m})$ quantifies how well the predicted data (equation~\ref{eq:forward}) fits the observed data given the PDF of the noise distribution. The prior distribution $p_\text{prior}(\bm{m})$ encodes prior beliefs on the unknown model. Comparing equations~\ref{eq:deterministic-inverse} and~\ref{eq:neg-log-posterior}, we observe that the regularization term $\mathcal{R}(\bm{m})$ in the deterministic formulation corresponds to the negative log-prior $-\log p_\text{prior}(\bm{m})$ in the Bayesian formulation---providing a probabilistic interpretation of regularization.

\subsection{Estimation with Bayesian inference}\label{sec:bayesian-estimation}

Acquiring statistical information regarding the variability in the solutions of the inverse problem requires computing expectations with respect to the posterior distribution. For most applications, the posterior PDF itself is not directly of interest; rather, we seek to evaluate expectations of some function $\psi(\bm{m})$ over the posterior:
\begin{equation}
\mathbb{E}_{\bm{m} \sim p_\text{post}(\bm{m}|\bm{d})}[\psi(\bm{m})] = \int \psi(\bm{m}) \, p_\text{post}(\bm{m} | \bm{d}) \, d\bm{m}.
\label{eq:expectation}
\end{equation}
The choice of $\psi$ determines what information is extracted from the posterior. For instance, setting $\psi(\bm{m}) = \bm{m}$ yields the posterior mean $\mathbb{E}[\bm{m}] = \int \bm{m} \, p_\text{post}(\bm{m} | \bm{d}) \, d\bm{m}$, which averages over all plausible models weighted by their posterior probability. Setting $\psi(\bm{m}) = (\bm{m} - \mathbb{E}[\bm{m}]) \circ (\bm{m} - \mathbb{E}[\bm{m}])$, where $\circ$ denotes elementwise (Hadamard) multiplication, provides pointwise variance estimates that characterize the spread among different models explaining the observed data. Other choices of $\psi$ include indicator functions $\psi(\bm{m}) = \mathbf{1}_{\mathcal{A}}(\bm{m})$, which allow the estimation of probabilities of events $\mathcal{A}$---e.g., the probability that the model exceeds a threshold at a given location---a quantity of particular relevance for risk assessment in geophysical applications.

Since the posterior distribution is generally intractable and the integral in equation~\ref{eq:expectation} cannot be evaluated analytically, these expectations must be approximated using Monte Carlo estimation \citep{RobertCasella2004, MosegaardTarantola1995}. Given $N_p$ samples $\{\bm{m}^{(j)}\}_{j=1}^{N_p}$ drawn from the posterior $p_\text{post}(\bm{m} | \bm{d})$, the expectation of $\psi(\bm{m})$ can be approximated as follows:
\begin{equation}
\mathbb{E}_{\bm{m} \sim p_\text{post}(\bm{m}|\bm{d})}[\psi(\bm{m})] \approx \frac{1}{N_p} \sum_{j=1}^{N_p} \psi(\bm{m}^{(j)}).
\label{eq:monte-carlo}
\end{equation}
This approximation converges to the true expectation as $N_p \to \infty$, regardless of the dimensionality of the problem. However, obtaining samples from the posterior is itself a challenging task, particularly in high-dimensional inverse problems where each evaluation of the forward operator is computationally expensive. Before discussing how to generate these samples, we briefly review key quantities of interest that can be computed from posterior samples.

\subsubsection{Conditional mean estimation}

The conditional mean (CM), also known as the posterior mean, is obtained by setting $\psi(\bm{m}) = \bm{m}$ in equation~\ref{eq:monte-carlo}:
\begin{equation}
\bm{m}_\text{CM} = \mathbb{E}_{\bm{m} \sim p_\text{post}(\bm{m}|\bm{d})}[\bm{m}] \approx \frac{1}{N_p} \sum_{j=1}^{N_p} \bm{m}^{(j)}.
\label{eq:conditional-mean}
\end{equation}
The conditional mean corresponds to the minimum-variance estimate \citep{AndersonMoore1979} and is generally less prone to overfitting \citep{MacKay2003} compared to point estimates obtained by optimization. Consequently, when the data are severely contaminated by noise, the conditional mean can provide enhanced robustness \citep{SiahkoohiRizzutiHerrmann2020a, SiahkoohiRizzutiHerrmann2020b, SiahkoohiRizzutiHerrmann2020c}. In contrast, single-mode estimates may become trapped in local minima or concentrate on narrow, noise-driven modes, whereas the conditional mean integrates over the full probability mass to yield a more stable and statistically representative solution.

\subsubsection{Pointwise standard deviation}

In its most rudimentary form, uncertainties in the model estimate can be assessed by computing the pointwise standard deviation, which expresses the spread among the different unknown models explaining the observed data. Setting $\psi(\bm{m}) = (\bm{m} - \bm{m}_\text{CM}) \circ (\bm{m} - \bm{m}_\text{CM})$ in equation~\ref{eq:monte-carlo}, this quantity can be computed via:
\begin{equation}
\bm{\sigma}_\text{post}^2 = \mathbb{E}_{\bm{m} \sim p_\text{post}(\bm{m}|\bm{d})}\left[ (\bm{m} - \bm{m}_\text{CM}) \circ (\bm{m} - \bm{m}_\text{CM}) \right] \approx \frac{1}{N_p} \sum_{j=1}^{N_p} (\bm{m}^{(j)} - \bm{m}_\text{CM}) \circ (\bm{m}^{(j)} - \bm{m}_\text{CM}),
\label{eq:std}
\end{equation}
The pointwise standard deviation $\bm{\sigma}_\text{post}$ provides a spatially resolved measure of uncertainty: regions where the data provide strong constraints on the model will exhibit small standard deviations, whereas poorly constrained regions---such as areas with limited illumination or complex geology---will show larger values.

\subsubsection{Credible intervals}

As described above, the pointwise standard deviation summarizes the spread among the likely estimates of the unknown. By means of this quantity, we can assign probabilities---i.e., credible levels---to the unknown being in a certain interval. The interval is obtained by treating the pointwise posterior distribution as approximately Gaussian with mean $\bm{m}_\text{CM}$ and standard deviation $\bm{\sigma}_\text{post}$. Given a desired credible level---e.g., $99\%$---the credible interval is $\bm{m}_\text{CM} \pm 2.576 \, \bm{\sigma}_\text{post}$, where $99\%$ of samples fall between the left ($\bm{m}_\text{CM} - 2.576 \, \bm{\sigma}_\text{post}$) and right ($\bm{m}_\text{CM} + 2.576 \, \bm{\sigma}_\text{post}$) tails of the Gaussian distribution. We emphasize that this Gaussian treatment is a pointwise summary statistic used for visualization; it places no distributional assumption on the joint posterior, which is represented nonparametrically by the particle ensemble and may be multimodal and non-Gaussian. In the results that follow we do not in fact rely on this summary: the profile figures display the full pointwise marginal density directly, so that any departure from Gaussianity is visible rather than assumed away.

Computing any of these quantities requires drawing samples from the posterior. In the next section, we describe SVGD, a particle-based method that generates approximate posterior samples by evolving an ensemble of models toward the target distribution.

\section{Stein variational gradient descent}\label{sec:svgd}

SVGD is a particle-based deterministic method for approximate Bayesian inference \citep{LiuWang2016}. Rather than sampling sequentially as in MCMC methods \citep{Hastings1970}, SVGD evolves an ensemble of $N_p$ particles $\{\bm{m}^{(j)}\}_{j=1}^{N_p}$ to approximate the target posterior distribution. This approach offers several practical advantages over MCMC for geophysical inverse problems \citep[e.g.,][]{ZhangCurtis2020, ZhaoCurtis2024a,Corrales_2025_ASV}: (i) the particles can be evolved in parallel; (ii) there is no burn-in period; and (iii) the method is deterministic given the initial particle positions.

The particles are updated iteratively according to:
\begin{equation}
\bm{m}^{(j)} \leftarrow \bm{m}^{(j)} + \eta \, \bm{\phi}(\bm{m}^{(j)}),
\label{eq:svgd-update}
\end{equation}
where $\eta$ is the step size and the optimal perturbation direction is:
\begin{equation}
\bm{\phi}(\bm{m}^{(j)}) = \frac{1}{N_p} \sum_{i=1}^{N_p} \left[ K(\bm{m}^{(i)}, \bm{m}^{(j)}) \nabla_{\bm{m}^{(i)}} \log p_\text{post}(\bm{m}^{(i)} | \bm{d}) + \nabla_{\bm{m}^{(i)}} K(\bm{m}^{(i)}, \bm{m}^{(j)}) \right].
\label{eq:svgd-phi}
\end{equation}

Here, $K(\cdot, \cdot)$ is a positive definite kernel, commonly chosen as the radial basis function (RBF) kernel:
\begin{equation}
K(\bm{m}, \bm{m}') = \exp\left( -\frac{\|\bm{m} - \bm{m}'\|^2}{2h^2} \right),
\label{eq:rbf-kernel}
\end{equation}
with bandwidth $h$ typically set using the median heuristic \citep{LiuWang2016}:
\begin{equation}
h = \frac{\text{median}\left(\left\{\|\bm{m}^{(i)} - \bm{m}^{(j)}\|_2 : 1 \leq i < j \leq N_p\right\}\right)}{\sqrt{2 \log N_p}},
\label{eq:median-heuristic}
\end{equation}
The
update in equation~\ref{eq:svgd-phi} consists of two terms that
together balance exploration and exploitation \citep{LiuWang2016}. The first term drives
particles toward regions of high posterior probability through the kernel-weighted gradient of the log-posterior---this term
is responsible for fitting the observed data and respecting the prior.
The second term acts as a repulsive force between nearby particles,
maintaining diversity among the ensemble and preventing all particles
from collapsing to a single mode of the posterior. This repulsion is
essential for Uncertainty Quantification (UQ), as it ensures that the
particle ensemble captures the spread of the posterior rather than
concentrating around a single point estimate. The bandwidth $h$ of the
kernel controls the range of interaction between particles: a larger
bandwidth encourages global exploration, while a smaller bandwidth
allows particles to resolve finer structure in the posterior. The median
heuristic provides an adaptive choice that scales with the current
spread of the ensemble. The step size $\eta$, by contrast, is held fixed
throughout the iterations and selected empirically in the stylized example that
follows; for the FWI experiments the Stein direction is
instead rescaled per particle, as described below. Neither setting employs a line
search or a step-size decay schedule.

SVGD requires only the gradient of the log-posterior, $\nabla_{\bm{m}}
\log p_\text{post}(\bm{m} | \bm{d})$, which can be computed using
adjoint methods in many geophysical applications \citep{Plessix2006, VirieuxOperto2009}. Additionally, the
particle-based nature of SVGD makes it particularly well-suited for
integration with iterative optimization schemes such as the ADMM. As we demonstrate below, SVGD can be seamlessly combined with
dual-space methods to handle constrained inverse problems, where the
physics constraints are incorporated through a Lagrangian formulation
rather than being enforced exactly at each iteration. The algorithm is
summarized in Algorithm~\ref{alg:svgd}, originally proposed by \citet{LiuWang2016}.

\begin{algorithm}[H]
\caption{Stein Variational Gradient Descent (SVGD)}
\label{alg:svgd}
\begin{algorithmic}[1]
\State \textbf{Input:} Target posterior $p_\text{post}(\bm{m} | \bm{d})$, number of particles $N_p$, step size $\eta$
\State \textbf{Initialize:} Sample $\bm{m}^{(j)} \sim p_\text{prior}(\bm{m})$ for $j = 1, \ldots, N_p$
\While{not converged}
    \For{$j = 1, \ldots, N_p$}
        \State Compute gradient: $\bm{g}^{(j)} = \nabla_{\bm{m}} \log p_\text{post}(\bm{m}^{(j)} | \bm{d})$
    \EndFor
    \State Compute kernel bandwidth $h$ via equation~\ref{eq:median-heuristic}
    \For{$j = 1, \ldots, N_p$}
        \State $\bm{\phi}^{(j)} \leftarrow \frac{1}{N_p} \sum_{i=1}^{N_p} \left[ K(\bm{m}^{(i)}, \bm{m}^{(j)}) \bm{g}^{(i)} + \nabla_{\bm{m}^{(i)}} K(\bm{m}^{(i)}, \bm{m}^{(j)}) \right]$
    \EndFor
    \For{$j = 1, \ldots, N_p$}
        \State $\bm{m}^{(j)} \leftarrow \bm{m}^{(j)} + \eta \, \bm{\phi}^{(j)}$
    \EndFor
\EndWhile
\State \textbf{Output:} Posterior samples $\{\bm{m}^{(j)}\}_{j=1}^{N_p}$
\end{algorithmic}
\end{algorithm}

\section{Constrained inverse problems}\label{sec:constrained-inverse}

In many inverse problems, the forward operator $\mathcal{F}_i$ is not given in closed form but is instead implicitly defined through a physical constraint that must be solved numerically. FWI is a prominent example \citep{Pratt1999, VirieuxOperto2009}: the forward model requires solving the wave equation for each source experiment. Recognizing this structure, the inverse problem can be cast as a constrained optimization problem \citep{Haber_2000_OTS}. For a medium characterized by the squared slowness $\bm{m} \in \mathbb{R}^N$, the seismic wavefields $\bm{u}_i \in \mathbb{C}^N$, and observed data $\bm{d}_i \in \mathbb{C}^{N_r}$, the constrained formulation reads:
\begin{equation}
\minimize_{\bm{m}, \{\bm{u}_i\}_{i=1}^{N_s}}~ \frac{1}{2\sigma^2} \sum_{i=1}^{N_s} \|\bm{P}\bm{u}_i - \bm{d}_i\|_2^2 \quad \text{subject to} \quad \bm{A}(\bm{m}) \bm{u}_i = \bm{b}_i, \quad i = 1, \ldots, N_s.
\label{eq:constrained-fwi}
\end{equation}
Here, the wavefields $\{\bm{u}_i\}_{i=1}^{N_s}$ are treated as independent optimization variables alongside the model $\bm{m}$, with the wave equation imposed as an explicit constraint rather than being solved to eliminate them. The objective measures the weighted data misfit between observed data $\bm{d}_i$ and predicted data at receiver locations, where $\bm{P} \in \mathbb{R}^{N_r \times N}$ is the receiver sampling operator ($N_r \ll N$) and $\sigma^2$ is the noise variance. The constraints $\bm{A}(\bm{m}) \bm{u}_i = \bm{b}_i$ encode the wave physics: $\bm{A}(\bm{m}) = (\omega^2 \text{diag}(\bm{m}) + \Delta) \in \mathbb{C}^{N \times N}$ is the Helmholtz operator at angular frequency $\omega$, $\Delta$ is the Laplacian with appropriate boundary conditions, and $\bm{b}_i \in \mathbb{C}^N$ is the source term for the $i$-th experiment. By eliminating $\bm{u}_i$ via the constraint, the forward operator from equation~\ref{eq:forward} takes the form $\mathcal{F}_i(\bm{m}) = \bm{P} \bm{A}(\bm{m})^{-1} \bm{b}_i$, which recovers the standard reduced formulation presented in equation~\ref{eq:deterministic-inverse}.

Viewing FWI as a constrained problem---rather than eliminating the constraint upfront---opens the door to solvers that treat both the model $\bm{m}$ and the wavefields $\{\bm{u}_i\}_{i=1}^{N_s}$ as independent variables. In this formulation, the wave equation need not be satisfied exactly at each iteration; instead, it is progressively enforced over the course of the iterations. This relaxation is advantageous for two reasons. First, it breaks the tight coupling between model and wavefield variables, significantly reducing the ill-conditioning that plagues the reduced-space approach where $\bm{u}_i = \bm{A}(\bm{m})^{-1}\bm{b}_i$ is enforced at every step. Second, it enables iterative solvers---such as the ADMM \citep{AghamiryGholamiOperto2019}---that approximately satisfy the constraint in early iterations and progressively dial in accuracy, yielding better conditioned subproblems and faster overall convergence.

\subsection{Challenges for Bayesian inference in constrained inverse problems}\label{sec:bayesian-challenges}

While the constrained formulation~\ref{eq:constrained-fwi} offers clear benefits for deterministic inversion, the core challenge is that it is unclear how to systematically incorporate such constraints into a Bayesian inference framework. The standard Bayesian approach of equation~\ref{eq:neg-log-posterior} operates exclusively in the reduced space, evaluating $\mathcal{F}_i(\bm{m}) = \bm{P}\bm{A}(\bm{m})^{-1}\bm{b}_i$ and implicitly enforcing exact constraint satisfaction at every posterior evaluation. Critically, this reduced-space formulation produces a posterior distribution that is highly nonlinear in the model parameters $\bm{m}$, because the forward map $\mathcal{F}_i$ inherits the full nonlinearity of the PDE solve. As demonstrated by \citet{FangEtAl2018}, this nonlinearity manifests as spurious local modes in the negative log-posterior landscape---modes that do not correspond to physically meaningful solutions but arise as artifacts of the tight coupling between model and wavefields in the reduced formulation. These spurious modes severely hinder both optimization (which may converge to a local minimum rather than the global MAP estimate) and sampling (which requires either traversing high-energy barriers between modes or running impractically long Markov chains).

A common strategy for handling the constraint is to relax it into a soft penalty \citep{vanLeeuwenHerrmann2013, vanLeeuwenHerrmann2016}, replacing the wave equation by an additive term $\frac{\mu}{2}\|\bm{A}(\bm{m})\bm{u}_i - \bm{b}_i\|_2^2$ in the objective. As shown by \citet{FangEtAl2018}, for an appropriate choice of the penalty parameter, the resulting extended posterior has fewer modes than the reduced formulation, because the relaxation breaks the tight nonlinear coupling between $\bm{m}$ and the data. However, the accuracy of the solution depends critically on the penalty parameter $\mu$: too small, and the wave equation is not sufficiently enforced; too large, and the problem recovers the multimodal landscape of the reduced formulation. Crucially, for the posterior samples to be physically meaningful, the constraint must eventually be satisfied to high accuracy---a soft penalty alone offers no guarantee of this.

To reap the benefits of the full-space formulation, we propose a dual-space approach based on the AL. By introducing Lagrange multipliers (dual variables), we derive a formulation that naturally balances data fidelity, prior information, and constraint enforcement, enabling posterior sampling without requiring exact constraint satisfaction at each step. The key advantage over a fixed penalty is that the multiplier updates provide a principled mechanism for progressively tightening the constraint, so that the extended posterior converges to the true posterior regardless of the initial penalty choice.

\section{Dual-space Bayesian inference for constrained inverse problems}\label{sec:dual-space}

To enable flexible Bayesian inference for the constrained problem (equation~\ref{eq:constrained-fwi}), we introduce a dual-space formulation based on the AL method.

\subsection{Lagrangian formulation}\label{sec:lagrangian}

The well-established method of Lagrange multipliers can be used to solve
the constrained inverse problem presented in equation~\ref{eq:constrained-fwi}.
This approach involves optimization of the Lagrangian function, defined
as follows:
\begin{equation}
\mathcal{L}(\bm{m}, \bm{u}, \bm{v}) = \frac{1}{2\sigma^2} \sum_{i=1}^{N_s} \|\bm{P}\bm{u}_i - \bm{d}_i\|_2^2 + \sum_{i=1}^{N_s} \text{Re}\langle \bm{v}_i, \bm{A}(\bm{m})\bm{u}_i - \bm{b}_i \rangle,
\label{eq:lagrangian}
\end{equation}
where $\bm{v}_i \in \mathbb{C}^{N}$ denotes the Lagrange multipliers associated with the wave equation constraint for source $i$, and $\langle \cdot, \cdot \rangle$ denotes the Hermitian inner product, of which only the real part enters the real-valued Lagrangian. The Lagrange multipliers play the role of adjoint wavefields: they encode the sensitivity of the objective to violations of the wave equation and provide the mechanism through which the physics constraints influence the optimization. The optimum point of the Lagrangian is a saddle point which attains a minimum over the primal variables $\bm{m}$ and $\bm{u}_i$ and a maximum over the dual variables $\bm{v}_i$.

\subsection{Augmented Lagrangian formulation}\label{sec:augmented-lagrangian}

While the standard Lagrangian provides the correct solution at its saddle point, it does not penalize constraint violations during the iterative process, which can lead to ill-conditioning. The AL method \citep{AghamiryGholamiOperto2019, GholamiAghamiryOperto2022, AghazadeGholami2025} addresses this by adding a quadratic penalty term that penalizes deviations from the wave equation:
\begin{equation}
\mathcal{L}_\mu(\bm{m}, \bm{u}, \bm{v}) = \frac{1}{2\sigma^2} \sum_{i=1}^{N_s} \|\bm{P}\bm{u}_i - \bm{d}_i\|_2^2 + \sum_{i=1}^{N_s} \text{Re}\langle \bm{v}_i, \bm{A}(\bm{m})\bm{u}_i - \bm{b}_i \rangle + \frac{\mu}{2} \sum_{i=1}^{N_s} \|\bm{A}(\bm{m})\bm{u}_i - \bm{b}_i\|_2^2,
\label{eq:augmented-lagrangian}
\end{equation}
where $\mu > 0$ is the penalty parameter. The solution to the constrained problem~\ref{eq:constrained-fwi} corresponds to a saddle point:
\begin{equation}
\minimize_{\bm{m}, \bm{u}} \maximize_{\bm{v}} \mathcal{L}_\mu(\bm{m}, \bm{u}, \bm{v}).
\label{eq:saddle-point}
\end{equation}

By introducing the scaled multiplier $\bm{\varepsilon}_i = \frac{1}{\mu}\bm{v}_i$
the AL can be rewritten in a more compact form:
\begin{equation}
\mathcal{L}_\mu(\bm{m}, \bm{u}, \bm{\varepsilon}) = \frac{1}{2\sigma^2} \sum_{i=1}^{N_s} \|\bm{P}\bm{u}_i - \bm{d}_i\|_2^2 + \frac{\mu}{2} \sum_{i=1}^{N_s} \|\bm{A}(\bm{m})\bm{u}_i - \bm{b}_i + \bm{\varepsilon}_i\|_2^2 - \frac{\mu}{2} \sum_{i=1}^{N_s} \|\bm{\varepsilon}_i\|_2^2.
\label{eq:augmented-lagrangian-scaled}
\end{equation}

This follows by completing the square in the multiplier and quadratic-penalty terms of equation~\ref{eq:augmented-lagrangian}: writing $\bm{r}_i = \bm{A}(\bm{m})\bm{u}_i - \bm{b}_i$ and substituting $\bm{v}_i = \mu\bm{\varepsilon}_i$, the identity $\text{Re}\langle \bm{v}_i, \bm{r}_i \rangle + \frac{\mu}{2}\|\bm{r}_i\|_2^2 = \frac{\mu}{2}\|\bm{r}_i + \bm{\varepsilon}_i\|_2^2 - \frac{\mu}{2}\|\bm{\varepsilon}_i\|_2^2$ holds for each source $i$, and summing over $i$ yields equation~\ref{eq:augmented-lagrangian-scaled} \citep{BoydADMM2011}.

This formulation provides several benefits over the reduced-space approach: (i) the quadratic penalty term adds curvature to the objective and enlarges the search space, improving the conditioning and convergence behavior relative to the standard Lagrangian or the reduced-space approach \citep{vanLeeuwenHerrmann2013, vanLeeuwenHerrmann2016}, as we demonstrate at FWI scale in the Marmousi~II example; (ii) the problem decomposes into subproblems over the model, wavefields, and multipliers that can be solved iteratively via the ADMM \citep{BoydADMM2011, AghamiryGholamiOperto2019, AghazadeGholami2025}, avoiding the need for simultaneous updates of all variables; and (iii) the wave equation constraint does not need to be satisfied exactly at each iteration---only in the limit of convergence---which allows the algorithm to take productive steps even when the current model is far from the true solution, as the evolution of the multiplier and constraint-residual norms reported for the Marmousi~II example makes visible.

\subsection{Probabilistic interpretation}\label{sec:probabilistic-interpretation}

A central advantage of the AL formulation is that it admits a natural probabilistic interpretation, which provides the bridge between constrained optimization and Bayesian inference. The idea is to treat the AL as a negative log-density---for fixed auxiliary variables $\bm{u}$ and multipliers $\bm{\varepsilon}$, the AL defines an unnormalized probability distribution over $\bm{m}$:
\begin{equation}
p(\bm{m} | \bm{d}, \bm{\varepsilon}) \propto \exp\left( -\mathcal{L}_\mu(\bm{m}, \bm{u}, \bm{\varepsilon}) \right) \, p_\text{prior}(\bm{m}),
\label{eq:augmented-posterior}
\end{equation}
where $\bm{u}$ is obtained by solving the optimality conditions of the AL given $(\bm{m}, \bm{\varepsilon})$. Crucially, this distribution over $\bm{m}$ depends on the current multipliers $\bm{\varepsilon}$, and therefore evolves as these are updated. Denoting the multipliers at iteration $k$ by $[\bm{\varepsilon}]_k$, the algorithm generates a sequence of target distributions
\begin{equation}
p_k(\bm{m} | \bm{d}) \;:=\; p(\bm{m} | \bm{d}, [\bm{\varepsilon}]_k) \;\;\xrightarrow{\;k \to \infty\;}\;\; p_\text{post}(\bm{m} | \bm{d}).
\label{eq:evolving-posterior}
\end{equation}
At each iteration, the SVGD update advances the model particles along the Stein transport direction toward the current target $p_k(\bm{m} | \bm{d})$; the auxiliary variables and multipliers are then updated, refining the target distribution for the next iteration. Because only a single transport step is taken per target, the ensemble tracks this slowly drifting sequence rather than equilibrating with each intermediate distribution, as we discuss following the algorithm description. As the multipliers converge to their optimal values and the constraint residuals vanish, the evolving distribution $p_k$ converges to the true posterior, and the model particles become approximate samples from it.

Taking the negative logarithm and using equation~\ref{eq:augmented-lagrangian}, we obtain
\begin{align}
-\log p(\bm{m} | \bm{d}, \bm{v}) &= \mathcal{L}_\mu(\bm{m}, \bm{u}, \bm{v}) - \log p_\text{prior}(\bm{m}) + \text{const} \nonumber \\
&= \underbrace{\frac{1}{2\sigma^2} \sum_{i=1}^{N_s} \|\bm{P}\bm{u}_i - \bm{d}_i\|_2^2}_{\text{data fidelity term}}
\;\underbrace{\vphantom{\frac{1}{2\sigma^2}}+ \sum_{i=1}^{N_s} \text{Re}\langle \bm{v}_i, \bm{A}(\bm{m})\bm{u}_i - \bm{b}_i \rangle}_{\text{multiplier term}} \nonumber \\
&\quad +\; \underbrace{\frac{\mu}{2} \sum_{i=1}^{N_s} \|\bm{A}(\bm{m})\bm{u}_i - \bm{b}_i\|_2^2}_{\text{penalty term}}
\;\underbrace{\vphantom{\frac{\mu}{2}}- \log p_\text{prior}(\bm{m})}_{\text{log-prior}}.
\label{eq:neg-log-augmented-posterior}
\end{align}

Each term in the above expression plays a distinct role. The data fidelity term measures the misfit between the predicted wavefield at receiver locations and the observed data---under the Gaussian noise assumption, this term corresponds to the negative log-likelihood $-\log p_\text{like}(\bm{d} | \bm{m})$. The linear multiplier term encodes accumulated information from the Lagrange multipliers---i.e., the adjoint wavefields---that steer the solution toward satisfying the wave equation; this term evolves over iterations as the multipliers are updated. The penalty term penalizes violations of the wave equation $\bm{A}(\bm{m})\bm{u} = \bm{b}$ through a quadratic term scaled by the penalty parameter $\mu$, where larger values of $\mu$ enforce the constraint more strictly but may lead to ill-conditioning. Finally, the negative log-prior incorporates prior knowledge about the model, e.g., smoothness or spatial correlation, regularizing the solution in a manner analogous to the deterministic setting \citep[see][ equation~\ref{eq:deterministic-inverse}]{EsserEtAl2018}.

By employing a moderate penalty parameter $\mu$, the resulting subproblems remain better conditioned, while the iterative multiplier updates progressively enforce the PDE constraint and steer the evolving distribution toward the true posterior. In early iterations, the auxiliary variables and multipliers are far from their optimal values, and the distribution in equation~\ref{eq:augmented-posterior} only coarsely approximates the true posterior; nevertheless, sampling from this approximate distribution still provides useful gradient information that moves the model particles in a productive direction. As the iterations progress and the constraint is increasingly enforced, the distribution sharpens and converges to the true posterior, so that the final particle positions represent approximate posterior samples that satisfy the physics constraints.

\subsection{ADMM-SVGD algorithm}\label{sec:admm-svgd}

Having established the probabilistic interpretation of the AL, we now combine ADMM with SVGD to sample from the posterior distribution in the dual space. The integration of these two methods is natural: ADMM provides the iterative structure for handling the physics constraints through dual variables, while SVGD enables UQ by evolving an ensemble of particles toward the posterior distribution. A convenient feature of this combination is that the gradient of the log-posterior required by SVGD (equation~\ref{eq:svgd-phi}) is read off directly from the dual iteration equations, without a \emph{separate} adjoint-state solve dedicated to the gradient. Each particle independently solves its own set of auxiliary equations, making the per-particle computations embarrassingly parallel, while the SVGD kernel interactions across particles maintain diversity and prevent mode collapse.

The algorithm maintains an ensemble of $N_p$ model particles $\{\bm{m}^{(j)}\}_{j=1}^{N_p}$, each with its own set of auxiliary wavefields $\bm{u}_i^{(j)}$ and scaled multipliers $\bm{\varepsilon}_i^{(j)}$ that are updated iteratively. The model particles are the primary unknowns---they represent the posterior samples at convergence---while the auxiliary variables serve as per-particle bookkeeping that drives each particle toward constraint satisfaction. The algorithm iterates through the following steps.

\textbf{Step 1: Auxiliary variables.} For each particle $j$, compute auxiliary variables by solving the dual iteration equations. The matrix $\bm{S}^{(j)} = \bm{P}\bm{A}(\bm{m}^{(j)})^{-1}$ represents the forward modeling operator for the current model estimate of particle $j$, mapping sources to predicted data at receiver locations. The data residual $\delta\bm{d}_i^{(j)} = \bm{d}_i - \bm{S}^{(j)}\bm{b}_i$ quantifies the misfit between observed and predicted data. The adjoint wavefields $\bm{\lambda}_i^{(j)}$ are obtained from the optimality condition of the wavefield subproblem---which balances the data fidelity (scaled by $1/\sigma^2$) against the constraint penalty (scaled by $\mu$):
\begin{align}
 \bm{\lambda}_i^{(j)} &= \bm{S}^{(j)H} (\bm{S}^{(j)}\bm{S}^{(j)H} + \mu\sigma^2 \bm{I})^{-1}\left(\delta\bm{d}_i^{(j)} + \bm{S}^{(j)} \bm{\varepsilon}_i^{(j)}\right), \label{adj_wave}\\
 \bm{u}_i^{(j)} &= \bm{A}(\bm{m}^{(j)})^{-1} (\bm{b}_i + \bm{\lambda}_i^{(j)} - \bm{\varepsilon}_i^{(j)}). \label{eq:wavefield-update}
\end{align}
The quantity $\mu\sigma^2$ controls the balance between data fitting and constraint enforcement: small values prioritize the data, while large values prioritize the wave equation constraint. In the limit $\sigma^2 \to 0$ (exact data), the system reduces to $\bm{S}^{(j)} \bm{\lambda}_i^{(j)} = \delta\bm{d}_i^{(j)} + \bm{S}^{(j)} \bm{\varepsilon}_i^{(j)}$, recovering exact data fitting at receiver locations. In this work, we adopt the residual-whiteness principle to adaptively determine the parameter $\mu\sigma^2$ at each iteration \citep[see][]{Aghazade_2025_APP}.

\textbf{Step 2: Gradient computation.} Once the auxiliary variables are available, the gradient of the log-posterior with respect to $\bm{m}^{(j)}$ can be computed. This gradient naturally decomposes into a likelihood term---derived from the data misfit and constraint penalty---and a prior term:
\begin{equation}
\bm{g}_m^{(j)} = \bm{g}_\text{like}^{(j)} + \bm{g}_\text{prior}^{(j)}.
\label{eq:gradient-total}
\end{equation}
The likelihood gradient is computed from the dual iteration as:
\begin{equation}
\bm{g}_\text{like}^{(j)} = -\frac{1}{\omega^2} \frac{\sum_{i=1}^{N_s} \text{Re}\left( (\bm{u}_i^{(j)})^* \circ \bm{\lambda}_i^{(j)} \right)}{\sum_{i=1}^{N_s} (\bm{u}_i^{(j)})^* \circ \bm{u}_i^{(j)}},
\label{eq:gradient-lik}
\end{equation}
In the above expression, ${}^*$ denotes complex conjugate. This expression has the familiar structure of a zero-lag cross-correlation between the forward wavefield $\bm{u}_i^{(j)}$ and the adjoint wavefield $\bm{\lambda}_i^{(j)}$, analogous to the gradient computation in adjoint-state methods for FWI \citep{Plessix2006}. The denominator normalizes this correlation by the diagonal Gauss--Newton term, so that equation~\ref{eq:gradient-lik} supplies a preconditioned gradient rather than the bare score of the augmented density. The prior gradient depends on the choice of prior:
\begin{equation}
\bm{g}_\text{prior}^{(j)} = \nabla_{\bm{m}} \log p_\text{prior}(\bm{m}^{(j)}).
\label{eq:gradient-prior}
\end{equation}

\textbf{Step 3: SVGD update for model.} With the gradients computed, the model particles are updated using the SVGD perturbation (equation~\ref{eq:svgd-phi}), which combines the gradient-driven term---that moves particles toward high-probability regions---with the repulsive kernel term that maintains ensemble diversity:
\begin{equation}
\bm{m}^{(j)} \leftarrow \bm{m}^{(j)} + \eta_m \, \bm{\phi}_m^{(j)},
\label{eq:svgd-m-update}
\end{equation}
where
\begin{equation}
\bm{\phi}_m^{(j)} = \frac{1}{N_p} \sum_{i=1}^{N_p} \left[ K(\bm{m}^{(i)}, \bm{m}^{(j)}) \bm{g}_m^{(i)} + \nabla_{\bm{m}^{(i)}} K(\bm{m}^{(i)}, \bm{m}^{(j)}) \right].
\label{eq:svgd-phi-m}
\end{equation}

\textbf{Step 4: Multiplier update.} Finally, the scaled Lagrange multipliers are updated by accumulating the wave equation residual. This step is the mechanism by which the algorithm progressively enforces the physics constraint---if the wave equation is not satisfied for a given particle, the multiplier grows, increasing the penalty on constraint violations in subsequent iterations:
\begin{equation}
\bm{\varepsilon}_i^{(j)} \leftarrow \bm{\varepsilon}_i^{(j)} + \bm{A}(\bm{m}^{(j)})\bm{u}_i^{(j)}-\bm{b}_i.
\label{eq:multiplier-update}
\end{equation}

The algorithm is summarized in Algorithm~\ref{alg:admm-svgd}.

\begin{algorithm}[H]
\caption{ADMM-SVGD for constrained Bayesian inverse problems}
\label{alg:admm-svgd}
\begin{algorithmic}[1]
\State \textbf{Input:} Data $\{\bm{d}_i\}_{i=1}^{N_s}$, sources $\{\bm{b}_i\}_{i=1}^{N_s}$, prior $p_\text{prior}(\bm{m})$, noise level $\sigma$, penalty $\mu$, particles $N_p$, step size $\eta_m$
\State \textbf{Initialize:} Sample $\bm{m}^{(j)} \sim p_\text{prior}(\bm{m})$, set $\bm{\varepsilon}_i^{(j)} = \bm{0}$ for $j = 1, \ldots, N_p$, $i = 1, \ldots, N_s$
\While{not converged}
    \For{$j = 1, \ldots, N_p$} \Comment{Embarrassingly parallel}
        \State Compute $\bm{S}^{(j)} = \bm{P}\bm{A}(\bm{m}^{(j)})^{-1}$, $\delta\bm{d}_i^{(j)} = \bm{d}_i - \bm{S}^{(j)}\bm{b}_i$ for all $i$
        \State Solve for $\bm{\lambda}_i^{(j)}$ for all $i$ via equation~\ref{adj_wave}
        \State Solve for $\bm{u}_i^{(j)}$ for all $i$ via equation~\ref{eq:wavefield-update}
        \State Compute $\bm{g}_\text{like}^{(j)}$ via equation~\ref{eq:gradient-lik}
        \State Compute $\bm{g}_m^{(j)} = \bm{g}_\text{like}^{(j)} + \nabla_{\bm{m}} \log p_\text{prior}(\bm{m}^{(j)})$
    \EndFor
    \State Compute kernel bandwidth $h$ via equation~\ref{eq:median-heuristic}
    \For{$j = 1, \ldots, N_p$} \Comment{Embarrassingly parallel}
        \State Compute $\bm{\phi}_m^{(j)}$ via equation~\ref{eq:svgd-phi-m}
    \EndFor
    \For{$j = 1, \ldots, N_p$} \Comment{Embarrassingly parallel}
        \State Update $\bm{m}^{(j)} \leftarrow \bm{m}^{(j)} + \eta_m \bm{\phi}_m^{(j)}$
        \State Update $\bm{\varepsilon}_i^{(j)} \leftarrow \bm{\varepsilon}_i^{(j)} +  \bm{A}(\bm{m}^{(j)})\bm{u}_i^{(j)}-\bm{b}_i$ for all $i$
    \EndFor
\EndWhile
\State \textbf{Output:} Posterior samples $\{\bm{m}^{(j)}\}_{j=1}^{N_p}$ approximating $p_\text{post}(\bm{m} | \bm{d})$
\end{algorithmic}
\end{algorithm}

The computational cost per iteration is dominated by the auxiliary variable solves in Step~1, each of which requires applying the inverse of the Helmholtz operator $\bm{A}(\bm{m}^{(j)})^{-1}$. Crucially, these solves are embarrassingly parallel across both particles and sources, making the algorithm well-suited for modern parallel computing architectures. The multiplier update (Step~4) and the SVGD perturbation computation (Step~3) are also independent across particles. This structure is akin to consensus optimization \citep{Zand2020}: each particle acts as an independent worker that solves its own auxiliary equations and accumulates its own multiplier updates, while the SVGD kernel step acts as a coordination mechanism that aggregates information across the ensemble to maintain diversity and prevent mode collapse. The only inter-particle communication occurs through the kernel evaluations in Step~3, which require access to all particle positions and gradients but can themselves be computed in parallel. In the limit of convergence, the constraint residual $\|\bm{A}(\bm{m}^{(j)})\bm{u}_i^{(j)} - \bm{b}_i\|$ vanishes for each particle---in practice it is reduced progressively over the frequency schedule---ensuring that the posterior samples satisfy the physics constraints while the SVGD kernel interactions maintain the diversity needed for reliable UQ.

\subsection{Convergence considerations}\label{sec:convergence}

A natural question is whether the evolving distribution $p_k(\bm{m} | \bm{d})$ converges to the true posterior. A complete proof for the general nonconvex setting remains open, but several lines of evidence provide strong support. The statement is a distributional one, concerning the sequence of target distributions rather than a single optimizer. The underlying mechanism is structural. As the multipliers converge, the constraint residual $\bm{A}(\bm{m})\bm{u} - \bm{b}$ becomes small. This suppresses the penalty and multiplier terms of equation~\ref{eq:neg-log-augmented-posterior} and leaves the data-fidelity and log-prior terms. The vanishing residual also forces the auxiliary wavefields to satisfy $\bm{u}_i \to \bm{A}(\bm{m})^{-1}\bm{b}_i$, so that the data-fidelity term coincides with that of the reduced-space negative log-posterior of equation~\ref{eq:neg-log-posterior}. Because the multiplier update of equation~\ref{eq:multiplier-update} acts on each particle independently, this suppression occurs across the ensemble, so that $p_k$ approaches $p_\text{post}$ throughout the region of model space explored by the particles rather than only at an isolated optimizer.

First, ADMM is known to tolerate approximate subproblem solves: \citet{Tapia1977} and \citet{MieleMoseleyLevyCoggins1972} established early on that multiplier methods converge even when the inner minimization is performed inexactly, and \citet{EcksteinBertsekas1992} proved that convergence is preserved as long as the approximation errors decrease over iterations. \citet{BoydADMM2011} showed that larger errors are permissible in early iterations when the iterates are far from the solution. In our setting, this is naturally satisfied through warm-starting---as the multipliers stabilize, $p_k$ changes slowly between iterations, so the particles from the previous step already approximate the current target well. A single SVGD transport step is therefore performed per target. The ensemble tracks the slowly drifting sequence rather than equilibrating with each intermediate $p_k$, as in homotopy-continuation and annealed particle methods. Second, for nonconvex objectives---typical in geophysical applications---\citet{HongLuo2016} and \citet{WangYinZeng2019} established that ADMM converges to stationary points provided $\mu$ is sufficiently large. Third, \citet{Vono2019} and \citet{VonoPaulinDoucet2022} proved that ADMM-type splitting is compatible with sampling (not just optimization), achieving dimension-free convergence rates for a split Gibbs sampler. Their analysis reveals a tradeoff: larger $\mu$ enforces the constraint faster but introduces temporary bias, while smaller $\mu$ reduces bias but slows convergence. 
Finally, \citet{Korba2020} bounds the SVGD approximation error against a fixed target, the regime that the warm-started iteration approaches once $p_k$ ceases to drift.

Our experiments are consistent with this picture. The convergence diagnostics in Figure~\ref{fig:rosenbrock-convergence}---in particular the settling of the ensemble-mean multiplier $\bar{\varepsilon}$ and the decay of the penalty term and constraint residual, alongside the stabilization of the particle mean, standard deviation, and kernel bandwidth---indicate that the target sequence stabilizes and that the ensemble tracks it. On the Rosenbrock problem, where the posterior is available in closed form, the particles match the analytic target and pass simulation-based calibration, consistent with the ensemble having, at termination, approximated the intended posterior without accumulating a persistent lag behind the drifting targets. In practice, care should be taken with the choice of $\mu$ \citep{BoydADMM2011} and with using enough particles to avoid underestimating the posterior spread \citep{LiuWang2016}.

\subsection{Memory and computational cost}\label{sec:cost}

As noted earlier, the per-iteration cost of ADMM-SVGD is dominated by the auxiliary-variable solves of Step~1; we now make this cost precise. Each particle $j$ carries its own model $\bm{m}^{(j)}$, and hence its own Helmholtz operator $\bm{A}(\bm{m}^{(j)})$. With a sparse direct solver, this operator is factorized once per particle at a cost of $O(N^{3/2})$ in two dimensions, where $N$ is the number of grid points; the factorization is then reused across all $N_s$ sources, each requiring a small number of triangular solves---including the adjoint apply $\bm{A}(\bm{m}^{(j)})^{-H}$ in equation~\ref{adj_wave}---at $O(N\log N)$ per solve. The wave-physics cost per outer iteration is therefore $O\!\left(N_p N^{3/2}\right)$ for the factorizations and $O\!\left(N_p N_s N\log N\right)$ for the back-substitutions; a matrix-free iterative solver instead forgoes the reusable factorization and performs each of the $O(N_p N_s)$ solves at $O(N)$ per inner iteration.

The adjoint wavefield $\bm{\lambda}_i^{(j)}$ (equation~\ref{adj_wave}) plays the role of the adjoint-state variable, and the gradient (equation~\ref{eq:gradient-lik}) is read off as a zero-lag cross-correlation of the forward and adjoint wavefields already computed in Step~1---so no \emph{separate} adjoint-state solve dedicated to the gradient is required. The per-iteration solve count is thus comparable to that of a reduced-space gradient evaluation; the benefit of the dual-space formulation lies not in fewer wave-equation solves but in the improved conditioning of the relaxed subproblems and the progressive enforcement of the constraint, which together can reduce the number of outer iterations. The remaining operations are subdominant: the data-space inverse $(\bm{S}^{(j)}\bm{S}^{(j)H} + \mu\sigma^2\bm{I})^{-1}$ in equation~\ref{adj_wave} acts on the $N_r$-dimensional data space and is applied through the per-particle Helmholtz factorization---each matrix--vector product with $\bm{S}^{(j)}\bm{S}^{(j)H}$ amounting to a forward and an adjoint triangular solve---so that $\bm{S}^{(j)} = \bm{P}\bm{A}(\bm{m}^{(j)})^{-1}$ need not be formed explicitly; the SVGD perturbation (equation~\ref{eq:svgd-phi-m}) forms an $N_p \times N_p$ kernel matrix at a cost of $O(N_p^2 N)$.

The memory footprint scales as $O(N_p N_s N)$, since each of the $N_p$ particles maintains its own auxiliary wavefields $\bm{u}_i^{(j)}$ and scaled multipliers $\bm{\varepsilon}_i^{(j)}$ for every source $i$, each a vector of dimension $N$. This per-particle bookkeeping is the price paid for the embarrassingly parallel structure described above; by contrast, the reduced-space approach stores only $O(N_p N)$ model vectors but recomputes the wavefields at every gradient evaluation. In practice, the memory cost is mitigated by the mutual independence of the auxiliary variables across particles and sources---they need not reside in memory simultaneously, and can instead be streamed, recomputed on the fly, or distributed across compute nodes. Because these wavefields enter the model update only through the gradient, which is accumulated as a sum over sources, they can be formed, used, and discarded one source---and one particle---at a time; the implementation therefore need not hold the full $N_p N_s$ collection in memory simultaneously, and the peak memory is in practice governed by the per-particle working set rather than the worst-case $O(N_p N_s N)$ bound. The two experiments presented next---a stylized conditional-inference problem and frequency-domain FWI---exercise these cost and memory tradeoffs at small and large scale, respectively.

\section{Numerical examples}\label{sec:numerical-examples}

The proposed ADMM-SVGD framework is validated on two classes of problems. First, we consider conditional inference on the Rosenbrock distribution as a stylized example that demonstrates the method's ability to perform Bayesian inference under nonlinear constraints. Second, we apply the method to FWI, where the wave equation constraint is enforced through the dual-space formulation. For the FWI experiments, we present results on a Gaussian anomaly model and on the Marmousi~II model, progressively increasing the complexity of the subsurface structure. All FWI computations are performed on a dual Intel Xeon Platinum 8176 system---56 cores operating at 2.10~GHz---with the per-particle updates distributed across 30 workers.
Code to partially reproduce these results is available at \url{https://github.com/luqigroup/SVGDADMMSampler.jl}.

\subsection{Conditional inference on the Rosenbrock distribution}\label{sec:rosenbrock}

The purpose of this example is to validate the constrained reformulation---to confirm that ADMM-SVGD recovers the same posterior that reduced-space SVGD samples directly, not to outperform it; the conditioning benefit of the splitting emerges in higher-dimensional problems, as we demonstrate at FWI scale in the Marmousi~II example. As a controlled demonstration of ADMM-SVGD for conditional inference, we consider sampling from the posterior distribution $p(\bm{x} | \bm{y})$ where the prior $p(\bm{x})$ follows the Rosenbrock distribution---a banana-shaped density defined by $p(\bm{x}) \propto \exp(-a(x_1 - \mu_0)^2 - (x_2 - x_1^2)^2)$---and the observations are generated as $\bm{y} = \bm{x} + \bm{n}$ with Gaussian noise $\bm{n} \sim \mathcal{N}(\bm{0}, \sigma^2 \bm{I})$. The nonlinear coupling through the term $(x_2 - x_1^2)^2$ makes direct sampling challenging. Following the constrained formulation presented above, we introduce an auxiliary variable $z \in \mathbb{R}$ and write the constrained problem
\begin{equation}
\min_{\bm{x}, z} \; \frac{1}{2\sigma^2} \|\bm{y} - \bm{x}\|^2 + a(x_1 - \mu_0)^2 + (x_2 - z)^2 \quad \text{subject to} \quad z = x_1^2,
\label{eq:rosenbrock-constrained}
\end{equation}
where $\bm{x}$ plays the role of the model $\bm{m}$, $z$ plays the role of the wavefield $\bm{u}$, and $z = x_1^2$ plays the role of the wave equation $\bm{A}(\bm{m})\bm{u} = \bm{b}$. The corresponding AL (cf.\ equation~\ref{eq:augmented-lagrangian}) is
\begin{equation}
\mathcal{L}_\mu(\bm{x}, z, \varepsilon; \bm{y}) = \frac{1}{2\sigma^2}\|\bm{y} - \bm{x}\|^2 + a(x_1 - \mu_0)^2 + (x_2 - z)^2 - \varepsilon(z - x_1^2) + \frac{\mu}{2}(z - x_1^2)^2,
\label{eq:rosenbrock-auglag}
\end{equation}
where $\varepsilon$ is the Lagrange multiplier and $\mu > 0$ the penalty parameter. Minimizing over $z$ for fixed $(\bm{x}, \varepsilon)$ gives the closed-form auxiliary update $z = (2 x_2 + \varepsilon + \mu x_1^2)/(2 + \mu)$, after which the gradient $\nabla_{\bm{x}} \mathcal{L}_\mu$ drives the SVGD particle update (equation~\ref{eq:svgd-phi-m}) and the multiplier is updated via $\varepsilon \leftarrow \varepsilon + \mu(x_1^2 - z)$, progressively enforcing the constraint. Here $\varepsilon$ is the unscaled multiplier $\bm{v}$ of equation~\ref{eq:augmented-lagrangian} rather than the scaled multiplier $\bm{\varepsilon} = \bm{v}/\mu$ of equation~\ref{eq:augmented-lagrangian-scaled} used in the general algorithm, so the penalty parameter $\mu$ appears explicitly in its update.

\begin{figure}[htbp]
\centering
\subfloat[]{\includegraphics[width=0.36\textwidth]{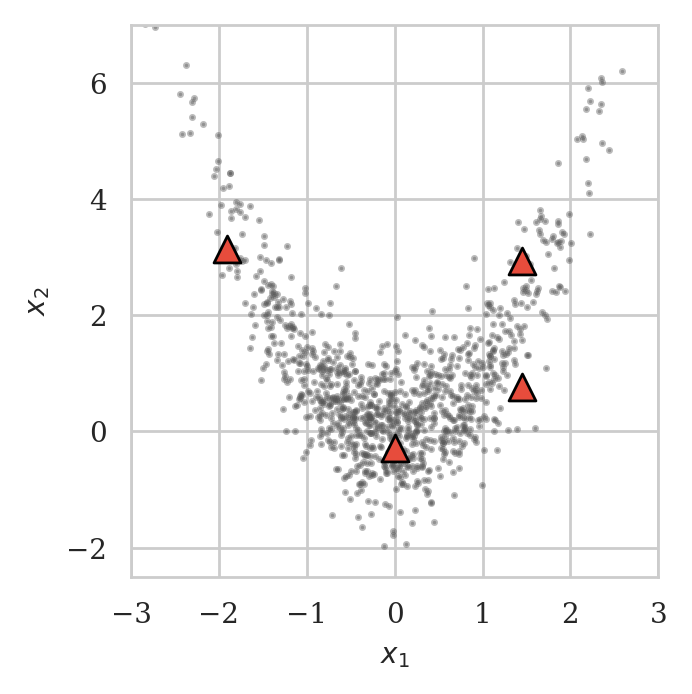}\label{fig:rosenbrock-prior}}
\quad
\subfloat[]{\includegraphics[width=0.36\textwidth]{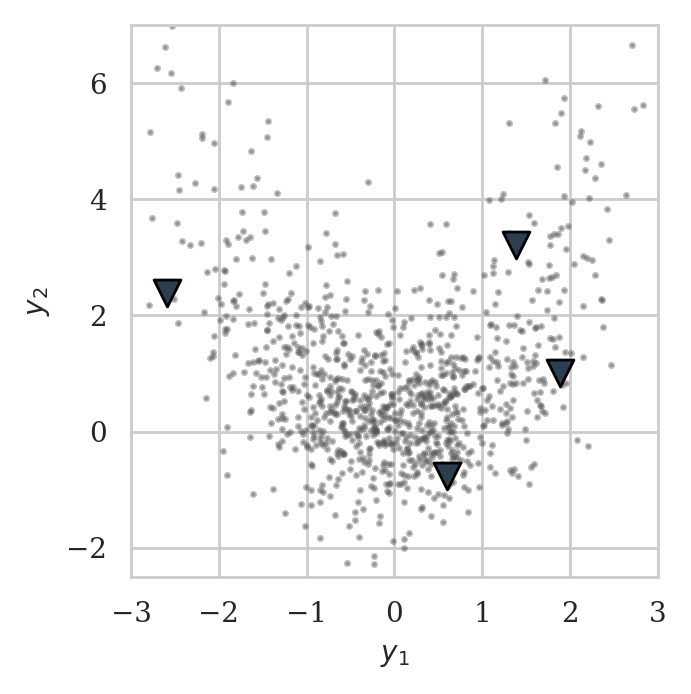}\label{fig:rosenbrock-data}}
\caption{Conditional sampling problem. (a) Rosenbrock prior $p(\bm{x})$ with four test instances (pink triangles). (b) Noisy observations, $\sigma = 0.5$.}
\label{fig:rosenbrock-overview}
\end{figure}

For the experiments, we set $a = 0.5$, $\mu_0 = 0$, $\sigma = 0.5$, $\mu = 1.0$, $N_p = 1000$ particles, step size $\eta = 0.30$, and run for 1500 iterations. To isolate the effect of the constraint splitting, we also run reduced-space SVGD (Algorithm~\ref{alg:svgd}) with the same number of particles, using the direct posterior gradient $\nabla_{\bm{x}} \log p(\bm{x} | \bm{y})$---without introducing auxiliary variables or multiplier updates. Because the coupled gradient is stiffer without the ADMM decomposition, reduced-space SVGD requires a smaller step size ($\eta = 0.20$) and more iterations (2500) to remain stable. Figure~\ref{fig:rosenbrock-overview} shows the prior distribution alongside the noisy observations for four test instances.

\begin{figure}[htbp]
\centering
\includegraphics[width=0.66\textwidth]{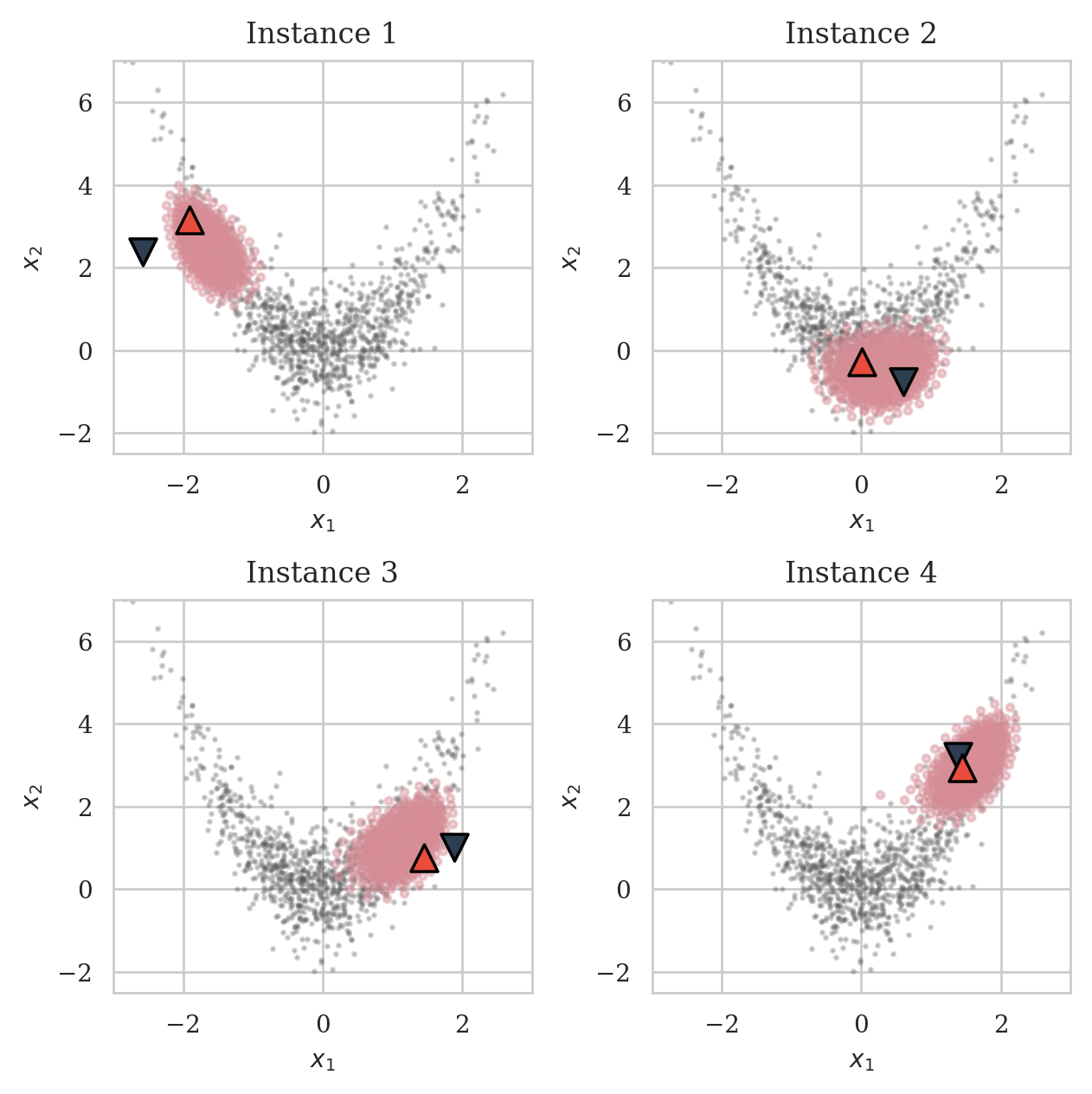}
\caption{ADMM-SVGD posteriors for the four test observations: prior samples (gray), posterior samples (pink), the noisy observation ($\triangledown$), and the true value ($\triangle$).}
\label{fig:rosenbrock-svgd-posteriors}
\end{figure}

Figure~\ref{fig:rosenbrock-svgd-posteriors} displays the posterior distributions obtained by ADMM-SVGD for all four test instances. In each panel, the gray points show prior samples illustrating the banana-shaped Rosenbrock manifold, while the pink points represent posterior samples. The posteriors concentrate around the observations while respecting the prior structure, and in all instances the true underlying values (red triangles) lie within the posterior support. 

\begin{figure}[htbp]
\centering
\includegraphics[width=0.78\textwidth]{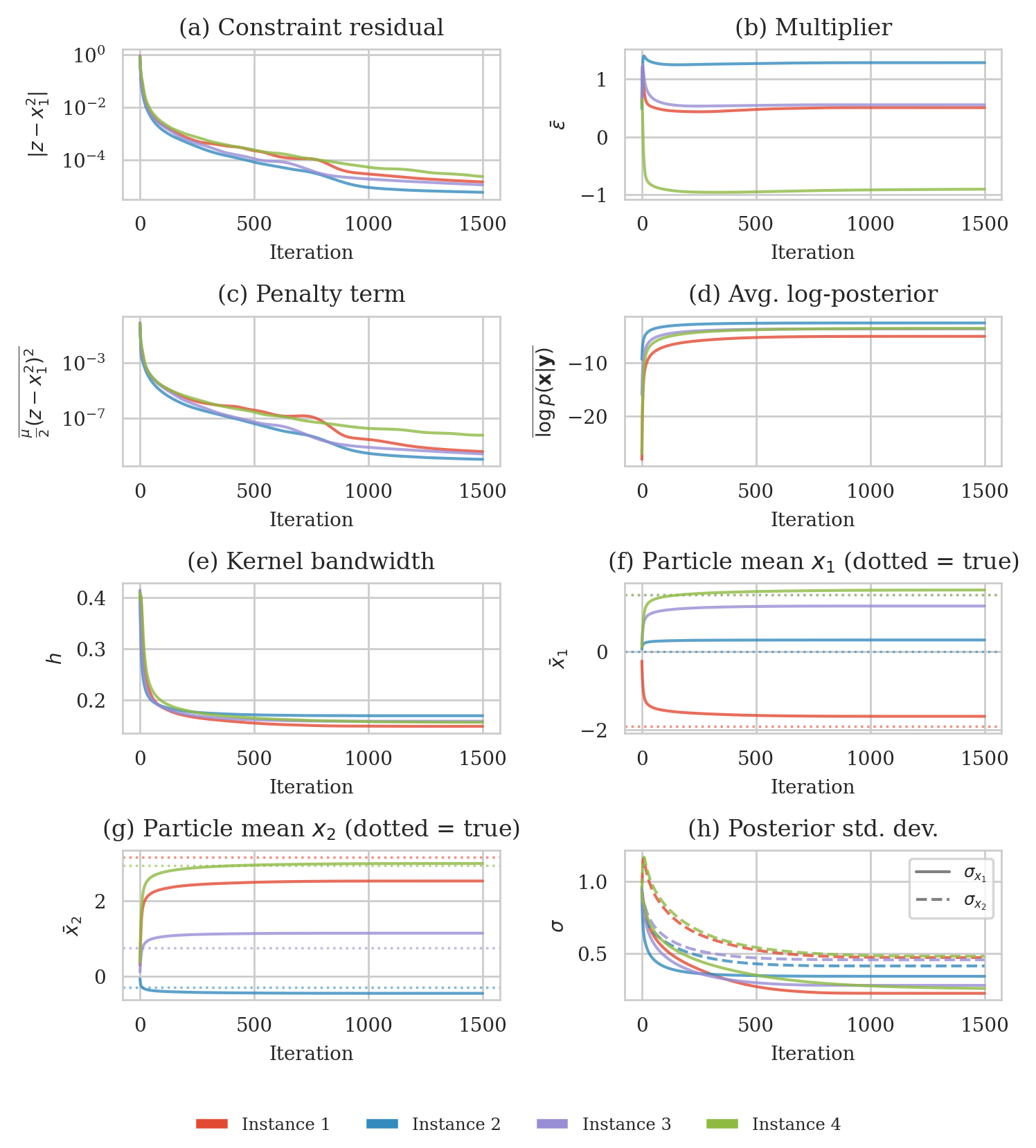}
\caption{ADMM-SVGD convergence diagnostics, four test instances. (a) Constraint residual $|z - x_1^2|$. (b) Ensemble-mean multiplier $\bar{\varepsilon}$. (c) Ensemble-mean penalty term. (d) Average log-posterior. (e) Kernel bandwidth. (f)--(g) Posterior mean, $x_1$ and $x_2$. (h) Posterior standard deviation.}
\label{fig:rosenbrock-convergence}
\end{figure}

Figure~\ref{fig:rosenbrock-convergence} shows the convergence diagnostics for ADMM-SVGD: the constraint residual $|z - x_1^2|$ and the AL penalty term decrease over iterations while the ensemble-mean multiplier $\bar{\varepsilon}$ settles to a stable value, confirming that the ADMM mechanism progressively enforces the nonlinear constraint; the average log-posterior increases as particles move toward higher-probability regions; and the kernel bandwidth, posterior mean, and standard deviation all stabilize, indicating convergence of the particle ensemble. The slowing drift of the multiplier and the decay of the penalty term also show that the intermediate target distribution stabilizes as the iterations proceed.

\begin{figure}[htbp]
\centering
\includegraphics[width=0.66\textwidth]{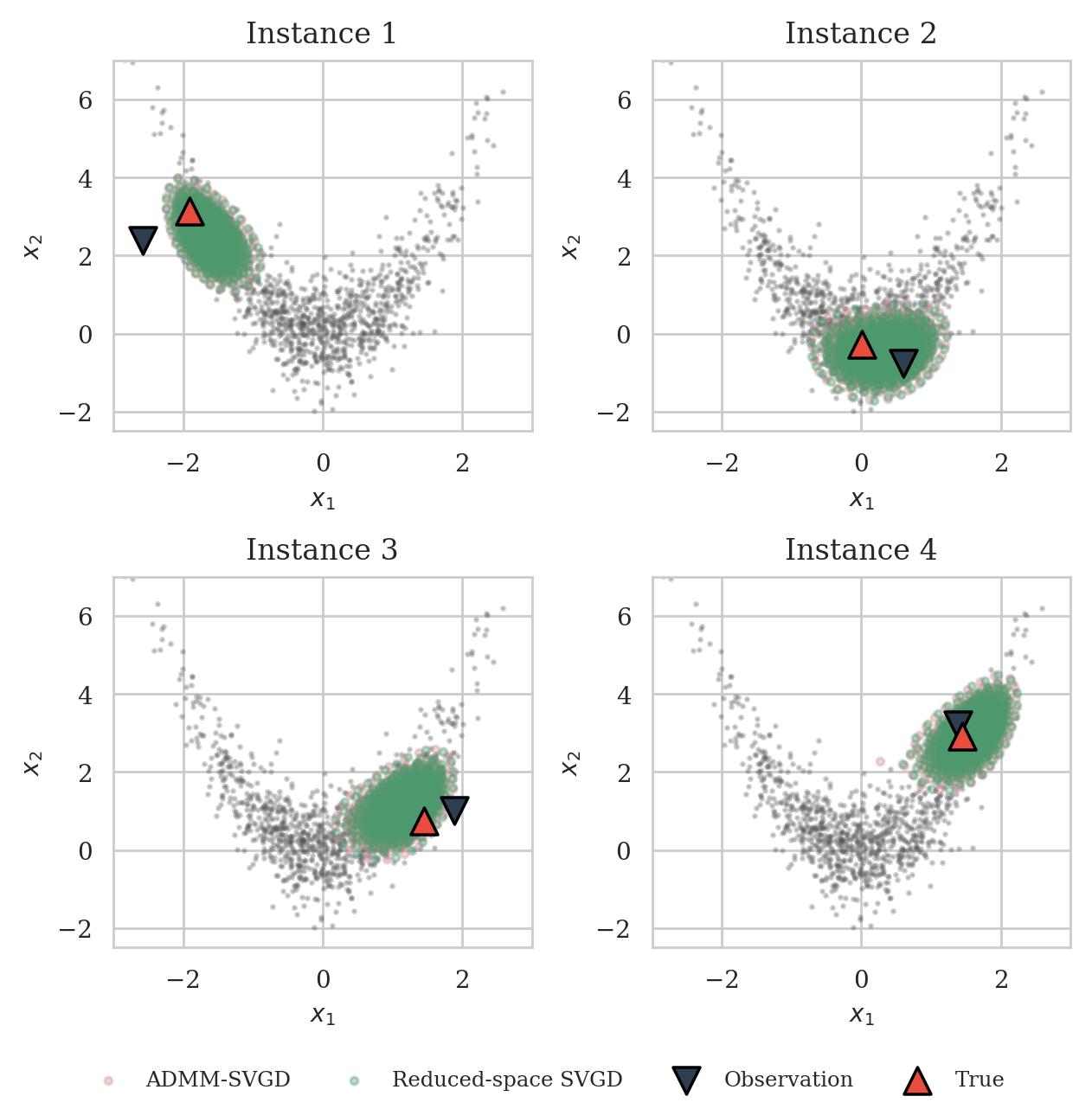}
\caption{Posteriors from both methods for the four test observations: ADMM-SVGD (pink), reduced-space SVGD (green), the noisy observation ($\triangledown$), and the true value ($\triangle$).}
\label{fig:rosenbrock-combined-all}
\end{figure}

\begin{figure}[htbp]
\centering
\includegraphics[width=0.78\textwidth]{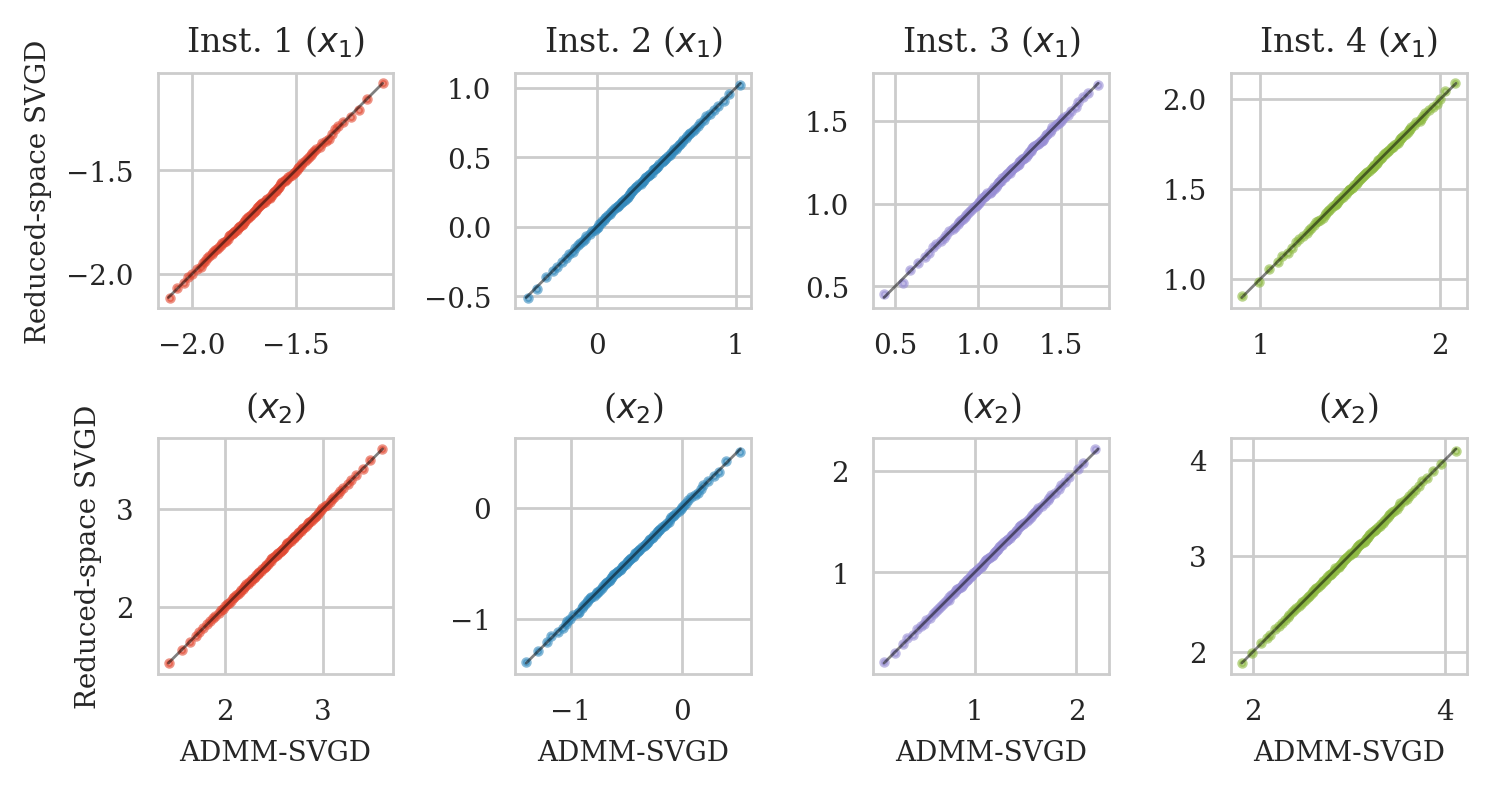}
\caption{Quantile--quantile plots of the reduced-space SVGD against the ADMM-SVGD marginals. Each column is one test instance; the top and bottom rows are the $x_1$ and $x_2$ marginals. Diagonal points indicate agreement.}
\label{fig:rosenbrock-qq}
\end{figure}

Figure~\ref{fig:rosenbrock-combined-all} overlays the posterior samples from both methods---ADMM-SVGD and reduced-space SVGD---for the four test instances. The two methods produce essentially the same posteriors, concentrating around the observations while respecting the banana-shaped Rosenbrock manifold $x_2 \approx x_1^2$. This agreement is the intended outcome of the experiment: because the Rosenbrock posterior can be sampled directly with reduced-space SVGD, recovering the same posterior through the constrained ADMM reformulation---in which the auxiliary constraint $z = x_1^2$ is introduced and progressively enforced---confirms that the dual-space splitting targets the correct distribution. The quantile--quantile plots in Figure~\ref{fig:rosenbrock-qq} corroborate this equivalence, showing comparable marginals for the two methods across all test instances. We note only that the stiffer gradient coupling of the unsplit posterior required reduced-space SVGD to use a smaller step size ($\eta = 0.20$ vs.\ $0.30$) and more iterations ($2500$ vs.\ $1500$) to remain stable; the conditioning advantage of the splitting is not the focus of this stylized example; it is demonstrated at FWI scale in the Marmousi~II example.

\begin{figure}[htbp]
\centering
\includegraphics[width=0.66\textwidth]{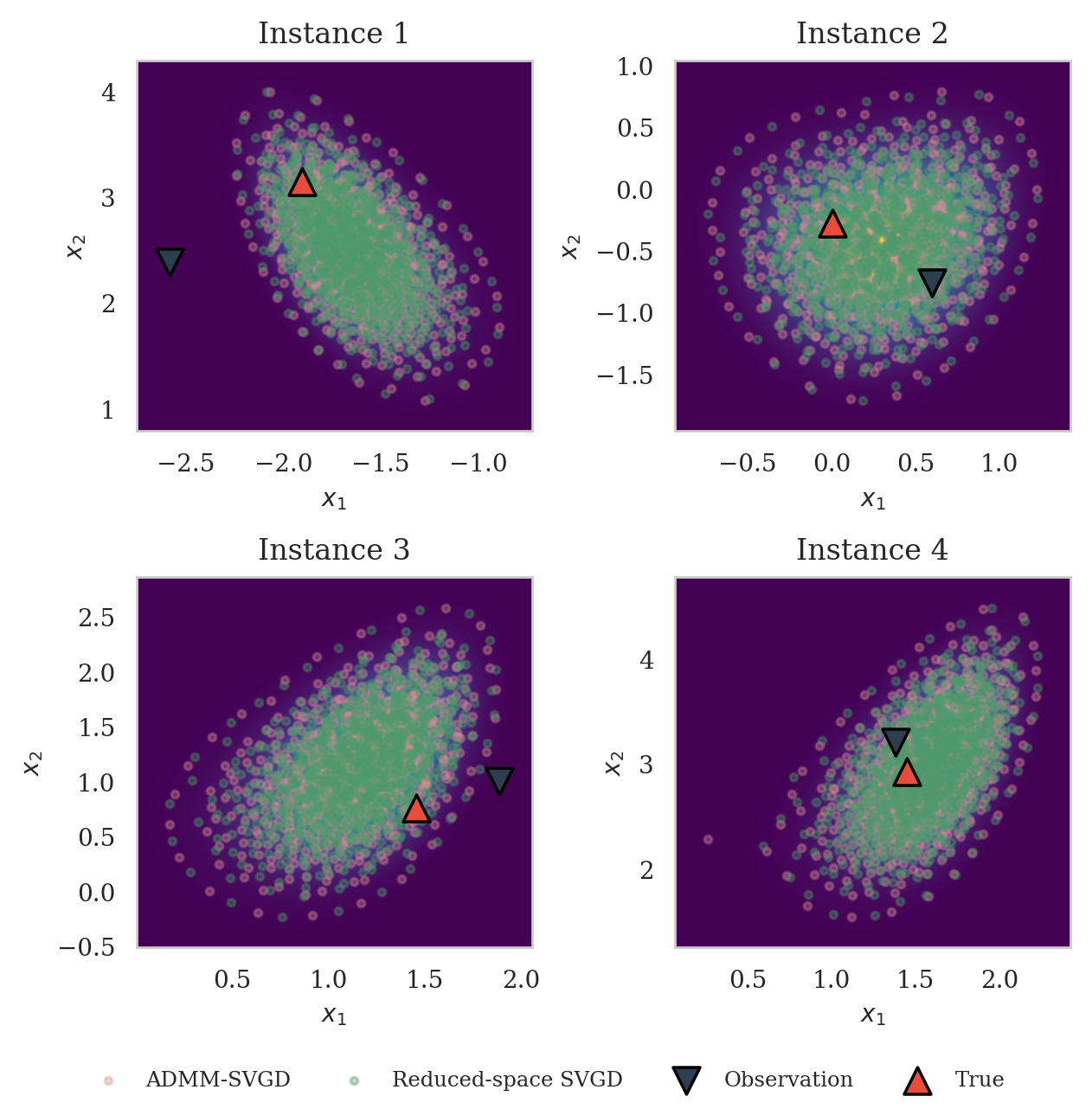}
\caption{Samples overlaid on the analytic conditional posterior $p(\bm{x} | \bm{y})$ for the four test instances; brighter background is higher density. ADMM-SVGD (pink), reduced-space SVGD (green), the noisy observation ($\triangledown$), and the true value ($\triangle$).}
\label{fig:rosenbrock-true-overlay}
\end{figure}

Because the conditional posterior $p(\bm{x} | \bm{y}) \propto p(\bm{y} | \bm{x})\,p(\bm{x})$ is available in closed form for this stylized problem, we can assess the particles against the true target rather than against one another alone. Figure~\ref{fig:rosenbrock-true-overlay} overlays the ADMM-SVGD and reduced-space SVGD samples on the analytic posterior density for the four test instances. Both ensembles populate the high-probability region of the true posterior, whose mass the prior pulls toward the Rosenbrock ridge $x_2 \approx x_1^2$ and away from the noisy observation---most visibly in the instances where the observation falls off the manifold. The agreement between the particles and the analytic density confirms that the dual-space reformulation samples the intended posterior.

\begin{figure}[htbp]
\centering
\includegraphics[width=0.78\textwidth]{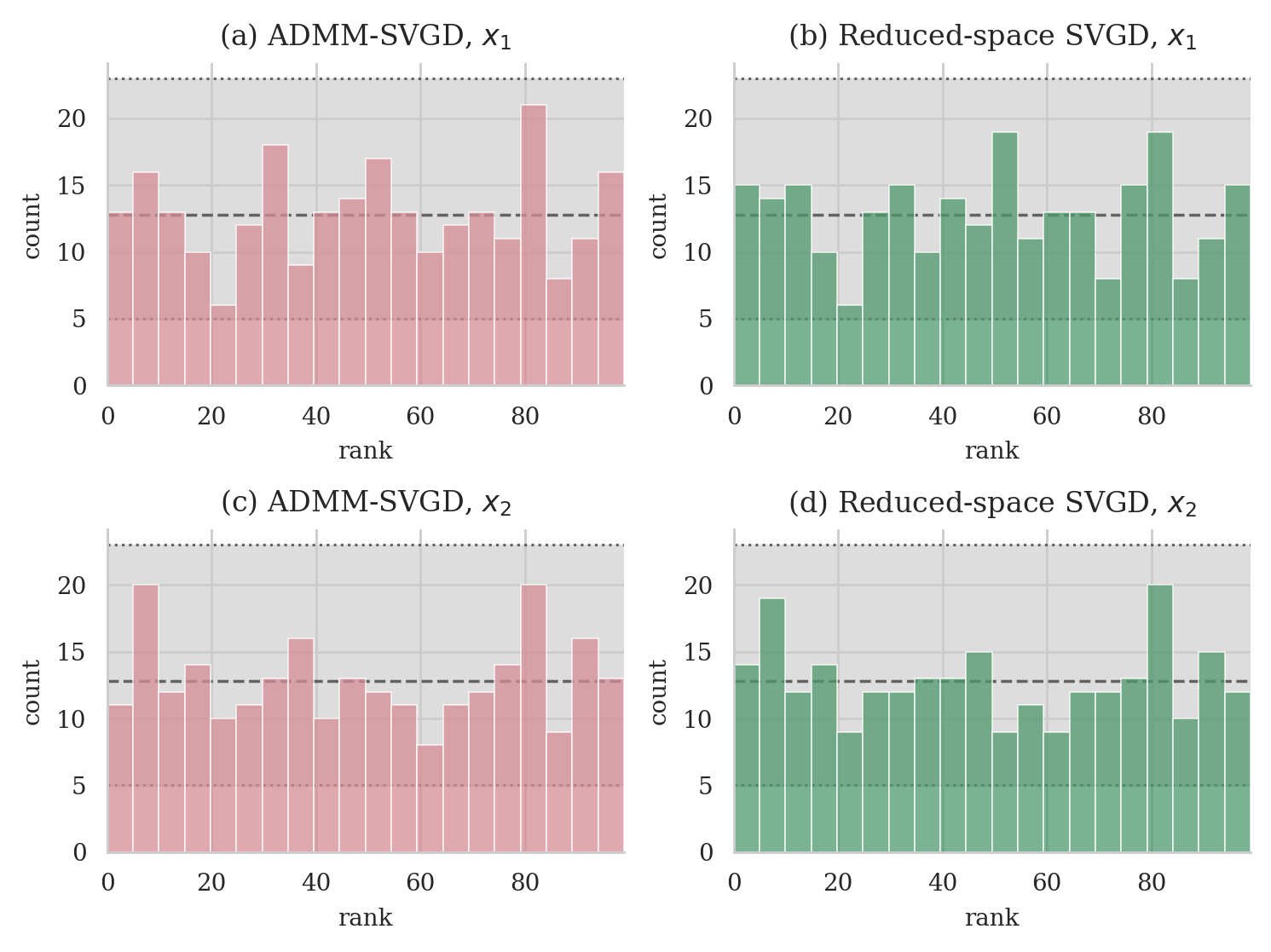}
\caption{Simulation-based calibration rank histograms; $L = 256$ replications, $N = 99$ posterior samples each. (a)--(b) Ranks of the true $x_1$ for ADMM-SVGD and reduced-space SVGD; (c)--(d) the same for $x_2$. Dashed: expected count under uniformity. Shaded band with dotted bounds: $99\%$ binomial interval.}
\label{fig:rosenbrock-sbc}
\end{figure}

The agreement examined so far is based on individual realizations and does not, by itself, establish that the reported uncertainties are well calibrated---i.e., that the posterior spread matches the frequency with which the target is actually recovered. To assess this directly, we perform simulation-based calibration \citep{TaltsEtAl2018}: we draw $L = 256$ parameter--observation pairs from the joint distribution $p(\bm{x})\,p(\bm{y} | \bm{x})$, sample the posterior for each pair, and record the rank of the true parameter among $99$ posterior samples. For a correctly calibrated sampler these ranks are uniformly distributed, so that systematic departures from uniformity diagnose miscalibration---a $\cup$-shaped histogram indicates under-dispersion, a $\cap$-shaped histogram over-dispersion, and a monotonic trend a bias in the posterior mean. As shown in Figure~\ref{fig:rosenbrock-sbc}, the rank histograms for both $x_1$ and $x_2$ are consistent with uniformity for ADMM-SVGD and reduced-space SVGD alike: every bin falls within the $99\%$ interval expected under uniformity, and the histograms exhibit no systematic shape beyond sampling fluctuation. Together, these diagnostics indicate that ADMM-SVGD is well calibrated on this controlled problem and reproduces the calibration of reduced-space SVGD. It is worth noting that this calibration is obtained with a large particle ensemble ($N_p = 1000$); at the scale of FWI, where far fewer particles are affordable, the finite-ensemble under-dispersion of SVGD \citep{LiuWang2016} becomes the limiting factor, and the reported uncertainties should accordingly be read as a lower bound on the posterior spread.

\subsection{Full waveform inversion}\label{sec:fwi}

We now apply ADMM-SVGD to UQ for FWI, where the wave equation constraint $\bm{A}(\bm{m})\bm{u}_i = \bm{b}_i$ is enforced through the dual-space formulation (Algorithm~\ref{alg:admm-svgd}). Although the model parameter $\bm{m}$ represents the squared slowness, we display all results in terms of velocity for ease of interpretation. We present results on two velocity models of increasing complexity. The relative model error, $\text{RME}=\|\bm{m}^{*}-\bm{m}\|_2/\|\bm{m}^{*}\|_2$ with $\bm{m}^{*}$ the true model, serves as the metric of model convergence, and in all examples a perfectly matched layer boundary condition is imposed along the four edges of the computational domain following \citet{Chen_2013_OFD}. 

The gradient computed by ADMM, $\bm{g}_\text{like}^{(j)}$, is naturally scaled by the penalty parameter $\mu$, which regularizes the data-space Hessian in equation~\ref{adj_wave} and its associated diagonal Gauss--Newton Hessian (equation~\ref{eq:gradient-lik}). We preserve this scaling in the Stein direction $\bm{\phi}_m^{(j)}$, which combines the ADMM updates across all particles (equation~\ref{eq:svgd-m-update}). Consequently, the proposed method retains the scaling of the ADMM update while replacing its direction with the SVGD direction:
$\eta^{(j)} = \kappa\,\frac{\| \bm{g}_\text{like}^{(j)} \|_2}{\| \bm{\phi}_m^{(j)} \|_2}$,
where $\kappa \leq 1$. We use $\kappa=1$ for the Gaussian-anomaly experiment and $\kappa=0.5$ for the Marmousi II example.

\subsubsection{Prior distribution}

We adopt a Gaussian random field (GRF) prior with a Mat\'{e}rn-type power spectrum \citep{RasmussenWilliams2006} to encode smoothness and spatial correlation in the velocity model. The prior is defined in the spectral domain through the eigenvalues
\begin{equation}
\lambda(\bm{k}) \propto \left(4\pi^2 |\bm{k}|^2 + \tau^2\right)^{-\alpha},
\label{eq:grf-spectrum}
\end{equation}
where $\bm{k}$ denotes the wavenumber vector, $\tau$ controls the correlation length, and $\alpha$ governs the smoothness of the realizations. Drawing samples from this prior reduces to generating independent standard normal coefficients in Fourier space, scaling them by $\sqrt{\lambda(\bm{k})}$, and applying the inverse fast Fourier transform.

To ensure that the prior samples represent physically plausible squared-slowness values, we linearly rescale the zero-mean GRF realizations so that their values fall within a prescribed range $[m_{\min}, m_{\max}]$. Specifically, the mean is set to $\bar{m} = (m_{\min} + m_{\max})/2$ and the standard deviation is scaled to $\sigma_m = (m_{\max} - m_{\min})/6$, so that approximately 99.7\% of the sampled squared-slowness values lie within the target bounds. The choice of prior can significantly influence the posterior, particularly in under-determined inverse problems \citep{ZhaoCurtis2024b}. The score function $\nabla_{\bm{m}} \log p_\text{prior}(\bm{m})$ required by SVGD is computed tractably in the spectral domain, where the covariance is diagonal, reducing the matrix inversion to elementwise division of the Fourier coefficients by the eigenvalues. Representative samples from this prior are shown in Figure~\ref{fig:gaussian-setup}b-c.

\begin{table}[htbp]
\centering
\caption{Settings for the Gaussian-anomaly and Marmousi~II experiments.}
\label{tab:settings_combined}
\footnotesize
\begin{tabular}{lll}
\toprule
\textbf{Parameter} & \textbf{Gaussian anomaly} & \textbf{Marmousi~II} \\
\midrule
\multicolumn{3}{l}{\textit{Model and domain}} \\
Domain size ($x \times z$)     & 2~km $\times$ 2~km        & 17.0~km $\times$ 3.5~km \\
Grid spacing                   & 10~m (200 $\times$ 200)   & 25~m (140 $\times$ 680) \\
\midrule
\multicolumn{3}{l}{\textit{Acquisition}} \\
Sources                        & 50, $\Delta s = 40$~m     & $N_s = 17, 34, 68$ \\
Receivers                      & 200, $\Delta r = 10$~m    & 114, $\Delta r = 150$~m \\
Source wavelet                 & Ricker, 6~Hz              & Ricker, 8~Hz \\
\midrule
\multicolumn{3}{l}{\textit{Prior / initial ensemble}} \\
Construction                   & {2~km/s + GRF perturbation} & \makecell[l]{Linear gradient (1.5--4.0~km/s) \\ + GRF perturbation} \\
GRF parameters                 & $\alpha = 2$, $\tau = 3$ (equation~\ref{eq:grf-spectrum})      & $\alpha = 2$, $\tau = 3$ (equation~\ref{eq:grf-spectrum}) \\
\midrule
\multicolumn{3}{l}{\textit{Sampler}} \\
Particles $N_p$                & 40                        & 50 \\
Frequency schedule             & 4~Hz (single-frequency)   & 2 cycles, 3--12~Hz, 0.5~Hz increment \\
Iterations                     & 60                        & 20 ($f \leq 7$~Hz), 10 ($f > 7$~Hz) \\
Velocity bounds                & $1.0 \leq v \leq 2.3$~km/s  & $0.9 \leq v \leq 6$~km/s \\
\midrule
\multicolumn{3}{l}{\textit{Noise robustness}} \\
Noise levels                   & ---                       & 10\%, 15\%, 20\% of $\max(|d|)$ ($N_s = 34$) \\
\bottomrule
\end{tabular}
\end{table}

\subsubsection{Gaussian anomaly model}\label{sec:Gaussian-model}

\begin{figure}[htbp]
\centering
\includegraphics[width=\textwidth]{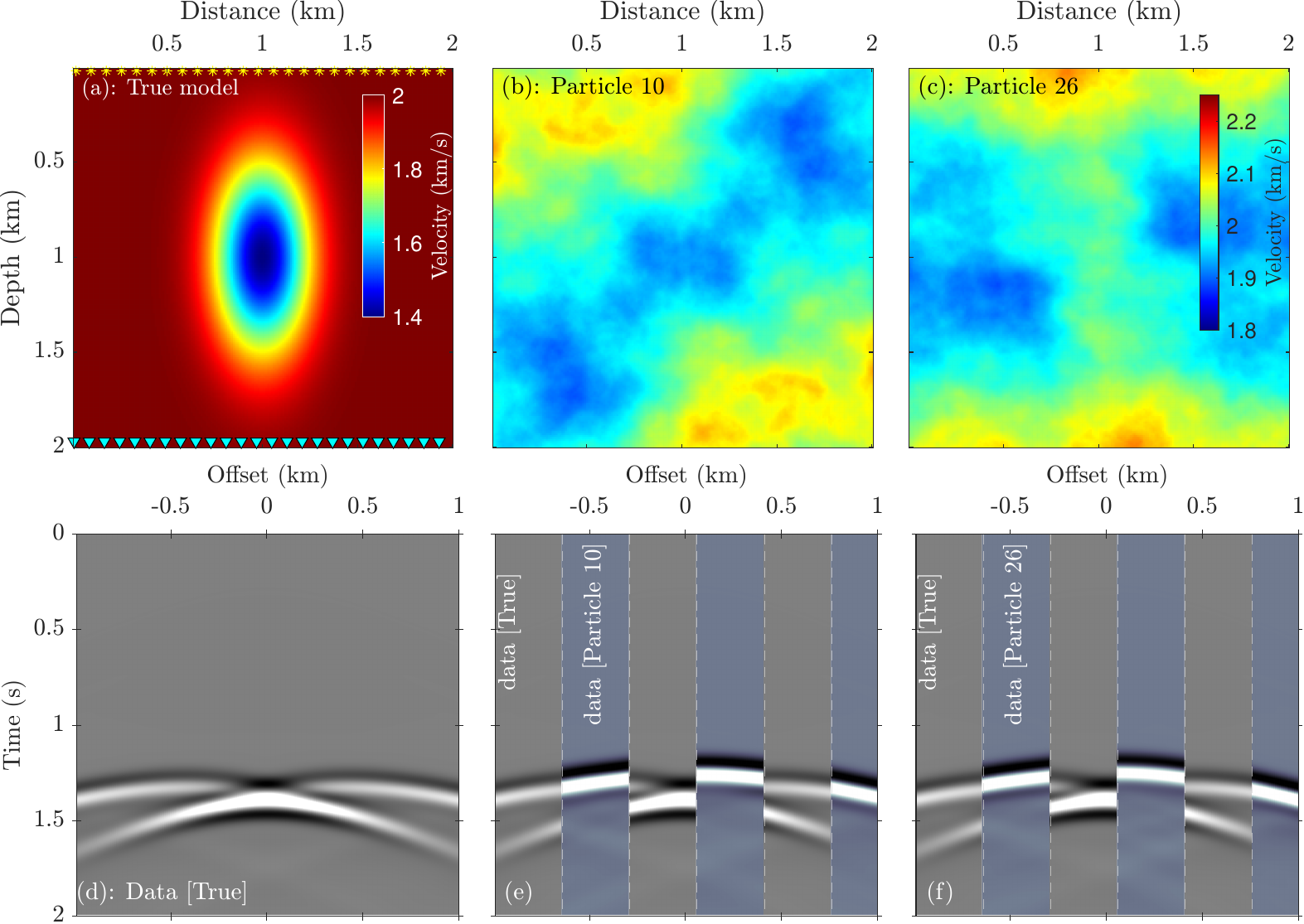}
\caption{Gaussian-anomaly setup. (a) True velocity; sources (colored dots) at the surface, receivers at the bottom. (b)--(c) Two prior particles. (d) Shot gather in the true model. (e)--(f) Its traces interleaved with those of (b) and (c).}
\label{fig:gaussian-setup}
\end{figure}

We first consider a controlled setting in which the true subsurface model consists of a Gaussian low-velocity anomaly embedded in a homogeneous background of 2~km/s \citep{Huang_2018_SEW}. The model spans 2~km in both horizontal distance $x$ and depth $z$ (Table~\ref{tab:settings_combined}), and the velocity is defined as  
\begin{equation*}
  v(x,z) = 2 - 0.6\,\exp\!\left(
      -\frac{(x-1)^2}{0.25^2} - \frac{(z-1)^2}{0.5^2}
  \right) \quad [\mathrm{km/s}].
\end{equation*}
%
The sources are placed along the top of the model and the receivers along the bottom (Figure~\ref{fig:gaussian-setup}a; Table~\ref{tab:settings_combined}).
Throughout the experiments, reduced-space SVGD refers to running the same sampler with the dual-space machinery switched off: the scaled multipliers $\bm{\varepsilon}_i^{(j)}$ and the adjoint wavefields $\bm{\lambda}_i^{(j)}$ are held at zero in equation~\ref{eq:wavefield-update} and never updated. The auxiliary wavefields then collapse to the exact solve $\bm{u}_i^{(j)} = \bm{A}(\bm{m}^{(j)})^{-1}\bm{b}_i$, so that the wave equation is satisfied exactly at every iteration and the sampler targets the reduced-space posterior of equation~\ref{eq:neg-log-posterior}. Both reduced-space SVGD and ADMM-SVGD are run with an ensemble of 40 particles, initialized from random velocity fields drawn from the prior distribution; while these initial fields are close to the homogeneous background, none of them contains the central low-velocity anomaly (Figures~\ref{fig:gaussian-setup}b and~\ref{fig:gaussian-setup}c). A seismogram computed in the true model is shown in Figure~\ref{fig:gaussian-setup}d, and Figures~\ref{fig:gaussian-setup}e and~\ref{fig:gaussian-setup}f interleave alternating trace segments of this seismogram with those computed in the two particles, the discontinuities at the segment boundaries making visible the substantial data misfit that the initial ensemble must overcome. The inversion is carried out on single-frequency data at 4~Hz over 60 iterations.
The other parameters are summarized in Table~\ref{tab:settings_combined}.

\begin{figure}[htbp]
    \centering
    \includegraphics[width=\textwidth]{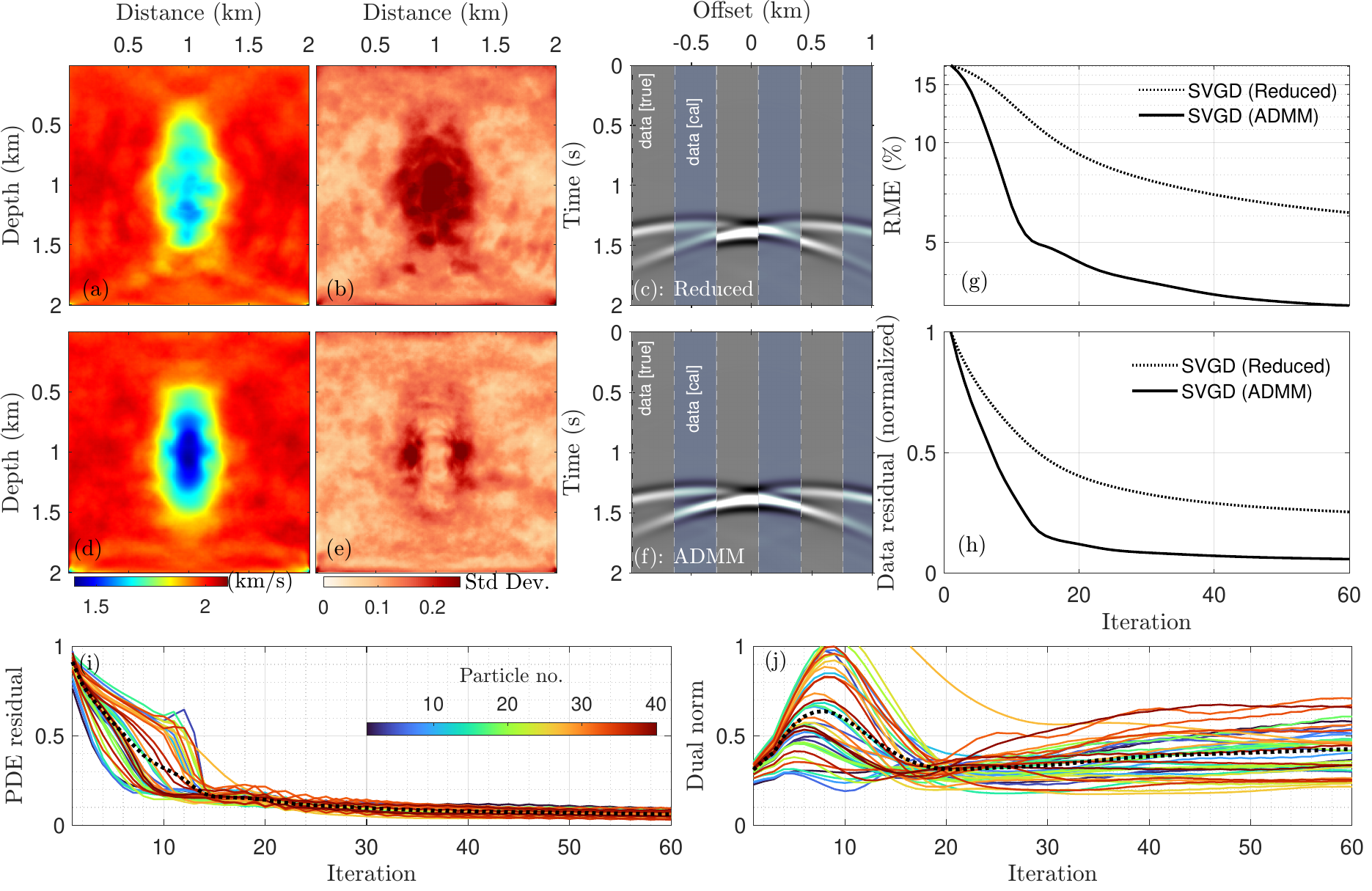}
    \caption{ADMM-SVGD against reduced-space SVGD, Gaussian-anomaly model; 4~Hz data, 60 iterations, identical initial ensemble. (a)--(c) Reduced-space SVGD: conditional mean, pointwise standard deviation, interleaved shot gather (middle source). (d)--(f) ADMM-SVGD, same quantities. (g) Relative model error, (h) normalized data residual. (i) PDE residual and (j) multiplier norm per ADMM-SVGD particle; ensemble mean dashed black.}
    \label{fig:gaussian-reduced-comparison}
\end{figure}

Figure~\ref{fig:gaussian-reduced-comparison} compares the two methods. Panels (a)--(c) show the conditional mean, the pointwise standard deviation, and the interleaved true-versus-reconstructed shot gather for reduced-space SVGD, and panels (d)--(f) the same quantities for ADMM-SVGD; the conditional mean recovered by ADMM-SVGD localizes the anomaly boundary more accurately than its reduced-space counterpart. The pointwise standard deviation provides a spatially resolved assessment of uncertainty: it is largest at the boundaries of the anomaly---where the velocity contrast creates the strongest sensitivity to perturbations---and toward the base of the model. Because the initial particles closely track the homogeneous background away from the anomaly, the ensemble spread concentrates where the true model departs from that background rather than along the weakly illuminated flanks of the domain; the spatial pattern of the posterior uncertainty is accordingly governed jointly by the acquisition geometry and by how informative the initial ensemble already is in each region. Panels (g) and (h) track the relative model error and the normalized data residual over the iterations, with ADMM-SVGD converging faster and to a lower floor on both measures, while panels (i) and (j) show the per-particle evolution of the constraint residual and the multiplier norm, whose ensemble-consistent trajectories indicate that the constraint is enforced uniformly across the particles. The total runtime and peak memory for this experiment are reported in Table~\ref{tab:cost-memory}.
\begin{figure}[htbp]
\centering
\includegraphics[width=0.75\textwidth]{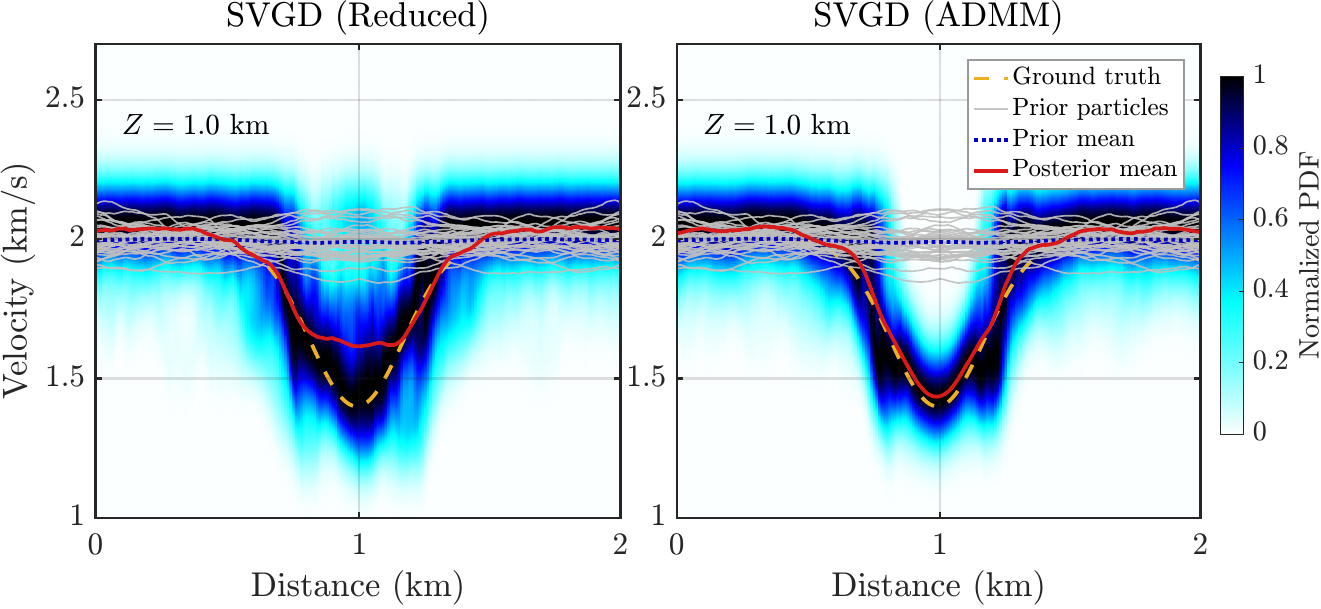}
\caption{Normalized posterior density (color) along the $z=1.0$~km profile through the Gaussian anomaly: reduced-space SVGD (left), ADMM-SVGD (right). Overlaid: conditional mean (solid red), prior mean (dotted blue), prior particles (gray), and true model (dashed orange).}
\label{fig:gaussian-wells-pdf}
\end{figure}

To further assess the quality of the uncertainty estimates, we extract the horizontal velocity profile at $z=1.0$~km through the center of the anomaly for both methods (Figure~\ref{fig:gaussian-wells-pdf}). The prior particles are broadly dispersed about the homogeneous background and carry no imprint of the low-velocity anomaly. Under ADMM-SVGD, the posterior density contracts sharply from this prior spread into a tight band that closely tracks the true velocity across the anomaly, and the conditional mean follows the true profile closely. Under the reduced-space formulation, by contrast, the posterior density remains considerably more diffuse: although it also contracts relative to the prior, its mass spreads more broadly about the true profile, and the conditional mean undershoots the anomaly. This contrast indicates that the dual-space reformulation conditions the sampling problem more favorably even in this simple transmission setting---the same initial ensemble, transported under the relaxed constraint, both localizes the anomaly more accurately and concentrates more decisively about the true profile.
%
%
\subsubsection{Marmousi~II model}\label{sec:Marmousi-II}

\begin{figure}[htbp]
\centering
\includegraphics[width=\textwidth]{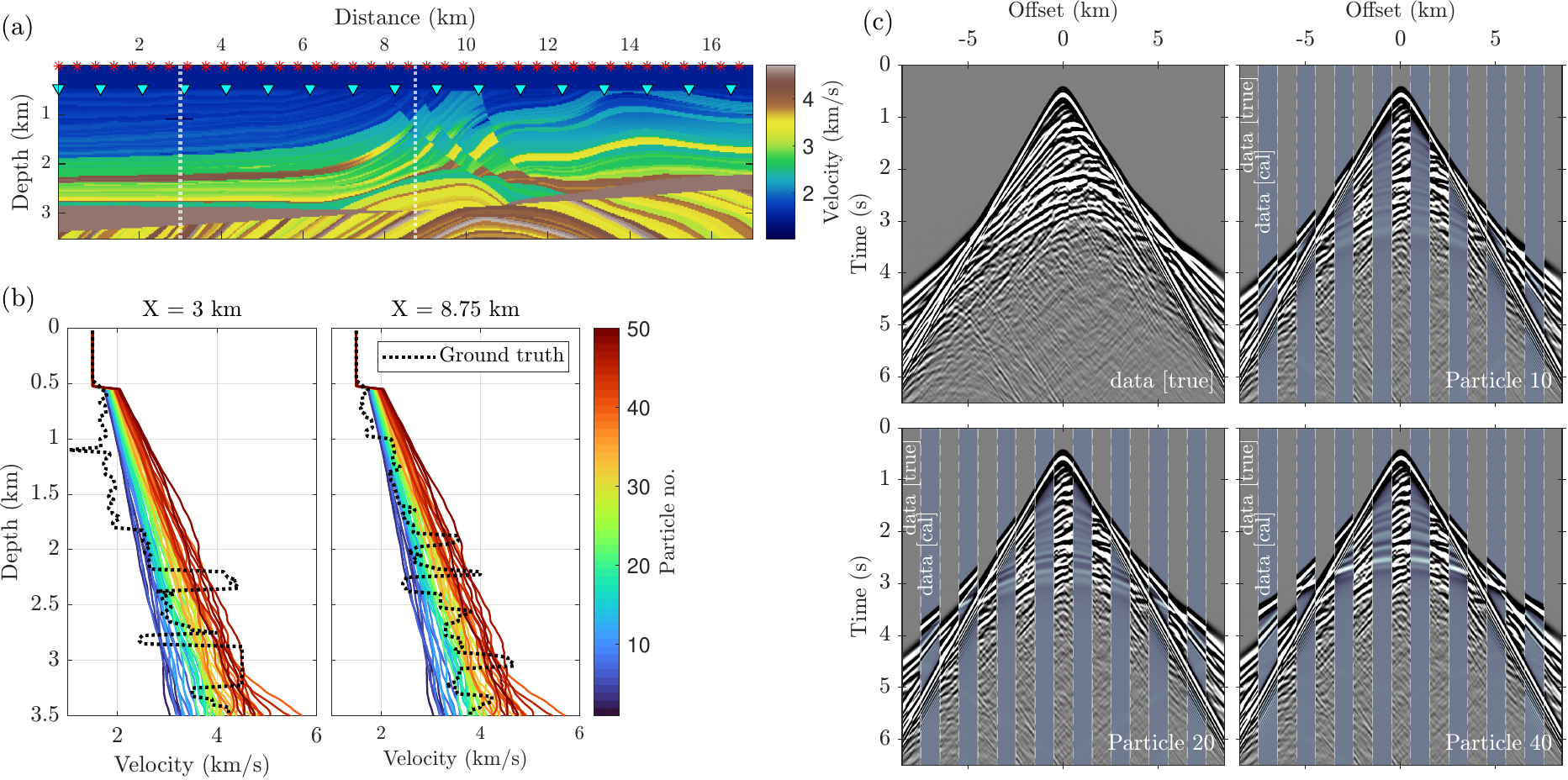}
\caption{Marmousi~II setup. (a) True P-wave velocity; receivers (inverted triangles), sources (red stars). (b) Velocity profiles at $x = 3$ and $8.75$~km (white dashed lines): initial particles (colored), true model (black dotted). (c) Shot gather in the true model, then its traces interleaved with those of particles 10, 20, and 40.}
\label{fig:marm-initial}
\end{figure}

We next apply ADMM-SVGD to the Marmousi~II velocity model \citep{MartinWileyMarfurt2006}, a widely used benchmark featuring complex geological structures including dipping layers, faults, anticlinal traps, and strong lateral velocity variations. The model is 17.0~km wide and 3.5~km deep (Figure~\ref{fig:marm-initial}a; Table~\ref{tab:settings_combined}).
The initial ensemble is constructed in two steps. First, a family of $N_p=50$ background models is obtained by varying the vertical velocity gradient beneath the water layer, whose velocity is held fixed at 1.5~km/s: the baseline gradient, corresponding to a velocity increase from 1.5 to 4.0~km/s over the 120 grid cells below the water layer, is perturbed by slope variations drawn uniformly from the interval $[-0.005, 0.01]$~km/s per grid cell, producing a diverse set of linear velocity trends whose bottom velocities span approximately 3.4--5.2~km/s; these are subsequently converted to the squared-slowness parameterization and denoted $\{\bm{m}^{(j)}_{\text{bg}}\}_{j=1}^{N_p}$. Second, an independent GRF realization is added directly to each squared-slowness background, $\bm{m}^{(j)} = \bm{m}^{(j)}_{\text{bg}} + \delta \bm{m}^{(j)}$, where each $\delta \bm{m}^{(j)}$ is drawn from the spectral formulation of equation~\ref{eq:grf-spectrum} with $\alpha=2$ and $\tau=3$. The resulting initial particles, represented by one-dimensional velocity profiles, are shown in Figure~\ref{fig:marm-initial}b, and Figure~\ref{fig:marm-initial}c compares the shot gathers simulated in the true model against those simulated in several representative initial particles. While the near-offset arrivals exhibit consistent kinematic behavior, noticeable phase discrepancies---indicating cycle-skipping between the observed data and the initial predictions---grow with offset as the traveltime differences between the initial models and the true model accumulate, illustrating the challenging nonlinear landscape the inversion faces.

The inversion is carried out in two multiscale frequency-domain cycles over the 3--12~Hz band (Table~\ref{tab:settings_combined}), with the final ensemble of the first cycle warm-starting the second.
During the model update, the water layer is assumed known and is held fixed in the gradient calculation. In addition, the updated particles are projected onto a feasible set defined by velocity bounds derived from prior information (Table~\ref{tab:settings_combined}), which prevents nonphysical updates.
To investigate the effect of data coverage on the posterior distribution, we run three acquisition configurations with $N_s = 17$, $34$, and $68$ sources uniformly distributed along the surface. The source signature and receiver geometry are common to all three configurations (Table~\ref{tab:settings_combined}).

\begin{figure}[htbp]
\centering
\includegraphics[width=\textwidth]{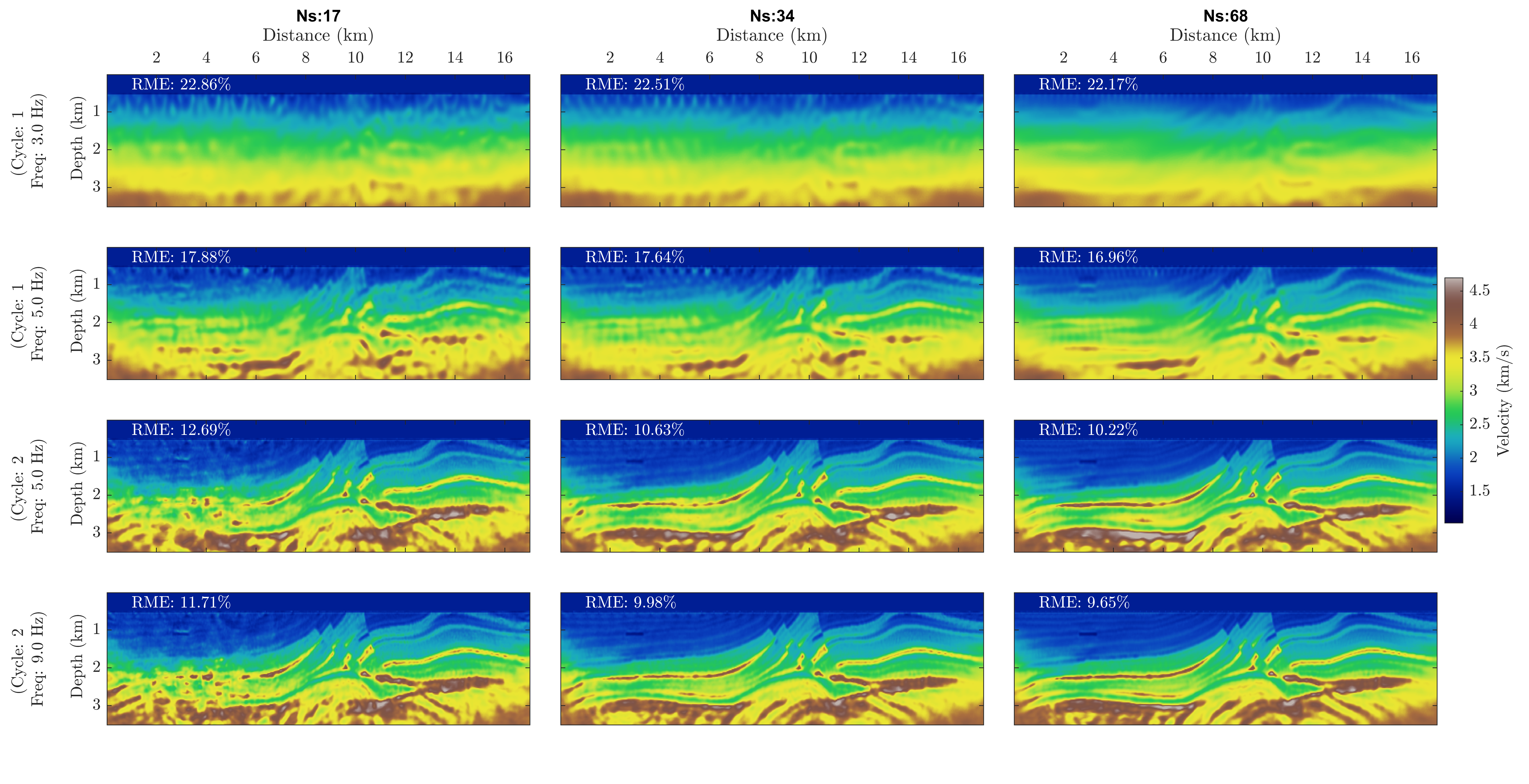}
\caption{Conditional mean velocity at successive cycle--frequency stages for $N_s=17$, $34$, and $68$ (columns). Rows are cycle--frequency pairs; RME is reported per panel.}
\label{fig:marm-results-Means}
\end{figure}

\begin{figure}[htbp]
\centering
\includegraphics[width=\textwidth]{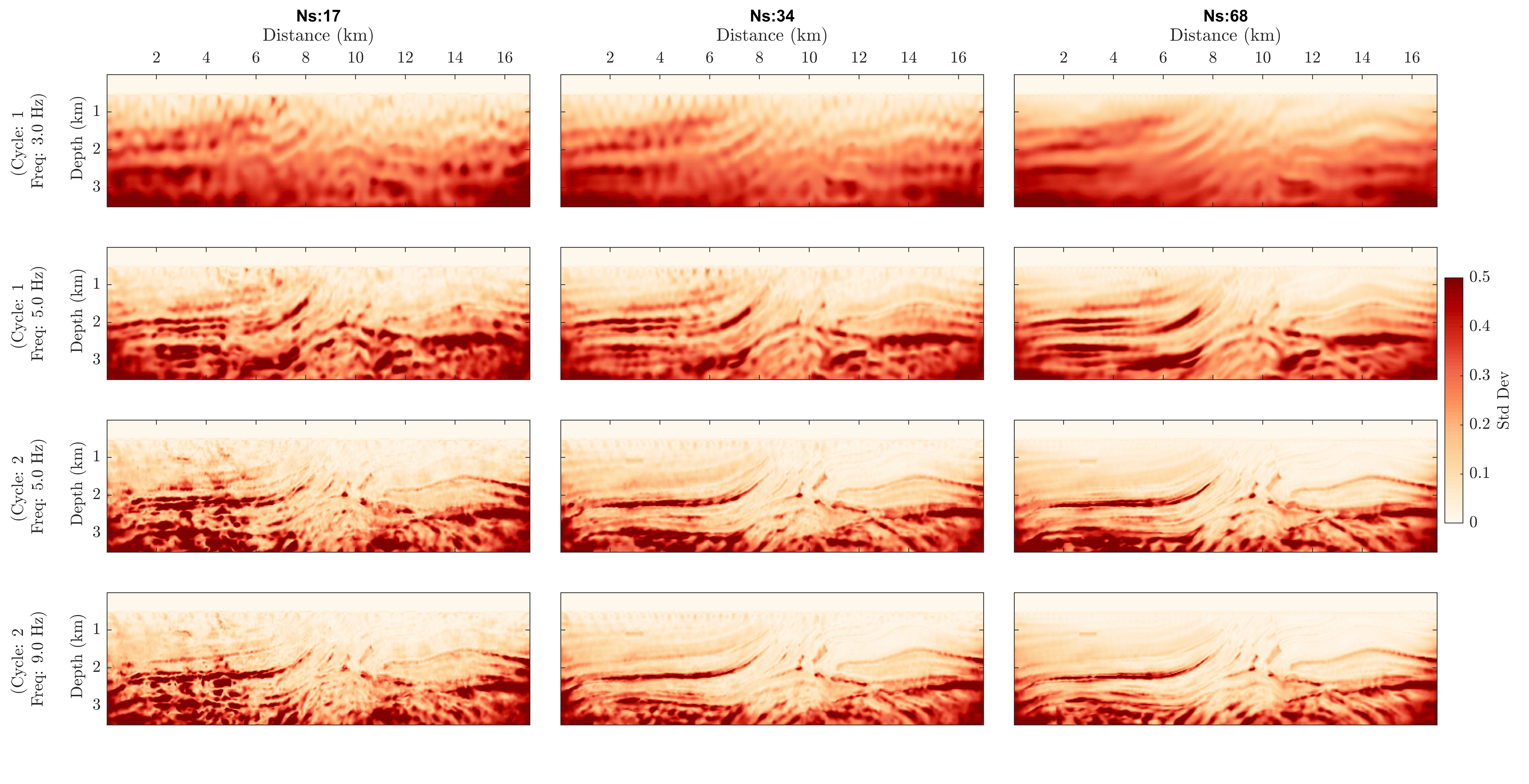}
\caption{Pointwise posterior standard deviation for the stages and ensemble sizes of Figure~\ref{fig:marm-results-Means}.}
\label{fig:marm-results-Stds}
\end{figure}

\begin{figure}[htbp]
\centering
\includegraphics[width=\textwidth]{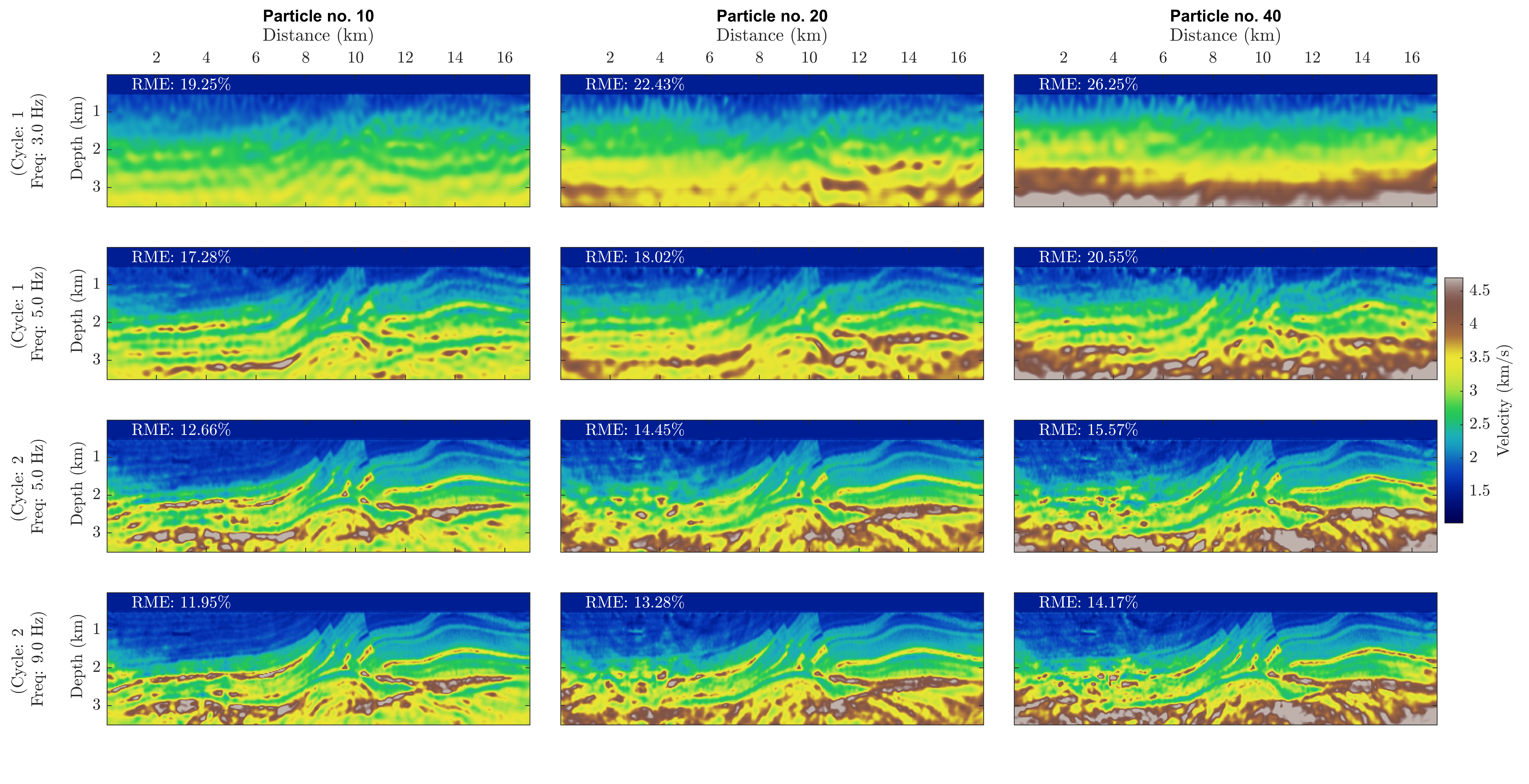}
\caption{Velocity estimates for particles 10, 20, and 40 (columns) with $N_s=17$ sources. Rows are successive cycle--frequency stages.}
\label{fig:marm-particles-ns17}
\end{figure}

\begin{figure}[htbp]
\centering
\includegraphics[width=\textwidth]{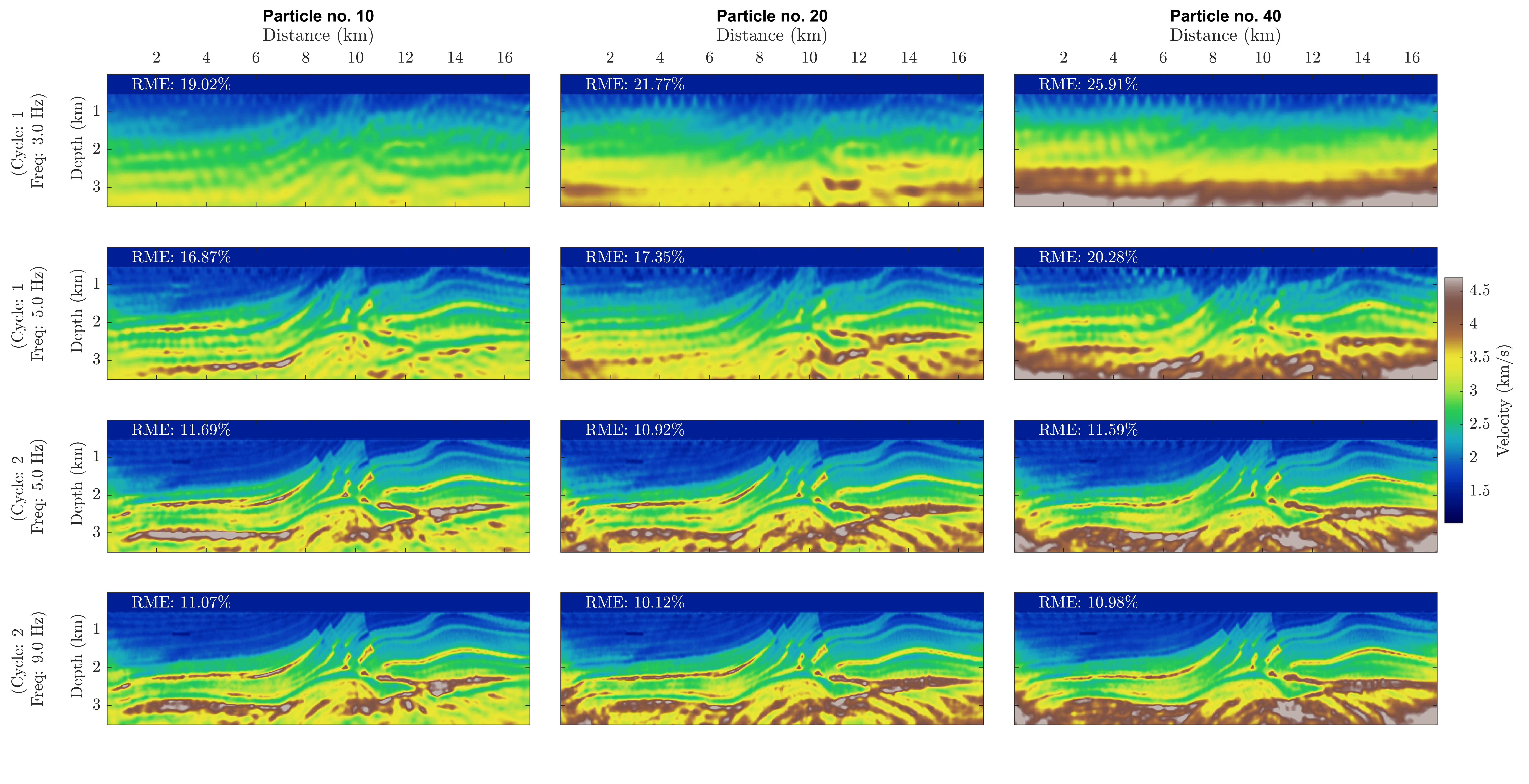}
\caption{Same as Figure~\ref{fig:marm-particles-ns17} but for $N_s=34$.}
\label{fig:marm-particles-ns34}
\end{figure}

\begin{figure}[htbp]
\centering
\includegraphics[width=\textwidth]{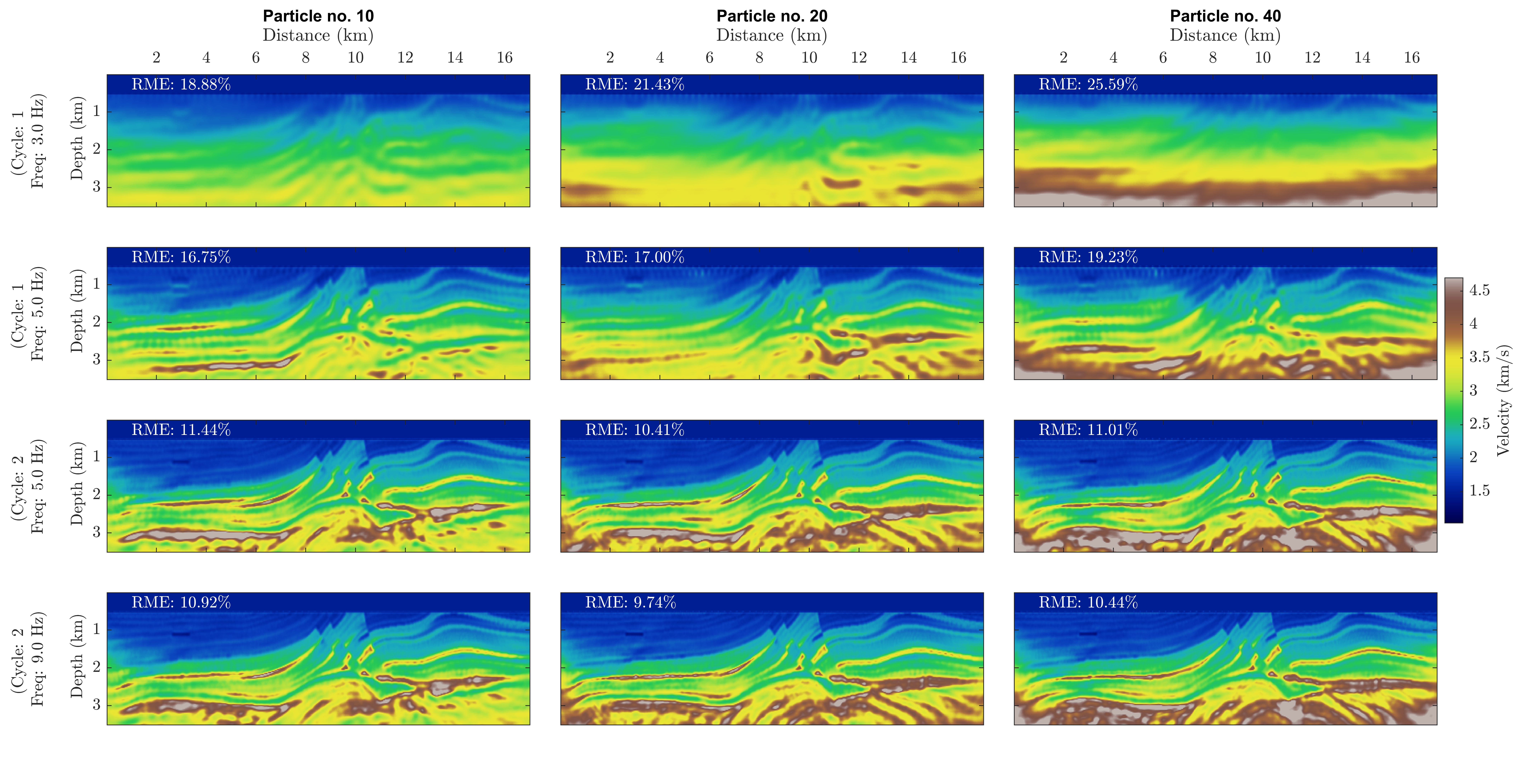}
\caption{Same as Figure~\ref{fig:marm-particles-ns17} but for $N_s=68$.}
\label{fig:marm-particles-ns68}
\end{figure}

Figure~\ref{fig:marm-results-Means} shows the evolution of the conditional mean velocity estimate across inversion cycles and frequencies, organized by cycle-frequency (rows) and source number (columns). In the first cycle at low frequencies (3--5~Hz), the conditional mean recovers only the large-scale background structure of the model. As higher frequencies are introduced and the inversion advances into the second cycle, structural details such as dipping reflectors, anticlines, and lateral velocity contrasts become progressively sharper. The RME decreases systematically across cycles for all source configurations. The corresponding pointwise posterior standard deviation is presented in Figure~\ref{fig:marm-results-Stds}. Early in cycle 1, the uncertainty is broadly distributed across the model. As the inversion progresses through the frequency schedule and into cycle 2, the standard deviation decreases throughout the model---most notably in the shallow section, where the surface acquisition provides strong illumination. The uncertainty remains persistently elevated at depth and near the model boundaries, reflecting the inherent limitations of surface-based seismic acquisition for resolving deep structure. Figures~\ref{fig:marm-particles-ns17}--\ref{fig:marm-particles-ns68} show the velocity estimates of three representative particles (indices 10, 20, and 40) across inversion stages for $N_s=17,~34,~\text{and}~68$ sources, respectively. These particle-level views complement the ensemble statistics shown in Figures~\ref{fig:marm-results-Means} and \ref{fig:marm-results-Stds} by illustrating the diversity of models that collectively represent the posterior. At the start of cycle 1, the individual particles differ markedly from one another and from the true model, reflecting the variability of the prior. As the inversion progresses, the particles converge toward consistent large-scale features---particularly in the shallow section---while retaining particle-specific differences in fine-scale structure and at depth. This residual spread is consistent with genuine posterior uncertainty rather than inversion instability, though the finite particle ensemble also contributes. Comparing across source configurations, interparticle variability decreases markedly as the number of sources increases, consistent with stronger posterior contraction under improved data coverage.

\begin{figure}[htbp]
\centering
\includegraphics[width=\textwidth]{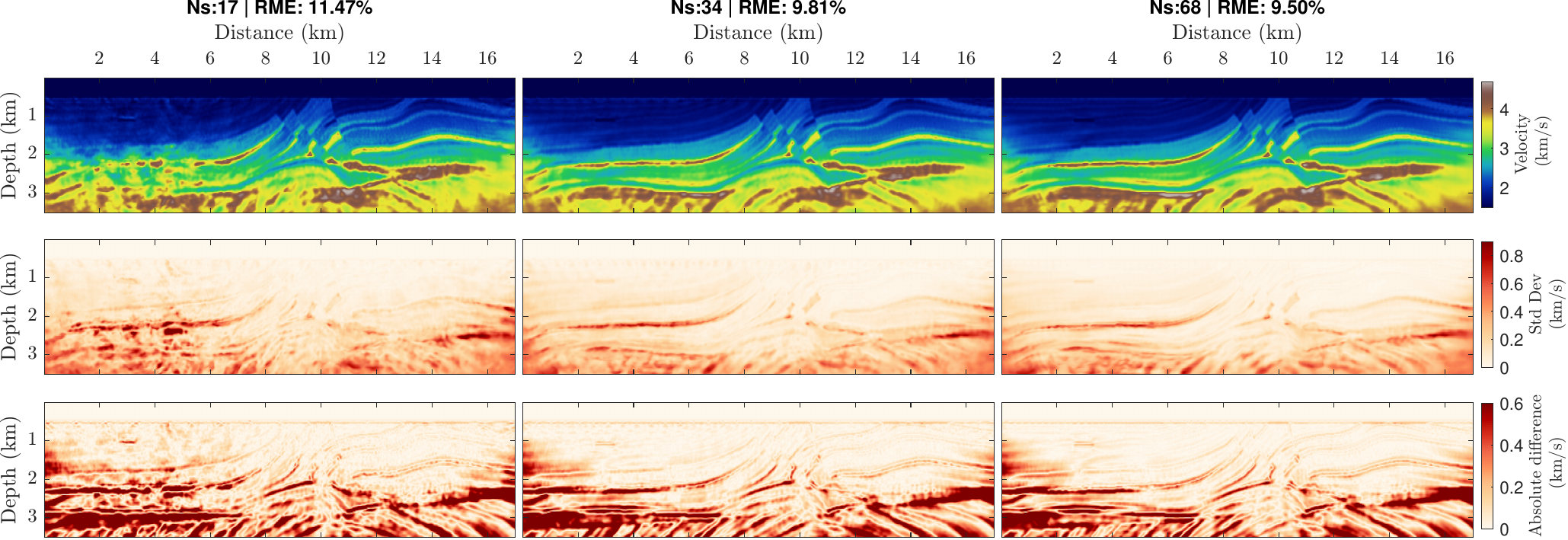}
\caption{Marmousi~II results for $N_s = 17$, $34$, and $68$ sources (left to right). Top: conditional mean velocity with RME. Middle: pointwise standard deviation. Bottom: absolute difference from the true model.}
\label{fig:marm-results}
\end{figure}

Figure~\ref{fig:marm-results} summarizes the final inversion results for the three source configurations. The top row displays the conditional mean velocity estimates, which improve as the number of sources increases, the relative model error falling monotonically across the three configurations. The conditional mean captures the major geological features of the Marmousi model, including the dipping reflectors, the anticlinal structures, and the velocity inversions at depth. The middle row shows the pointwise standard deviation, which serves as a spatially resolved uncertainty map. With $N_s = 17$ sources, the uncertainty is broadly distributed, reflecting the limited data coverage. As the number of sources increases to 34 and 68, the pointwise standard deviation decreases---particularly in the shallow section where the denser source coverage provides better illumination. The uncertainty remains elevated in the deep section of the model and near the lateral boundaries, consistent with the reduced sensitivity of surface measurements to deep structures. The bottom row depicts the absolute difference between the conditional mean and the true velocity model. Notably, the error and standard deviation maps are strongly correlated, and both increase systematically with depth. This spatial agreement is consistent with well-calibrated uncertainty, although it does not by itself constitute a formal calibration test.

\begin{figure}[htbp]
\centering
\includegraphics[width=0.7\textwidth]{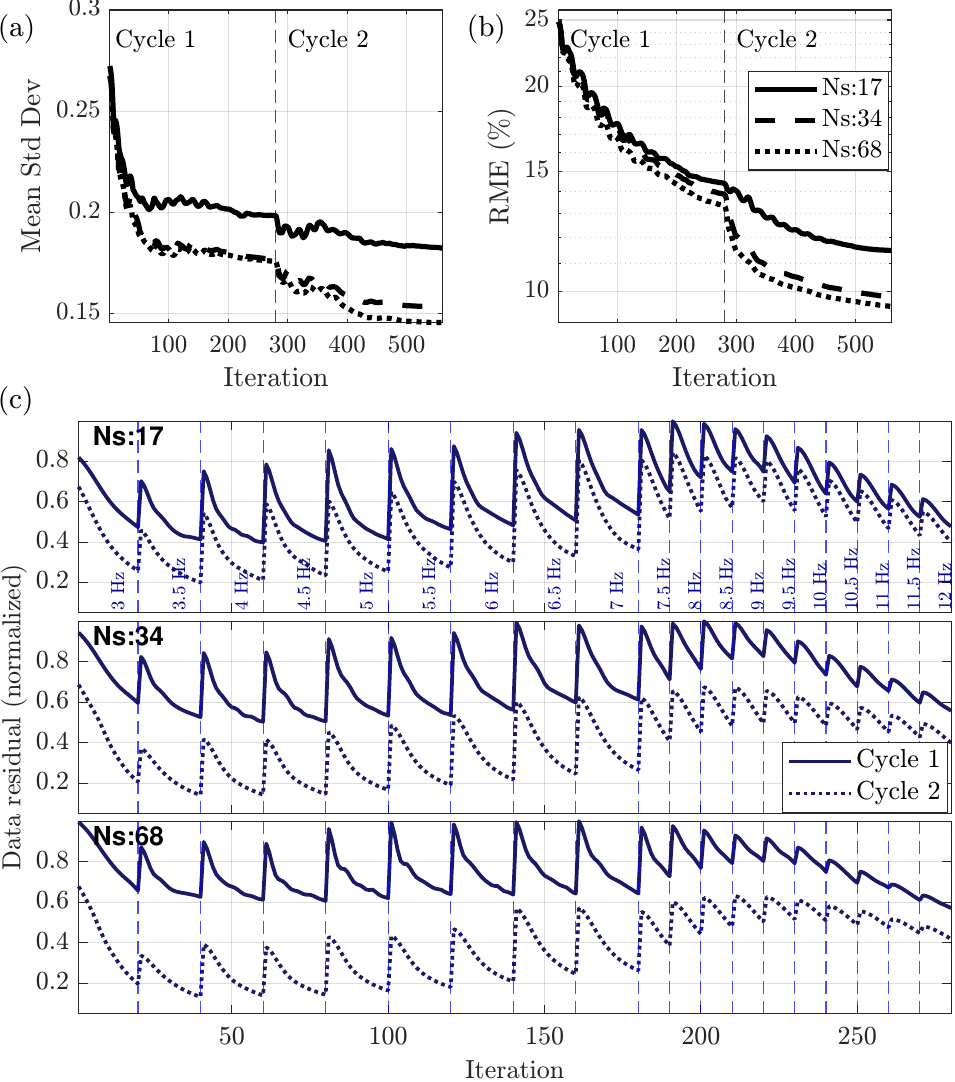}
\caption{Marmousi~II convergence. (a) Mean pointwise standard deviation and (b) relative model error versus iteration; the vertical dashed line separates cycles. (c) Normalized data residual for $N_s=17$, $34$, and $68$ (top to bottom), cycle 1 solid and cycle 2 dotted, with frequency transitions dashed.}
\label{fig:marm-convergence}
\end{figure}

The convergence behavior underlying these results is summarized in Figure~\ref{fig:marm-convergence}. The mean pointwise standard deviation (Figure~\ref{fig:marm-convergence}a) decreases over iterations for all source configurations, indicating progressive posterior contraction as the data constraints are assimilated. The RME curves (Figure~\ref{fig:marm-convergence}b) show a consistent decreasing trend, with the rate of decrease being faster for larger numbers of sources. For $N_s = 68$, the RME drops below 10\% by roughly 500 iterations, whereas with $N_s = 17$ the error plateaus around 11.5\%. This behavior is consistent with the expected Bayesian posterior contraction property: more observed data leads to a more concentrated posterior and a more accurate conditional mean estimate. The normalized data residual (Figure~\ref{fig:marm-convergence}c) decreases within each frequency stage and across the two cycles for all three configurations. The total runtime and peak memory of the three runs are also reported in Table~\ref{tab:cost-memory}.


\begin{figure}[htbp]
    \centering
    \includegraphics[width=0.8\linewidth]{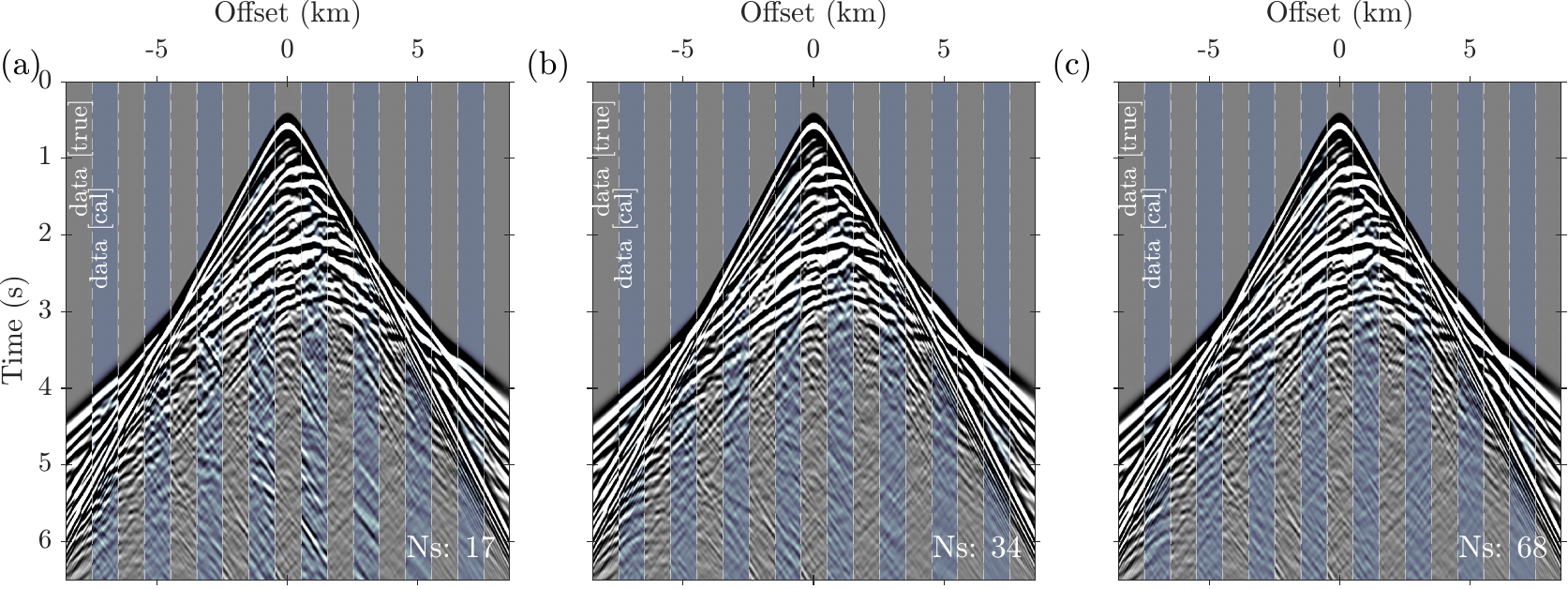}
    \caption{Interleaved shot gathers for (a) $N_s=17$, (b) $34$, and (c) $68$ sources, alternating trace segments from the true seismogram and from the conditional-mean model.}
    \label{fig:marm_data_ns}
\end{figure}

Because the relative model error presupposes knowledge of the true model and is therefore unavailable in field applications, we complement it with a direct assessment of the data fit. Figure~\ref{fig:marm_data_ns} interleaves alternating trace segments of the observed seismogram with the seismogram computed in the conditional-mean model, for each of the three acquisition configurations. Across all three, the reflection events continue across the boundaries between adjacent segments without a visible break in traveltime or phase, indicating that the conditional mean reproduces the recorded wavefield---including the later reflections that follow the direct arrival---rather than merely its kinematic envelope. The segment boundaries are least well matched at the largest offsets, where the surface acquisition constrains the model least and where the posterior uncertainty reported above is correspondingly largest.

\begin{figure}[htbp]
    \centering
    \includegraphics[width=\textwidth]{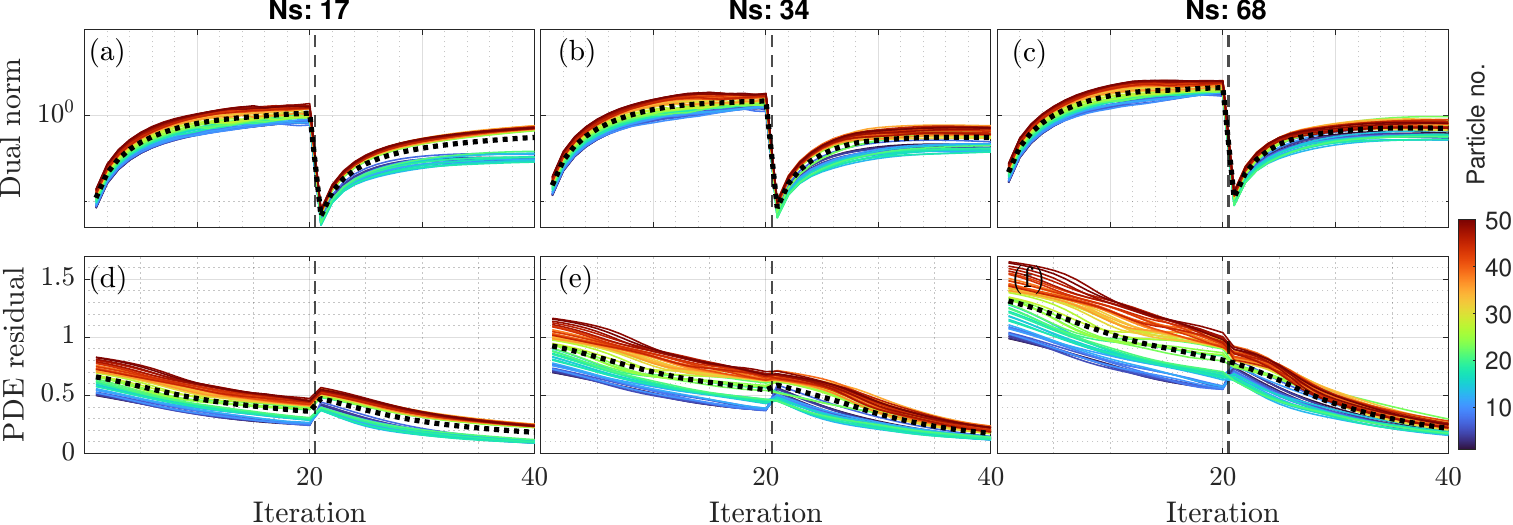}
    \caption{AL terms for Marmousi~II at the lowest frequency: dual norm (a)--(c) and constraint-residual norm (d)--(f), for $N_s=17$, $34$, and $68$ (columns). Individual particles colored, ensemble mean black dotted. The vertical dashed line marks the cycle boundary, where the multipliers are reinitialized.}
    \label{fig:marm-dual-evolution}
\end{figure}

The convergence argument developed earlier rests on the multiplier and penalty terms being progressively suppressed as the PDE constraint is enforced, and the Marmousi~II experiment allows this mechanism to be observed directly at realistic scale. Figure~\ref{fig:marm-dual-evolution} tracks the multiplier norm and the PDE-residual norm over the iterations at the lowest inversion frequency, for the three acquisition configurations. Within each cycle the two quantities move in opposite directions: the residual decays as the primal updates draw the wavefields toward the wave equation, while the multiplier accumulates the corresponding correction and rises toward a plateau. This is precisely the behavior the dual iteration is constructed to produce. The pronounced drop at the cycle boundary reflects the reinitialization of the multipliers at the start of the second cycle, after which the same pattern repeats from the warm-started ensemble. The individual particles follow this trajectory closely---no member of the ensemble departs from it---indicating that the constraint is enforced consistently across the ensemble rather than for a favored subset of particles. It is worth noting that the residual decreases substantially but remains finite at the end of the low-frequency stage, so that the extended density approaches its reduced-space counterpart progressively rather than attaining it exactly; the reported posterior is the one associated with this level of constraint satisfaction, a level that the multiplier updates tighten as the iterations proceed.

\begin{figure}[htbp]
\centering
\includegraphics[width=0.68\textwidth]{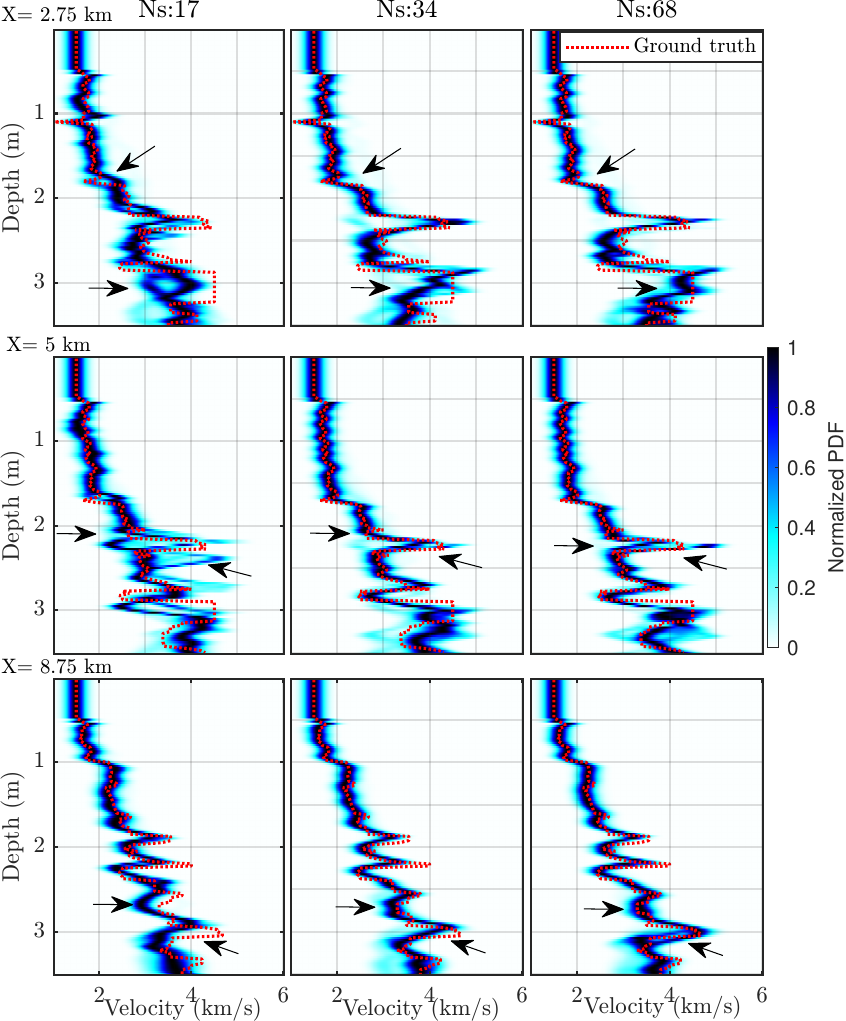}
\caption{Vertical profiles of the normalized posterior density at three lateral positions for $N_s = 17$, $34$, and $68$ sources (left to right). True velocity profile dotted red.}
\label{fig:marm-pdf-vertical}
\end{figure}

To examine the depth-dependent structure of the posterior, Figure~\ref{fig:marm-pdf-vertical} shows vertical profiles of the normalized posterior probability density at three lateral positions across the model ($X=2.75,~5.0,~8.75~\text{km}$). At each depth, the color intensity represents the probability density over velocity, with the true velocity overlaid as a red dotted curve. In the shallow section, the posteriors are tightly concentrated around the true velocity for all source configurations, reflecting the strong data constraints near the surface. With increasing depth, the posterior densities broaden and, in geologically complex regions, develop multimodal structure---indicating that several velocity values are consistent with the observed data. As the number of sources increases from $N_s = 17$ to $68$, the posteriors sharpen and the probability mass concentrates more tightly around the true velocity. This posterior contraction with increasing data coverage is the behavior expected as the data become more informative: each additional source provides independent constraints on the subsurface velocity, reducing the volume of model space consistent with the data and yielding tighter credible intervals. The persistent broadening at depth, irrespective of source count, reflects the intrinsic resolution limits of surface-based seismic acquisition geometry.

\begin{figure}[htbp]
\centering
\includegraphics[width=0.95\textwidth]{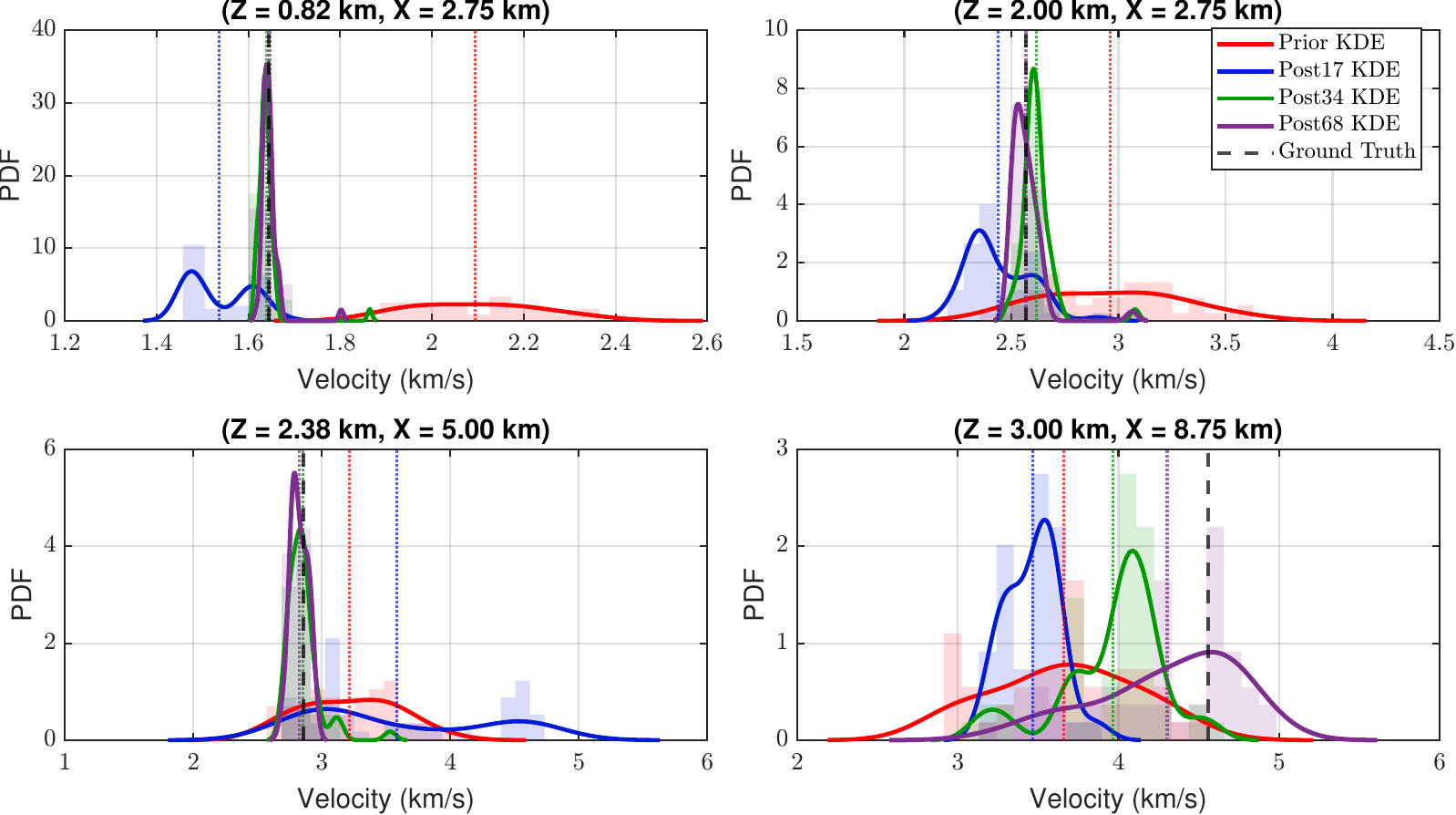}
\caption{Pointwise posterior density at four locations in the Marmousi~II model for $N_s = 17$, $34$, and $68$ sources. Prior in red, true velocity dashed black.}
\label{fig:marm-pdf}
\end{figure}

Zooming into specific locations, Figure~\ref{fig:marm-pdf} displays pointwise posterior probability density estimates at four positions in the model, each corresponding to a different geological setting. At the shallow location $(z = 0.82$~km, $x = 2.75$~km$)$, the posteriors are unimodal and tightly concentrated around the true velocity (dashed black), with increasing sharpness as the number of sources grows. The prior (red) is broad and uninformative, while the posteriors contract progressively toward the true velocity. At deeper and more geologically complex locations, the posterior distributions become broader and, in some cases, multimodal. While such multimodality is consistent with genuine ambiguity in the velocity structure that the available data cannot resolve, part of it may also reflect the finite particle ensemble. For instance, at $(z = 2.4$~km, $x = 5.0$~km$)$, the posteriors exhibit multiple peaks at plausible velocity values for this depth, with the multimodality diminishing as more sources are added. Where it is robust, this ability to reveal multimodal uncertainty at geologically complex locations is valuable for risk assessment in geophysical applications, where identifying the full range of plausible models is as important as obtaining a single point estimate.

\begin{table}[htbp]
\centering
\caption{Total runtime and peak memory for reduced-space SVGD and ADMM-SVGD in the Gaussian-anomaly (Figure~\ref{fig:gaussian-reduced-comparison}) and Marmousi~II (Figure~\ref{fig:reduced-vs-admm}) experiments, with ADMM-SVGD resolved by source count $N_s$.}
\label{tab:cost-memory}
\footnotesize
\begin{tabular}{llccc}
\toprule
\textbf{Experiment} & \textbf{Method} & \textbf{$N_s$} & \textbf{Runtime (min)} & \textbf{Peak memory (GB)} \\
\midrule
\multirow{2}{*}{Gaussian anomaly}
    & Reduced-space SVGD & 50 & 20.83  & 9.29  \\
    & ADMM-SVGD          & 50 & 32.76  & 12.00 \\
\midrule
\multirow{4}{*}{Marmousi~II}
    & \multirow{3}{*}{ADMM-SVGD} & 17 & 329.87 & 10.031 \\
    &                            & 34 & 463.62 & 14.934 \\
    &                            & 68 & 829.09 & 25.968 \\
    & Reduced-space SVGD         & 34 & 363.67 & 10.098 \\
\bottomrule
\end{tabular}
\end{table}

\begin{figure}[htbp]
\centering
\includegraphics[width=\textwidth]{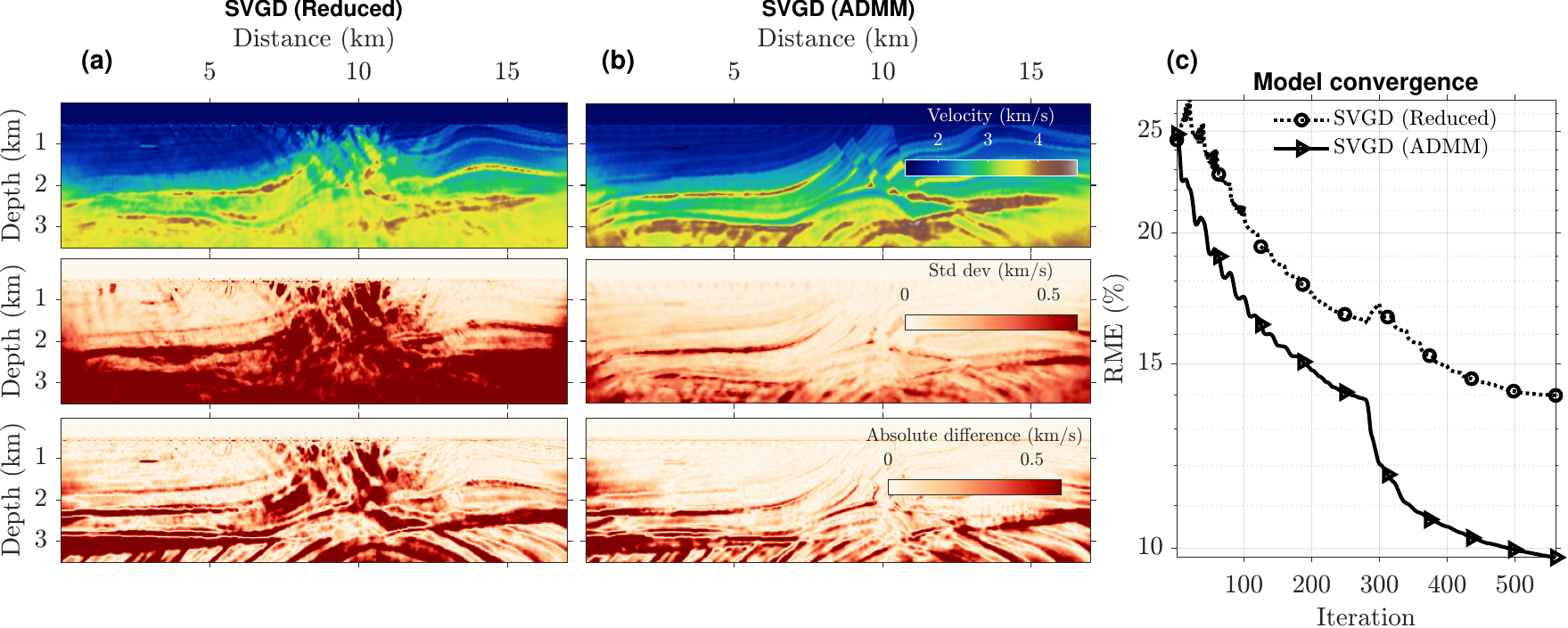}
\caption{Reduced-space SVGD (a) and ADMM-SVGD (b), Marmousi~II; $N_s=34$, $N_p=50$, identical initial ensemble. Top to bottom: conditional mean, pointwise standard deviation, absolute error against the true model. (c) Relative model error versus iteration.}
\label{fig:reduced-vs-admm}
\end{figure}

\paragraph{Comparison with reduced-space sampling and deterministic inversion.}
The dual-space formulation is motivated by the favorable conditioning of the extended problem, and the Marmousi~II setting---in which the initial ensemble is obtained by perturbing a one-dimensional background model, so that a subset of its members is already cycle-skipped relative to the observed data---offers a direct test of whether that advantage survives at realistic scale. To isolate it, we run reduced-space SVGD from the identical initial ensemble, with the same source configuration and frequency schedule as ADMM-SVGD (Figure~\ref{fig:reduced-vs-admm}). Reduced-space SVGD does not recover from this initialization: its conditional mean retains pronounced, spatially localized artifacts through the shallow-to-mid section, and both the pointwise standard deviation and the absolute error remain elevated over the same region, indicating that a subset of the particles remains trapped near cycle-skipped minima for the duration of the inversion. Initialized identically, ADMM-SVGD yields a considerably cleaner conditional mean with markedly lower spread and error, and its relative model error lies below that of the reduced-space run at every iteration, terminating near 10\% against roughly 14\%. This contrast is the FWI counterpart of the conditioning argument developed above: relaxing the wave equation allows each particle to fit the observed data before it is consistent with its own model, and the multiplier updates subsequently draw the ensemble back onto the constraint---a path that is unavailable to a sampler confined to the reduced space.

\begin{figure}[htbp]
    \centering
    \includegraphics[width=0.9\linewidth]{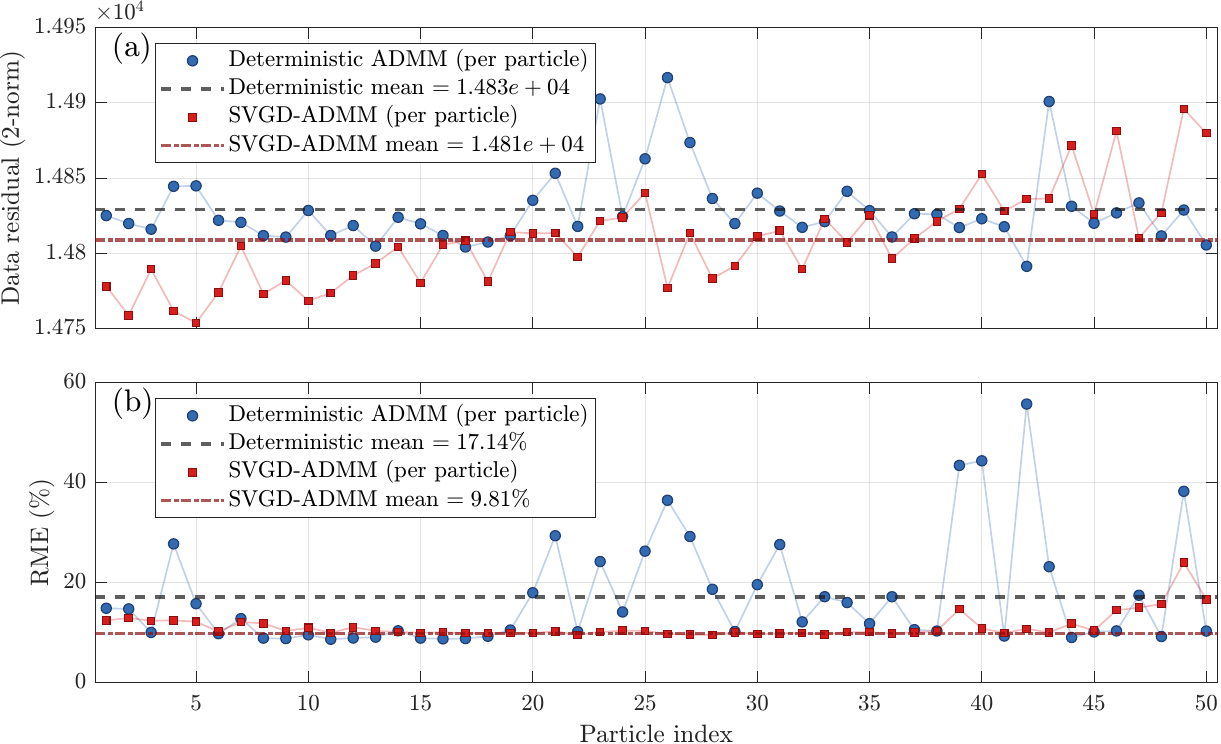}
    \caption{Deterministic ADMM against ADMM-SVGD, Marmousi~II ($N_s = 34$). (a) Final per-particle data residual (2-norm): deterministic (blue circles), ADMM-SVGD (red squares); ensemble means black dashed and red dash-dot. (b) Per-particle relative model error, with means.}
    \label{fig:det_vs_svgd}
\end{figure}

A complementary question is what the ensemble affords relative to deterministic inversion at a matched cost per particle. We therefore run deterministic ADMM independently from each of the initial particles used to seed ADMM-SVGD, so that the two are compared under identical initial conditions on a per-particle basis (Figure~\ref{fig:det_vs_svgd}). The two ensembles fit the observed data essentially equally well---their final data residuals are essentially equal---yet their model errors are not: the deterministic runs scatter widely, with several particles stalling at relative model errors of 40--56\% and an ensemble mean of 17.14\%, whereas the ADMM-SVGD particles remain tightly clustered---with a single exception---about a mean of 9.81\%. This is a signature of non-uniqueness rather than of data fit---many models explain the observations equally well---and the kernel interaction in the SVGD update allows each particle to exploit information carried by its neighbors, tempering the tendency of an isolated optimization trajectory to stall in a poor local minimum, an avenue that is not available to a set of independent deterministic inversions.

\begin{figure}[htbp]
\centering
\includegraphics[width=0.6\textwidth]{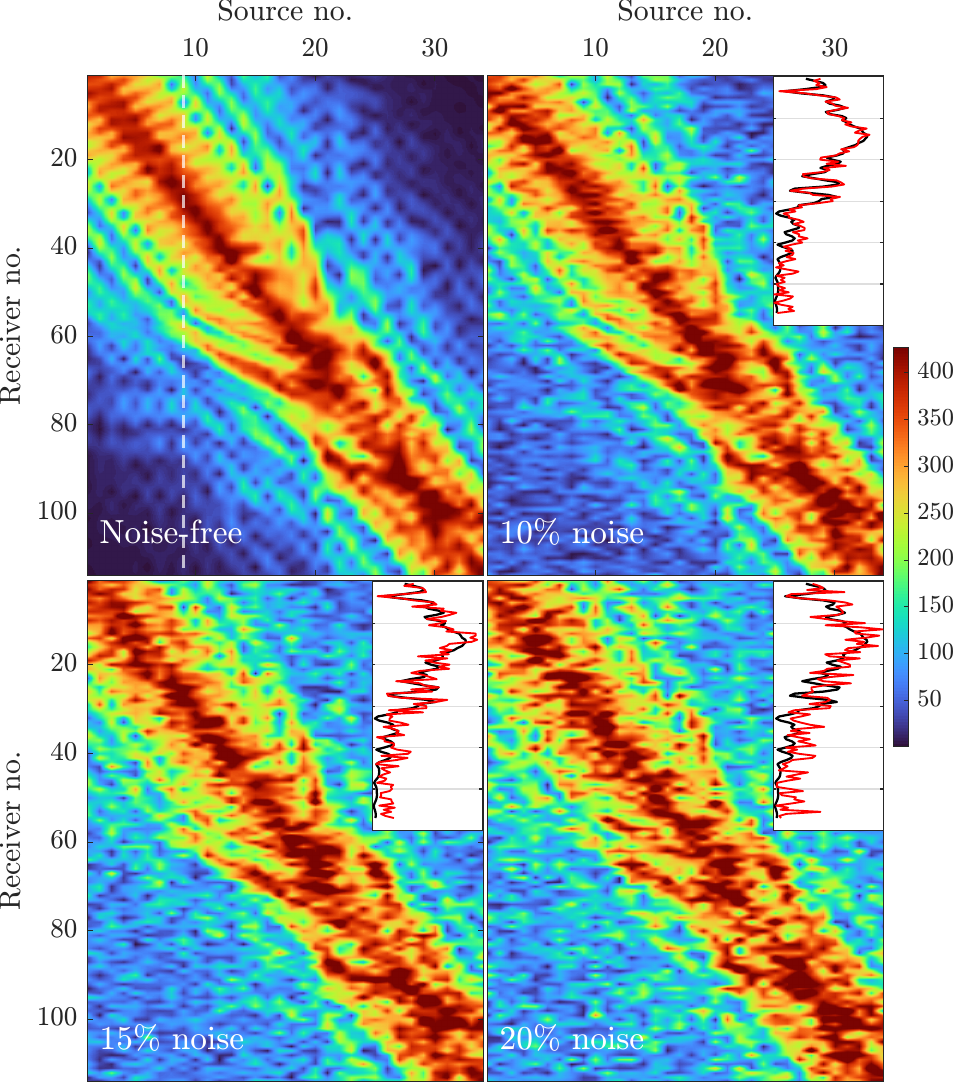}
\caption{Magnitude of the 3~Hz data in the source--receiver domain. Top left: noise-free; the rest add 10\%, 15\%, and 20\% complex Gaussian noise per frequency. White dashed line: source gather no.~9, inset showing its magnitude versus receiver number, noise-free (black) and noisy (red).}
\label{fig:marm-data}
\end{figure}

\begin{figure}[htbp]
\centering
\includegraphics[width=\textwidth]{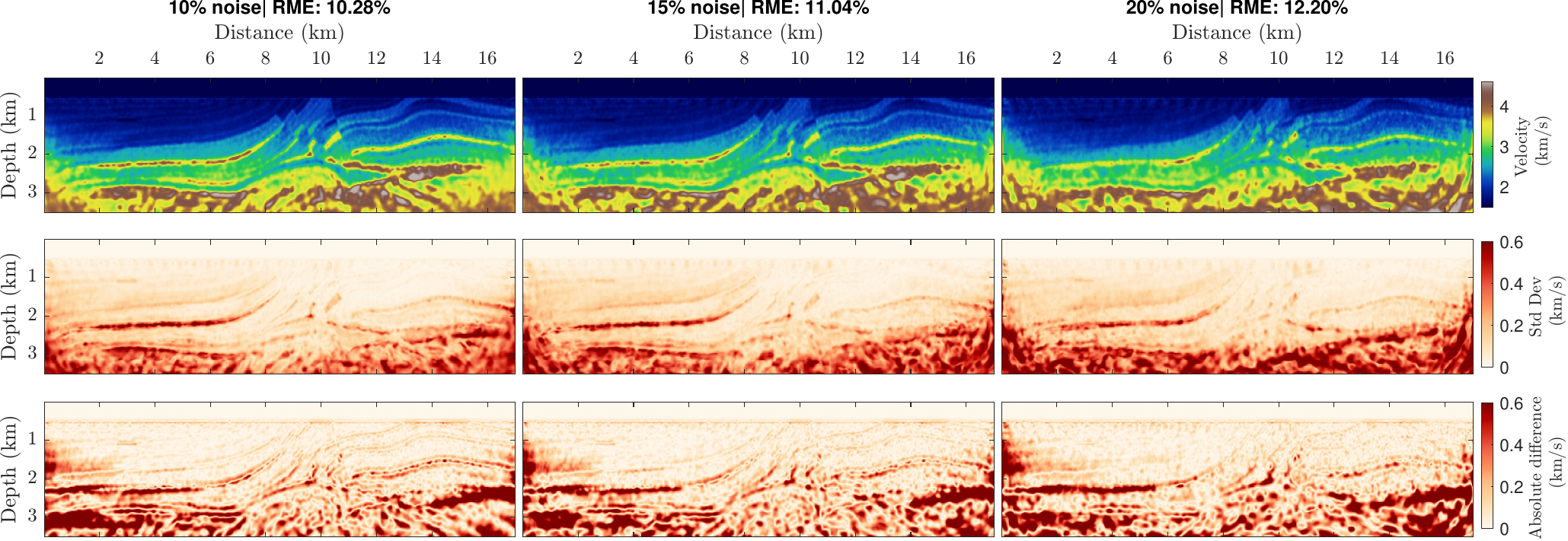}
\caption{Marmousi~II inversion under increasing noise (left to right). Top: conditional mean velocity with RME. Middle: pointwise posterior standard deviation. Bottom: absolute difference from the true model.}
\label{fig:marm-noise-image}
\end{figure}

\begin{figure}[htbp]
\centering
\includegraphics[width=0.70\textwidth]{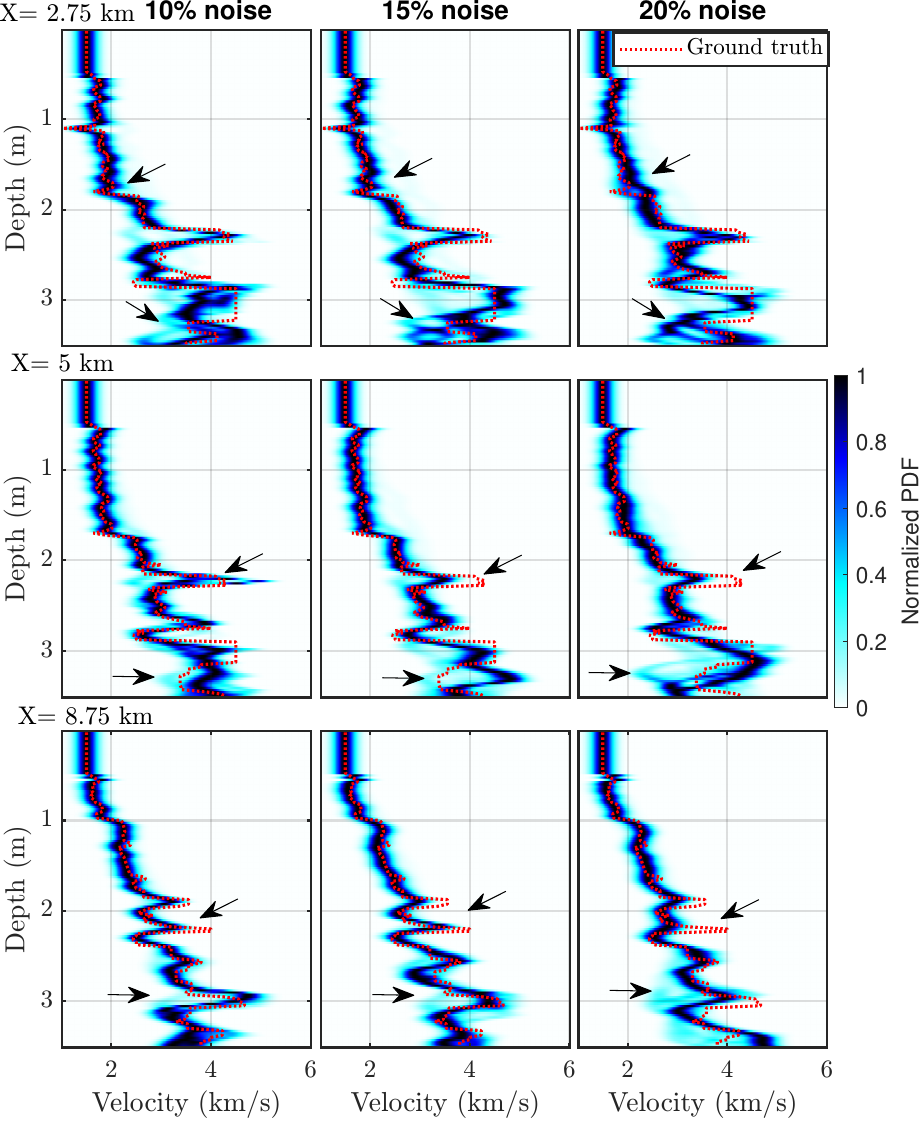}
\caption{Vertical profiles of the normalized posterior density at three lateral positions for $N_s=34$, with 10\%, 15\%, and 20\% noise (left to right). True velocity profile dotted red.} 
\label{fig:marm-noise-PDF}
\end{figure}

\paragraph{Noise case.}
To investigate the effect of noise, we repeat the Marmousi~II experiment with complex-valued Gaussian noise added to the frequency-domain data independently at each frequency. For a given noise level, the noise standard deviation is defined as a fixed percentage of the maximum absolute value of the data, $\sigma=\alpha~\text{max}(|d|)$, where $\alpha=0.10, 0.15~\text{and}~0.20$ correspond to 10\%, 15\%, and 20\% noise levels, respectively. The noise realization is generated as 
\begin{equation*}
    e=\frac{\sigma}{\sqrt{2}}\left(\mathcal{N}(0,1)+i\mathcal{N}(0,1)\right),
\end{equation*}
which ensures equal variance in the real and imaginary components. All other experimental settings---acquisition geometry, prior specification, frequency schedule, and number of particles---are kept identical to the noise-free case with $N_s=34$ sources, allowing the effect of noise to be isolated. 
Figure~\ref{fig:marm-data} compares the magnitude of the 3~Hz data in the source--receiver coordinate system for the noise-free case and for increasing noise levels. In the absence of noise, the magnitude plot shows a clear and coherent diagonal energy pattern, reflecting the dominant source--receiver response at this low frequency. As the noise level increases from 10\% to 20\%, the background becomes progressively more contaminated by random fluctuations, leading to a gradual reduction in contrast and continuity of the coherent energy trend. In a Bayesian framework, the noise level enters the sampler through the compound parameter $\mu\sigma^2$, which balances data fidelity against constraint enforcement and is determined at each iteration by the residual-whiteness principle rather than by manual tuning.
Figure~\ref{fig:marm-noise-image} presents the inversion results under the three noise levels, organized in the same format as Figure~\ref{fig:marm-results}: the top row shows the conditional mean velocity estimate with its corresponding RME, the middle row shows the pointwise posterior standard deviation, and the bottom row shows the absolute difference between the conditional mean and the true model. The conditional mean estimates remain geologically coherent across all noise levels, successfully recovering the major structural features of the Marmousi~II model, including the dipping reflectors and anticlinal structures. However, the RME increases monotonically with noise level from 10.28\% at 10\% noise to 11.04\% at 15\% and 12.20\% at 20\%, reflecting the reduced information content of the contaminated data and the increased difficulty of distinguishing signal from noise.
Figure~\ref{fig:marm-noise-PDF} shows vertical profiles of the normalized posterior probability density at the same three lateral positions as Figure~\ref{fig:marm-pdf-vertical} ($X=2.75,~5.0~,~\text{and}~8.75~\text{km}$), now for data contaminated with noise. The structure of these profiles mirrors the depth-dependent behavior observed in the noise-free case, but with systematically broader posterior bands at all depths and noise levels. In the shallow section, the posteriors remain approximately unimodal and concentrated near the true velocity across all noise levels, though the concentration is visibly looser than in the noise-free case. With increasing depth, the posterior densities broaden further and the multimodal structure observed in the noise-free case (cf.\ Figure~\ref{fig:marm-pdf-vertical}) becomes more pronounced, particularly at 20\% noise where the probability mass spreads over a wider range of velocities. This progressive degradation with noise level is consistent with the reduced signal-to-noise ratio and confirms that the method responds appropriately to the diminished information content of the data.

\begin{figure}[htbp]
\centering
\includegraphics[width=\textwidth]{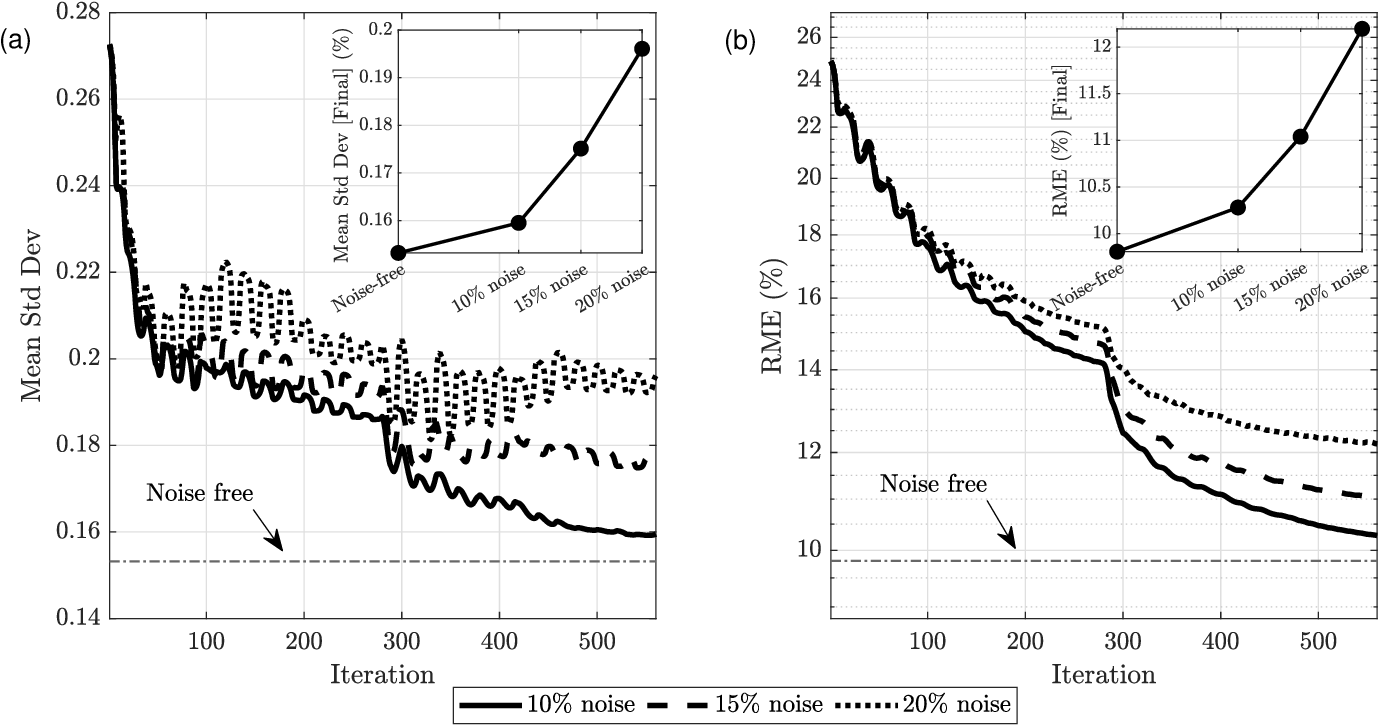}
\caption{Marmousi~II convergence at different noise levels. (a) Mean pointwise posterior standard deviation and (b) relative model error versus iteration.}
\label{fig:marm-noise-MSE}
\end{figure}

Figure~\ref{fig:marm-noise-MSE} summarizes the convergence behavior of the inversion under the three noise levels. The mean pointwise standard deviation (Figure~\ref{fig:marm-noise-MSE}a) decreases over iterations for all noise levels, confirming that the ADMM multiplier updates and frequency continuation schedule successfully drive posterior contraction regardless of the noise regime. This curve is a spatial average of the per-grid-point ensemble spread, and is therefore a global diagnostic of the aggregate spread rather than a location-specific uncertainty measure---much as the RME curve tracks the average model error rather than the error at any one point. The visible oscillation, most pronounced at higher noise, reflects the growing contribution of poorly constrained regions to this spatial average as the noise increases, together with the shift of the target posterior at each frequency transition of the multiscale schedule. However, the rate of contraction is slower and the asymptotic level is higher for noisier data: at 20\% noise, the standard deviation plateaus at a value substantially above the noise-free case, reflecting higher uncertainty. This moderate impact is a consequence of the adjoint-wavefield computation of equation~\ref{adj_wave}: the observed data are fitted only up to the estimated noise level through the compound parameter $\mu\sigma^2$, so the inversion is discouraged from fitting the noise itself---a behavior also reported by \citet{Aghazade_2025_APP}. The inset boxes in Figure~\ref{fig:marm-noise-MSE} provide a quantitative summary of the asymptotic behavior by reporting the final RME and mean pointwise standard deviation at convergence for each noise level. Examining these terminal values reveals a consistent trend: both quantities increase approximately linearly with the noise level. Since the posterior precision scales inversely with the noise variance $\sigma^2$ under a Gaussian likelihood, higher noise levels lead to broader posteriors and, consequently, larger pointwise standard deviations. The approximately linear relationship between noise level and terminal RME further suggests that the conditional mean degrades gracefully under increasing noise contamination---rather than exhibiting abrupt failure modes---corroborating the robustness of the conditional mean estimator discussed earlier.

\begin{figure}[htbp]
\centering
\includegraphics[width=\textwidth,trim={0cm 0cm 2cm 0cm},clip]{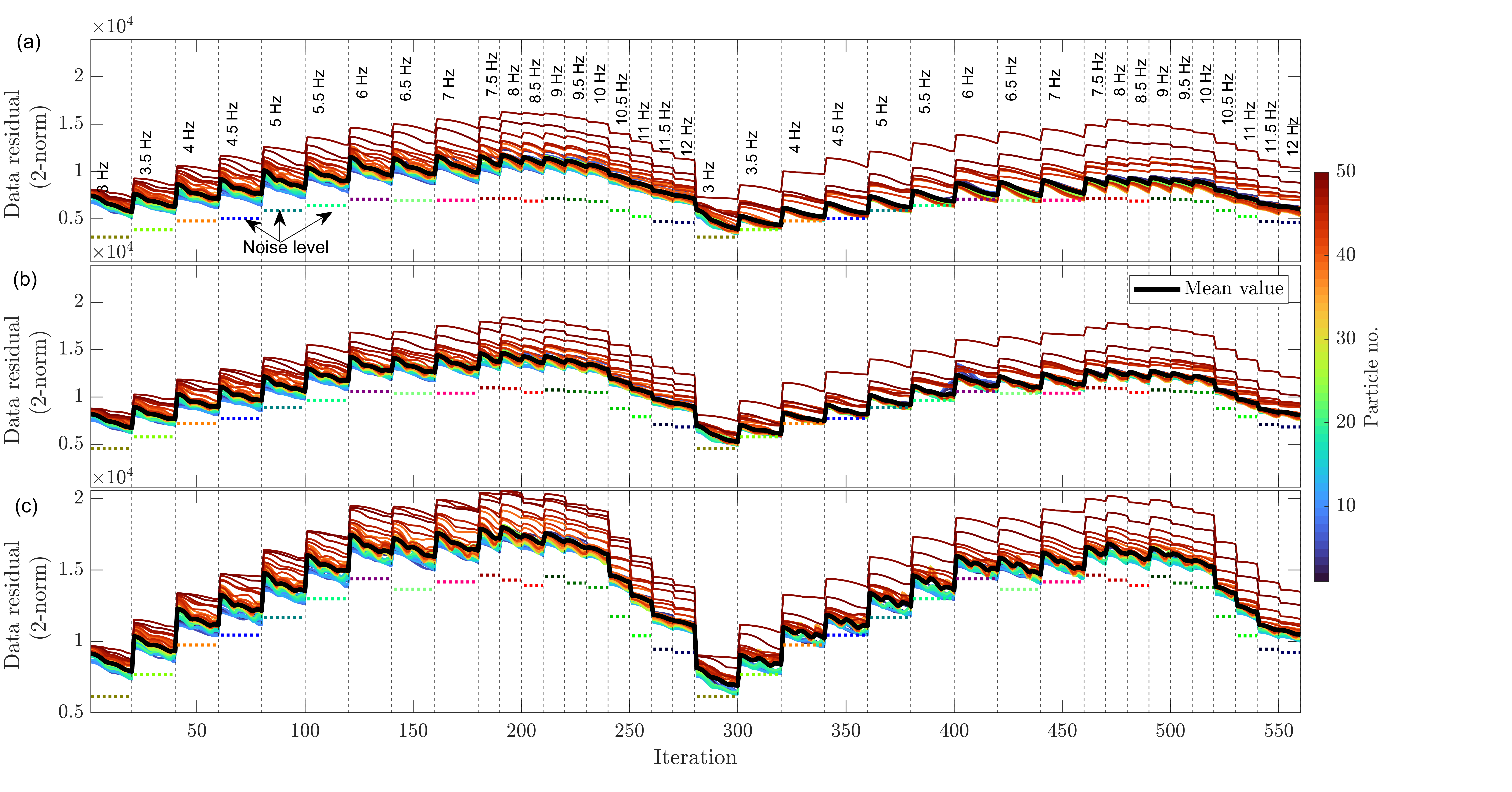}
\caption{Per-particle data residual (2-norm) versus iteration over two cycles, for (a) 10\%, (b) 15\%, and (c) 20\% noise. Particles colored by index (colorbar), ensemble mean black. Vertical dashed lines: frequency transitions, labeled atop (a). Horizontal dotted lines: per-frequency noise level, colored to match.}
\label{fig:marm-noise-misfit}
\end{figure}

Figure~\ref{fig:marm-noise-misfit} provides a detailed view of per-particle 2-norm data residual evolution across iterations and frequencies for the three noise levels, offering insight into the fitting behavior of the individual particles and the ensemble as a whole. Each colored line tracks an individual particle's residual---with color encoding the particle index---while the black solid line denotes the ensemble mean.  The residual curves exhibit a characteristic sawtooth-like pattern arising from the multiscale frequency continuation strategy: at each frequency transition the residual increases abruptly as higher-frequency data are introduced, then decreases monotonically within each stage as the particles adapt their velocity models accordingly. Across all noise levels and both cycles, the ensemble mean residual approaches but does not systematically undercut the noise level lines, confirming that ADMM-SVGD achieves a data fit that avoids overfitting. The broader particle distributions at higher noise levels are consistent with the wider posteriors reported in Figures~\ref{fig:marm-noise-image}--\ref{fig:marm-noise-MSE}. The markedly lower residual starting point at the beginning of cycle 2 across all panels further demonstrates the benefit of warm-starting from the cycle 1 ensemble, accelerating convergence within each subsequent frequency stage.

\section{Discussion}\label{sec:discussion}

The numerical results presented above demonstrate that ADMM-SVGD can produce physically consistent posterior samples for constrained inverse problems across a range of problem sizes and complexities. The progressive, multiplier-driven enforcement of the wave-equation constraint is evident in the convergence diagnostics, where the penalty term and constraint residual decay while the multiplier stabilizes; the nonparametric particle ensemble characterizes the multimodal, non-Gaussian posterior structure that emerges at depth on the Marmousi~II model and that a Gaussian approximation would miss; and the controlled Rosenbrock experiment confirms that the constrained reformulation recovers the correct posterior. The improved conditioning of the relaxed subproblems---long documented for the extended, AL formulation in deterministic inversion---carries over to the sampling setting: on Marmousi~II, reduced-space SVGD does not recover from an initial ensemble that is partly cycle-skipped, whereas ADMM-SVGD, started from the same ensemble, converges to a markedly more accurate posterior. Several aspects of the method merit further discussion.

The per-iteration computational and memory cost of ADMM-SVGD, its scaling with the number of particles and sources, and its relation to reduced-space methods are detailed in the Memory and computational cost subsection. We recall only that the embarrassingly parallel structure---all particle-source combinations can be computed independently---makes the method well-suited to modern parallel computing architectures.

The penalty parameter $\mu$ plays a dual role: it controls the rate at which the wave equation constraint is enforced and, through the probabilistic interpretation, determines the width of the augmented posterior relative to the true posterior. Large values of $\mu$ enforce the constraint rapidly but can make the augmented posterior too narrow in early iterations, potentially trapping particles in local modes. Small values slow convergence but allow broader exploration. As discussed earlier, the compound parameter $\mu\sigma^2$ is adaptively determined at each iteration using the residual whiteness principle \citep{Aghazade_2025_APP}, which removes the need for manual tuning and provides a robust balance between constraint enforcement and posterior exploration in all of our experiments.

Deterministic particle methods such as SVGD can struggle with multimodal posteriors when modes are separated by low-probability regions. The SVGD kernel repulsion maintains local diversity but cannot transport particles across distant modes once they have settled. This is an inherent limitation of deterministic transport methods compared to stochastic samplers such as MCMC \citep{MosegaardTarantola1995, MartinEtAl2012, ZhaoSen2021}. The comparison with reduced-space SVGD on the Rosenbrock example confirms that, for this low-dimensional problem, both SVGD variants produce comparable posterior approximations. The conditioning benefit of the ADMM splitting instead becomes apparent at the scale of FWI, where the tight coupling between model and wavefield variables renders the reduced-space posterior landscape substantially more ill-conditioned---as the comparison against reduced-space SVGD on Marmousi~II makes evident. For the FWI examples, where the posterior is expected to be more concentrated, the mode-crossing limitation did not noticeably affect the results. In applications where strong multimodality is anticipated, combining ADMM-SVGD with stochastic perturbations or tempering schemes may be beneficial.

The Marmousi~II experiment demonstrates that ADMM-SVGD can be applied to models with realistic complexity. The memory cost scales linearly with $N_p$, as each particle maintains its own set of auxiliary variables and multipliers. The number of particles must be chosen large enough to adequately represent the posterior---too few particles leads to underestimation of the posterior spread due to SVGD's mode-seeking behavior \citep{LiuWang2016}, as also observed in the frugal SVGD-based FWI framework of \citet{IzzatullahEtAl2024}. The particle counts used here represent a practical compromise between posterior fidelity and the cost of the per-particle wavefield solves; the reported uncertainties should accordingly be regarded as a lower bound on the true posterior spread. A systematic study of the particle count versus posterior calibration for large-scale FWI applications is an interesting avenue for future work. Alternatively, neural control variates based on Stein's identity \citep{SiahkoohiOh2025} could reduce the variance of posterior expectations computed from the particle ensemble, partially mitigating the effect of using a limited number of particles.

The performance of SVGD depends on the initial particle ensemble, a sensitivity that is intrinsic to finite-ensemble particle methods rather than to the constrained formulation. We follow the established practice in geophysical applications of initializing the particles from the prior \citep{ZhangCurtis2020, IzzatullahEtAl2024}---here, Gaussian random fields, augmented with a depth-increasing background for Marmousi~II---so that the initialization is a draw from the model we condition on rather than an arbitrary tuning choice. Several aspects of the method temper this dependence: the kernel bandwidth is re-estimated by the median heuristic at every iteration, rescaling the repulsion to the current spread of the ensemble; and the multiscale frequency continuation together with warm-starting places the particles in the appropriate basin before high-frequency refinement. The dependence nevertheless persists: because the particles are initialized by perturbing a one-dimensional background model rather than a converged deterministic estimate, a subset of them begins the inversion already cycle-skipped, and---since a deterministic transport method cannot move particles between disconnected basins---the ensemble characterizes the posterior associated with the basin that its initialization reaches. A systematic study at FWI scale, together with mitigations such as annealing or stochastic SVGD, is left for future work.

The experiments presented here are two-dimensional, but the formulation is dimension-agnostic and carries over to three dimensions without modification. The principal obstacle to a 3-D deployment is the per-particle Helmholtz solve: the sparse direct factorization that we exploit in 2-D---computed once per particle and reused across all sources---becomes prohibitively expensive in 3-D, where one would instead turn to matrix-free iterative solvers with dedicated preconditioning, such as shifted-Laplacian, sweeping, or domain-decomposition schemes. The remainder of the method is well suited to this regime: the particles and sources are mutually independent and embarrassingly parallel, and, as discussed above, the auxiliary variables can be streamed or distributed rather than held in memory simultaneously. That reduced-space SVGD has already been demonstrated for 3-D Bayesian variational FWI in the reduced formulation \citep{ZhangEtAl2023} indicates that particle-based inference is feasible at this scale; combined with the improved conditioning and progressive constraint enforcement of the dual-space formulation, this makes ADMM-SVGD a promising candidate for 3-D application, which we leave to future work.

The closest related work is that of \citet{FangEtAl2018}, who share our starting point of relaxing the PDE constraint for Bayesian inference. The key difference is that their Gaussian approximation centered at the MAP estimate \citep{BardsleyEtAl2014} is limited to unimodal posteriors, whereas our particle-based approach captures arbitrary posterior geometries, including the multimodality observed on Marmousi~II. Additionally, our ADMM multiplier updates progressively enforce the constraint without requiring careful a priori calibration of the penalty parameter. The split-and-augmented Gibbs sampler of \citet{Vono2019} uses a similar ADMM-type splitting but relies on exact conditional sampling; their dimension-free convergence analysis \citep{VonoPaulinDoucet2022} provides indirect theoretical support for our approach. More broadly, ADMM-SVGD connects to the growing body of work bridging optimization and sampling \citep{JordanEtAl1999, BleiEtAl2017, TrillosHosseiniSanzAlonso2023} and on Stein methods for geophysical inference \citep{LiuWang2016, ZhangCurtis2020}. Alternative approaches to Bayesian inference in seismic problems \citep{BuiThanhEtAl2013, ZhuEtAl2016} include deep-learning based methods \citep{AdlerOktem2018, SiahkoohiRizzutiHerrmann2022, BaldassariEtAl2023, YangSaadAlkhalifahWu2024, ZengEtAl2025}, amortized variational inference with normalizing flows \citep{RezendeMohamed2015, RadevEtAl2022, SiahkoohiEtAl2023, OrozcoEtAl2025}, and transport-map accelerated sampling \citep{ParnoMarzouk2018}. In contrast to the constrained, dual-space formulation developed here, a complementary line of work samples the reduced-space posterior directly---without relaxing the wave-equation constraint---through Markov chain Monte Carlo, including trans-dimensional \citep{GuoVisserSaygin2020}, Langevin-dynamics \citep{IzzatullahVanLeeuwenPeter2021}, and adaptive \citep{HuZhaoSen2025} samplers, as well as computationally efficient \citep{BertiAleardiStucchi2024} and learned-prior \citep{HuSenZhaoElmeliegyZhang2025} variants. The AL formulation has also been extended to time-domain FWI \citep{GholamiAghamiryOperto2022}, dual formulations \citep{AghazadeGholami2025, RizzutiEtAl2021}, time-lapse monitoring \citep{ZhangCurtis2024}, and differentiable programming frameworks \citep{LouboutinEtAl2023}.

Several directions remain for future work. First, preconditioning the SVGD updates with matrix-valued kernels \citep{WangEtAl2019svgd} or projected Newton directions \citep{ChenWuChen2019} could reduce the number of particles and iterations needed, alleviating the linear cost scaling with ensemble size. Second, extending the formulation to time-domain FWI---where broadband data can be processed in a single solve rather than sweeping over discrete frequencies---would reduce the number of outer iterations. Third, replacing the particle ensemble with normalizing-flow posteriors or neural-network-based priors could amortize the sampling cost, providing instant posterior samples for new data without iterative updates.

\section{Conclusion}\label{sec:conclusion}

We have presented ADMM-SVGD, a method for posterior sampling in constrained inverse problems that combines the ADMM with Stein variational gradient descent. By exploiting the probabilistic interpretation of the AL, the method translates hard PDE constraints into a sequence of evolving target distributions that are amenable to particle-based sampling, while the ADMM multiplier updates ensure that the constraints are progressively enforced. This dual-space approach inherits the favorable conditioning properties of wavefield reconstruction inversion while enabling principled UQ through the particle ensemble.

Numerical experiments on the Rosenbrock distribution, where the posterior is available in closed form, validated the method against reduced-space SVGD as a baseline without constraint splitting: the particle ensemble matches the analytic posterior and passes a simulation-based calibration test, which the unsplit sampler reproduces. At the scale of FWI, a comparison against reduced-space SVGD, and against deterministic inversion started from the same particles, showed that the dual-space formulation recovers from an initialization from which reduced-space sampling does not, and attains a substantially lower model error at an equal fit to the observed data. Application to frequency-domain FWI---for both a Gaussian anomaly model and the Marmousi~II benchmark---demonstrated that the method produces physically meaningful uncertainty estimates: pointwise standard deviations that are spatially correlated with estimation error, and posterior contraction with increasing data coverage---with the reported spread to be read as a lower bound on the posterior uncertainty on account of the finite particle ensemble. On the Marmousi~II model, the method revealed multimodal posterior structure at geologically complex locations, highlighting the value of full posterior characterization over single-point estimates.

\section*{Acknowledgments}

Ali Siahkoohi acknowledges support from the Institute for Artificial Intelligence at the University of Central Florida.

\bibliography{refs}

\end{document}